\documentclass{aa}
\usepackage{graphicx}
\usepackage{txfonts}
\usepackage[squaren]{SIunits}
\usepackage{makecell}
\usepackage{natbib}
\usepackage{booktabs}
\usepackage{tikz}
\usepackage{rotating}
\usepackage{ulem}
\usepackage{multirow}
\usepackage{placeins}
\usepackage{hyperref}
\usepackage[switch]{lineno}
\bibpunct{(}{)}{;}{a}{}{,}

\newcommand{\qnESix}{(2_{-1}-1_{-1})}
\newcommand{\qnESeven}{(2_{0}-1_{0})}
\newcommand{\qnEEight}{(2_{1}-1_{1})}
\newcommand{\qnTwHFSix}{(5_{0}-4_{0})}
\newcommand{\qnTwHFSeven}{(5_{-1}-4_{-1})}
\newcommand{\qnThHE}{(7_{0}-6_{0})}
\newcommand{\qnThHEOne}{(7_{-1}-6_{-1})}
\newcommand{\ratioESevenESix}{\ensuremath{I\qnESeven / I\qnESix}}
\newcommand{\ratioEEightESeven}{\ensuremath{I\qnEEight / I\qnESeven}}
\newcommand{\ratioEEightESix}{\ensuremath{I\qnEEight / I\qnESix}}
\newcommand{\ratioTwHFSevenTwHFSix}{\ensuremath{I\qnTwHFSeven / I\qnTwHFSix}}
\newcommand{\ratioThHEOneTHE}{\ensuremath{I\qnThHEOne / I\qnThHE}}

\begin{document}
    \title{CH$_3$OH as a User-Friendly Density Probe: Calibration and Beyond}

    \author{A. Giannetti\inst{1}
    \and S. Leurini\inst{2}
    \and E. Schisano\inst{3}
    \and V. Casasola\inst{1}
    \and T. G. S. Pillai\inst{4}
    \and C. Sanna\inst{5,2}
    \and S. Ferrada-Chamorro\inst{1}
    }

    \institute{INAF - Istituto di Radioastronomia di Bologna, Via Gobetti 101, 40129 Bologna, Italy\\
    \email{andrea.giannetti@inaf.it}
    \and
    INAF - Osservatorio Astronomico di Cagliari, Via della Scienza 5, I-09047 Selargius (CA), Italy
    \and
    INAF - Istituto di Astrofisica e Planetologia Spaziali, Via Fosso del Cavaliere 100, I-00133, Rome, Italy
    \and
    Haystack Observatory, Massachusetts Institute of Technology, 99 Millstone Road, Westford, MA 01886, USA
    \and
    Dipartimento di Fisica, Università degli Studi di Cagliari, S.P.Monserrato-Sestu km 0,700, I-09042 Monserrato (CA), Italy }

    \date{\today}

    \abstract
    {
        Almost all the physics of star formation critically depends on the number density of the molecular gas involved.
        However, the methods to estimate this keystone property often rely on very uncertain assumptions about the geometry of the molecular fragment, or depend on overly simplistic, uniform models, or require time-expensive observations to simultaneously constrain the gas temperature as well.
        An easy-to-use method to observationally derive the number density that is valid under realistic conditions is conspicuously absent, causing an evident asymmetry in how accurately the volume density is estimated, and how often dedicated tracers are used, compared to the gas temperature.
    }
    {
        To fill this gap, we propose and calibrate a versatile diagnostic tool based on methanol spectral lines that greatly simplifies the inference of molecular number density.
        Methanol is abundant in both cold- and hot gas, and has a dense spectrum of lines, maximising observational efficiency.  Therefore it can be applied to a wide variety of scales, from entire clouds to protostellar disks, and both in our Galaxy and beyond. Moreover, this tool does not need to be tailored to the specific source properties (such as distance, temperature, and mass).
    }
    {
        We construct large grids of clump models and perform radiative transfer calculations to investigate the robustness of different line ratios as density probes with different assumptions, also in the presence of density and temperature gradients.
    }
    {
        We find that the line ratios of the ($2_K-1_K$) band transitions around $96.7$~GHz are able to fully constrain the average number density along the line-of-sight within a factor of two-three in the range $\sim 5 \times 10^4 - 3 \times 10^7\usk\mathrm{cm^{-3}}$.
        The range can be extended down to a few times $10^3\usk\mathrm{cm^{-3}}$, when also using line ratios from the ($5_K-4_K$) and/or ($7_K-6_K$) bands, around $241.7\usk\mathrm{GHz}$ and $338.1\usk\mathrm{GHz}$, respectively.
        We provide the reader with practical analytic formulas and a numerical method for deriving volume density and its uncertainty from observed values of the line ratios.
    }
    {
        Thanks to our calibration of line ratios, we make the estimate of the number density much simpler, with an effort comparable or inferior to deriving excitation temperatures.
        By providing directly applicable recipes that do not need the creation of a full large velocity gradient model grid, but are equally accurate, we contribute to offsetting the disparity between these two fundamental parameters of the molecular gas.
        Applying our method to a sub-sample of sources from the ATLASGAL TOP100 we show that the material in the clumps is being compressed, accelerating in the latest stages.
    }


    \maketitle

    \section{Introduction}\label{sec:intro}
    To access the details of the star-formation (SF) process, we must know the physical properties (density and temperature) of the molecular gas and of the dust that will form new stars, how they are distributed in three dimensions \citep{Tassis07_mnras379_50, LiGoldsmith12_apj756_12, Tritsis+16_mnras458_789}, and how they evolve in time.

    Temperature is crucial to estimate the amount of gas and dust in the interstellar medium and to characterise its heating and cooling.
    On the other hand, the joint distributions of temperature, and gas number density play a major role in governing crucial processes in SF.
    For instance, chemical reaction rates are density- and temperature dependent.
    Density also governs the Jeans mass and the free-fall time of the fragment, that are crucial factors for its stability against collapse and its dynamical timescales.
    Furthermore, energy losses for the particles of cosmic rays depend on density, changing their efficiency in ionising the gas, and initiating chemistry in cold gas.
    The velocity field is measured through molecular lines, the intensity of which is determined by both density and temperature influencing measurements of the turbulence spectrum.
    Similarly, both quantities influence measurements of accretion onto molecular fragments, because the velocity field and the occurrence of self-absorption depend on density and temperature.
    These connections and effects have been discussed in a number of studies \citep[e.g.][]{Roberts+2000_aa361_388, Caselli+1999_apj_523_165, Caselli+2002_apj565_344, Padovani+2009_aa501_619, RomanDuval+2011_apj740_120, Krumholz+2012_ApJ745_69, VazquezSemadeni+2019_mnras490_3061, VazquezSemadeni+2024_mnras530_3445}.
    Moreover, the comparison between independent measurements of the column density and the average number density of molecular condensations provide a way to estimate the size of these structures along the line-of-sight \citep[LOS; e.g.][]{LiGoldsmith12_apj756_12}, and permit to reconstruct the 3D distribution of the gas.
    This information and the collapse timescale are relevant to determine the dominant mode of SF \citep[e.g.][]{Tassis07_mnras379_50, Bovino+21_aa654_34, Sabatini+21_aa652_71}.
    If collapse is hindered by magnetic fields or strong turbulence, long timescales compared to the free-fall time are expected \citep[e.g.][]{Mouschovias+1991_apj373_169}.
    Instead, if gravity is dominating the collapse should proceed on timescales comparable to the local free-fall time \citep[e.g.][and references therein]{VazquezSemadeni+2019_mnras490_3061}.
    Analogously, different formation mechanisms have different predictions for the intrinsic shape of molecular fragments.
    The shape of molecular fragments is predicted to be triaxial and prolate if they form from turbulent converging flows \citep[e.g.][]{Li+04_apj605_800}, whereas they should be mostly oblate when magnetic fields are important \citep[e.g.][]{CiolekBasu2006_apj652_442}.
    Therefore, accurate measurements of both temperature and density are essential, because they determine the free-fall timescale, and they can be used to infer the degree of support against gravity and the intrinsic shape of molecular fragments.

    Despite their importance, accurately measuring temperature and density in molecular fragments presents observational challenges.
    Temperature is the most straightforward parameter to measure.
    Lines of symmetric-top molecules, like NH$_3$, with different excitation properties are commonly used for this \citep[e.g.][]{WalmsleyUngerechts1983_aa122_164, Cummins+1983_apj266_331, Askne+1984_aa130_311}.
    On the other hand, estimates of the number density of molecular hydrogen ($n_\mathrm{H_2}$) often require a measure of the temperature \citep{Walmsley87_ASIC210_161}.

    The usual procedure to infer $n_\mathrm{H_2}$ is to derive the gas column density of molecular hydrogen from dust continuum observations (under the assumption of a given temperature), assume a geometry for the cloud to derive its depth along the line-of-sight, and divide the column density by this estimated size \citep[e.g.][]{Rigby+2019_aa632A_58, Sanhueza+2019_apj886_102, Redaelli+2021_aa650_202}.
    Density can also be estimated with molecular lines.
    For simple linear species (CS, HC$_3$N, etc.) observations of several transitions at very different wavelengths are required.
    In this case, number density can be estimated by identifying the critical value of the rotational quantum number $J$ corresponding to the brightest transition, for which number of collisions and spontaneous radiative de-excitations roughly balance.
    However, observing multiple transitions at different wavelengths implies the need to consider calibration uncertainties, comparatively large integration times, and excitation effects \citep[see, e.g., the discussion in][]{LiGoldsmith12_apj756_12}.

    Slightly-asymmetric rotors, like formaldehyde (H$_2$CO) or methanol (CH$_3$OH), have spectral lines whose relative intensities depend both on $T$ and $n_\mathrm{H_2}$ \citep[e.g.][]{Mangum+93_apjs89_123, Leurini+04_aa422_573, Tang+18_aa611_6}.
    Line combinations can be strategically selected to minimise the dependence on temperature or density, by leveraging on forbidden transitions as in the case of symmetric tops, and fast radiative de-excitation and/or similar excitation properties.
    Moreover, these diagnostics are more time-efficient and robust than linear molecules, because multiple lines often fall within a narrower frequency range, reducing the number of required observational setups and the impact of calibration uncertainties.

    Methanol is a widespread and abundant molecule, that traces both cold- and hot material \citep[e.g.,][]{Giannetti+17_aa603_A33, Sanna+2021_aa655_72}, in the Milky Way and in external galaxies \citep[e.g.][]{Henkel+1987_aa188_1, Shimonishi+18_apj862_102, Sewilo+2018_apjl853_19, Humire+2020_aa633_106, McCarthy+2020_mnras491_4642, Humire+2022_aa663_33}.
    Combinations of CH$_3$OH lines can be used to probe both density and temperature.
    The consequence is that this species can be used to conveniently explore the physical properties of a significant variety of environments.
    \citet{Leurini+04_aa422_573} proposed specific combinations of CH$_3$OH lines within the same spectral band that exhibit ratios with a strong density dependence, while showing a very weak dependence on the gas temperature.
    These lines (see Table~\ref{tab:lines_list}) are carefully selected to have similar excitation conditions (energies above the ground state, and critical densities), ensuring that they originate from the same emitting region.
    As a result, their ratios are unaffected by differences in gas distribution and in the specific line excitation conditions.
    However, the diagnostic power of the ratios was assessed by \citet{Leurini+04_aa422_573} using a simplified model, consisting of a spherical clump with constant density, temperature, and abundance, only for the optically-thin case.

    With the aim of providing an easy-to-use tool to analyse the emission of specific CH$_3$OH transitions, we extended the work of \citet{Leurini+04_aa422_573} to more realistic conditions, including radial variations of density and temperature.
    This allows us to test if these ratios preserve their observational diagnostic potential, and to assess their robustness, in case of density and temperature stratification along the LOS.
    We also calibrated the ratios against the average value of the volume density along the LOS, and discussed the differences with the uniform and isothermal case.
    To improve the accuracy of the final density estimates we efficiently increased the sampling of the parameter space explored by our grid of models four-fold through emulation of the radiative transfer results using machine learning (ML).
    We specifically trained decision tree-based algorithms, that are very robust and efficient.
    They can be trained even with relatively small training sets and generally take only a fraction of the time needed to train deep neural networks, while still providing state-of-the-art predictions \citep[e.g.][]{xgboost}.

    In Sect.~\ref{sec:methanol_lines_modeling} we show how we built realistic clump models and simulate their emission in methanol lines, the details of the clump model, the procedure followed to compute the line ratios, and how we derive the H$_2$ number density from the ratios.
    Section~\ref{sec:results} summarises the results that we obtained interpreting and calibrating the line ratios with a series of different experiments, comparing them to simpler modeling. We also discuss their robustness as number density indicators, and how we used machine learning to improve the final density estimates.
    In Sect.~\ref{sec:analytic_representation} we provide empirical analytic approximations to directly estimate the volume density from the observed line ratios.
    In Sect.~\ref{sec:poc} we show the application of our analysis to selected high-mass star-forming clumps from the ATLASGAL TOP100 sample \citep{Giannetti+2014_aa570_65}, for which we estimate the number density and investigate its time evolution during the first stages of the process of high-mass SF.
    Finally, Sect.~\ref{sec:conclusions} lists our conclusions.

    \section{Simulating CH$_3$OH lines ratios: models and tools}\label{sec:methanol_lines_modeling}

    We selected methanol lines from accessible regions of the millimeter spectrum.
    These lines are very close in frequency ($\sim 5-200 \usk\mathrm{MHz}$, cf. Table~\ref{tab:lines_list}), so that they can be observed with a single spectral setup by most telescopes.
    To minimise the impact of temperature on the ratios, and ensure that line emission originates from the same volume of gas we consider ratios between transitions with similar energies above the ground state and critical densities.
    Our first objective is to calibrate the \ratioESevenESix\footnote{$I(J_K-J'_{K'})$ is the integrated-line intensity of the  $J_K-J'_{K'}$ methanol transition}, \ratioEEightESix, \ratioEEightESeven, \ratioTwHFSevenTwHFSix, and \ratioThHEOneTHE\ ratios against the average density along the LOS for realistic models of molecular condensations.

    To pursue this objective we have produced a tool to simplify the creation of realistic molecular fragment models, simulate and analyse their molecular line emission.
    We exploited this tool to model the line ratios listed above.
    We will refer to it as the Swiss Army Knife, or SAK for short, following the nickname given to methanol by \citet{Leurini+05_astrochemistry231_99}, to emphasize the tool versatility in modeling and analyzing molecular line emission, similar to the versatility of methanol in tracing molecular gas properties.
    A more detailed description of its capabilities, of its infrastructure and automation is given in Appendix~\ref{app:automation} and in the repository documentation\footnote{\url{https://www.ict.inaf.it/gitlab/andrea.giannetti/swiss_army_knife_stable}}.

    \subsection{SAK infrastructure}\label{subsec:infrastructure}

    We began by developing a model generator which handles physical model setup, performs radiative transfer computations, and summarizes the results for  easy inspection.
    The system for model building and post-processing, as well as for line ratio computation, is designed to guarantee reproducibility, data conservation, and structuring.
    This design also minimizes the risk of errors during data retrieval and analysis.

    We created a container to run SAK, so that the environment for execution is easily replicable and accessible on different machines, both locally and on distributed systems.
    Containers are a way to isolate software, increase its portability and the reproducibility of the results, ensuring that users can run SAK without needing to install or manage any dependencies.

    \subsection{Model generation}\label{subsec:model_generation}

    The system for generating the physical grids can currently handle single-cell models, or spherical clumps, with power-law distributions in number density and temperature.
    For spherical sources, the density and temperature profiles can be described by \citep[e.g.][]{vanderTak+2000_apj537_283, Lin+21_aa658_128}:
    \begin{equation}
        n_\mathrm{H_2} = n_{\mathrm{H_2},0} \left ( \frac{r}{r_0} \right ) ^ p,\label{eq:pl_density_profile}
    \end{equation}
    \begin{equation}
        T = T_{0} \left ( \frac{r}{r_0} \right ) ^ q,\label{eq:pl_temperature_profile}
    \end{equation}
    where $r$ is the distance from the fragment centre, $r_0$ is a scaling radius, $n_{\mathrm{H_2},0}$ and $T_0$ are the values of the number density and temperature at the scaling radius, respectively.
    The central grid cell is replaced with the maximum value in the neighbour cells to avoid the singularity at clump centre, present in case of a negative exponent in Eq.\ref{eq:pl_density_profile} and Eq.\ref{eq:pl_temperature_profile}.
    Our choice of the temperature profile was motivated by the fact that clumps do not host single stars, but a cluster.
    A power-law reproduces well the radial temperature variation observed in these regions \citep{Lin+21_aa658_128}.
    Simulating an embedded cluster by sampling the initial mass function, assuming the spectrum of the individual protostars, and then computing the temperature profile consistently is beyond the scope of the present paper.

    The radiative transfer is performed with RADMC 3D V2.0 \citep{radmc3d}.
    SAK automates the radiative transfer computation by generating all necessary input files for RADMC 3D and then executing the code.
    Because we want to explore the sensitivity to number density, we perform non-LTE computations, using the large velocity gradient (LVG)+escape probability algorithm as described in the RADMC 3D documentation.
    To use this method, we have assumed a solid-body rotation of the clump, with a gradient of $2 \usk \mathrm{km\usk s^{-1}\usk pc^{-1}}$, and a typical length scale for escape probability equal to the diameter of the clump, as listed in Table~\ref{tab:clump_properties}.
    We adopted the collisional excitation rates for H$_2$ \citep{Rabli+10_mnras406_95} provided by the LAMDA database \citep{lamda}.
    The gas and dust are assumed to be thermally coupled.
    Because the collisional coefficients are only available for para-H$_2$ in LAMDA, we assumed that no difference exists between ortho- and para-H$_2$.
    Also, if the temperature is higher than the maximum one listed ($200\usk\mathrm{K}$), the coefficients were assumed to remain constant.

    \begin{table}[tb]
        \caption{\label{tab:lines_list}Properties of the methanol lines considered in this work.}
        \small
        $$
        \begin{array}{ccccc}
            \hline
            \hline
            \mathrm{Transition}      & \mathrm{Rest \usk Frequency} & A_{ij}                & E_{up}    & n_{crit,ij} \\
                                     & [\giga\hertz]              & [\second^{-1}]        & [\kelvin] & [\mathrm{cm^{-3}}] \\
            \hline
            (2_{-1} - 1_{-1})        & 96.739358                  & 2.557794\times10^{-6} & 12.5      & 2.8\times10^4 \\
            (2_{0} - 1_{0})          & 96.744545                  & 3.407341\times10^{-6} & 20.1      & 3.1\times10^4 \\
            (2_{1} - 1_{1})          & 96.755501                  & 2.624407\times10^{-6} & 28.0      & 2.7\times10^4 \\
            (5_{0} - 4_{0})          & 241.700159                 & 6.036633\times10^{-5} & 47.9      & 5.5\times10^5 \\
            (5_{-1} - 4_{-1})        & 241.767234                 & 5.806215\times10^{-5} & 40.4      & 4.8\times10^5 \\
            (7_{0} - 6_{0})          & 338.124488                 & 1.696120\times10^{-4} & 78.1      & 1.2\times10^6 \\
            (7_{-1} - 6_{-1})        & 338.344588                 & 1.666997\times10^{-4} & 70.6      & 1.2\times10^6 \\
            \hline
        \end{array}
        $$
        \tablefoot{The transitions are indicated as $(J_K - J'_{K'})$. The Einstein coefficient for spontaneous emission, the energy of the upper level above the ground state, and the critical density of the transitions are indicated as $A_{ij}$, $E_{up}$, and $n_{crit,ij}$, respectively.}
    \end{table}

    \subsection{Clump model details}\label{subsec:model_details}
    The parameters for our clump model are reported in Table~\ref{tab:clump_properties} and \ref{tab:grids}.
    The values of these parameters resemble the typical properties of clumps identified in Galactic plane surveys \citep[e.g.][]{Urquhart+18_mnras473_1059, Elia+2021_mnras504_2742}.
    We chose to focus our analysis on the case of high-mass clumps, since providing a more precise estimate of the dynamical timescales for these structures have immediate consequences for cluster formation.
    Extensive datasets targeting these systems are available for our Galaxy  \citep[e.g.,][]{Urquhart+18_mnras473_1059, Elia+2021_mnras504_2742} but also for external galaxies, where high-resolution observations start to reach linear scales that approach the size of clumps \citep[$\approx 0.1-1\usk\mathrm{pc}$, e.g.][]{Indebetouw+2013_apj774_73, Muraoka+2020_apj903_94}.

    \begin{table*}[tb]
        \caption{\label{tab:clump_properties}Basic properties of the simplified model of clumps.}
        \centering
        \small
        \begin{tabular}{cccccccc}
            \hline
            \hline
            \thead{Grid \\size} & \thead{Grid\\elements}      & \thead{Outer \\radius} & \thead{Solid-body \\rotation gradient}                         & \thead{GTD \\ratio} & \thead{Image \\pixels} & \thead{Velocity \\resolution}            & Channels \\
            $\mathrm{[pc]}$     &                             & [pc]                   & $\mathrm{[\kilo\metre\usk\second^{-1} \usk pc^{-1}]}$ &                     &                        & $\mathrm{[\kilo\metre\usk\second^{-1}]}$ &          \\
            \hline
            2                   & 101                         & 0.9                    & 2                                                     & 100                 & 101                    & 0.1                                      & 100 \\
            \hline
        \end{tabular}
        \tablefoot{The grid size, number of grid elements, and image pixels are given per dimension; all dimensions have equal properties. The gas-to-dust ratio is indicated as GTD ratio.}
    \end{table*}

    \begin{table*}[tb]
        \caption{\label{tab:grids}Parameter space covered by the grids in our experiments.}
        \centering
        \small
        \begin{tabular}{cccccccc}
            \hline
            \hline
            $T_0$              & $n_{\mathrm{H_2},0}$           & $\chi$                   & $\chi$ profile      & $p$                & $q$       & \thead{FWHM \\microturb.}                       & \thead{$L_{bol}$\\cluster} \\
            $[K]$              & $log_{10}(\mathrm{[cm^{-3}]})$ &                          &                     &                    &           & $\mathrm{[\kilo\metre\usk\second^{-1}]}$ & $log_{10}(\mathrm{[L_\odot]})$ \\
            \hline
            $10 - 32.5; (2.5)$ & $3 - 6.82; (0.28)$             & $10^{-10}-10^{-8}; (10)$ & Const., Step funct. & $-1.2, -1.5, -1.8$ & $0, -0.5$ & $3.5, 7.0$                                                   & $3.6-6.3$ \\

            \hline
        \end{tabular}
        \tablefoot{Where a regular grid is used, number between parentheses indicates the step size, otherwise all the values considered are indicated as a list. The abundance is indicated as $\chi$ and the bolometric luminosity of the embedded cluster as $L_{bol}$.}
    \end{table*}

    We assumed a fiducial value of $p=-1.5$ for the density structure (see Eq.~\ref{eq:pl_density_profile}), as observed in galactic clumps \citep{Beuther+02_apj566_945, Giannetti+13_aa556_16}, and  scaling radius $r_0=0.5\usk\mathrm{pc}$.
    We explored regularly-spaced values of the scaling density $n_{\mathrm{H_2},0}$ between $10^3\usk\mathrm{cm^{-3}}$ and $\sim 6.6 \times 10^6\usk\mathrm{cm^{-3}}$, spaced by a factor of three.
    A ceiling is set for the central density of $10^8 \usk\mathrm{cm}^{-3}$, compatible with the modeling of clumps where clusters are forming.
    Grid cells for which the density value would exceed this threshold, are set to $10^8 \usk\mathrm{cm}^{-3}$ instead.

\begin{figure}
        \centering
        \includegraphics[width=0.95\hsize]{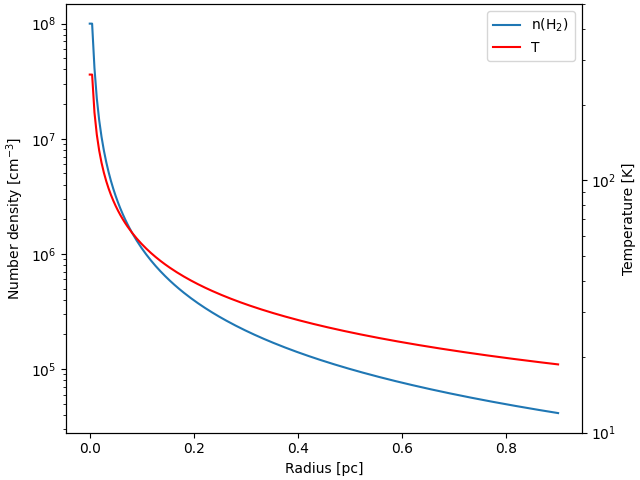}
        \caption{Example of the radial density and temperature profiles inside the clump, for $n_{\mathrm{H_2},0}=10^5\usk\mathrm{cm^{-3}}$ and $T_0=25\usk\mathrm{K}$.}
        \label{fig:profiles}
    \end{figure}

    The presence of a luminous Young Stellar Object (YSO) embedded at the centre of the clump, and heating the material from within was simulated by introducing the temperature gradient of Eq.~\ref{eq:pl_temperature_profile}, with a slope $q=-0.5$, that assumes optically-thin dust.
    The value chosen for the exponent has been observed in the inner regions of high-mass cluster-forming regions \citep{Lin+21_aa658_128}.
    Again we set a ceiling, for the temperature value to $2000\usk\mathrm{K}$.
    An example of the assumed density and temperature profiles is shown in Figure~\ref{fig:profiles}.
    The assumed temperature profile is not independent of the bolometric luminosity of the embedded cluster \citep{Wilner+95_apjl449_l73}.
    Adapting Equation~1 of that work for our parameters, the luminosity of the embedded cluster can be expressed as:
    \begin{equation}
        \frac{L}{L_\odot} = \left ( \frac{T}{1.25} \right )^4 \left ( \frac{R}{0.5 \usk \mathrm{pc}} \right )^2;
    \end{equation}
    Considering the grid limits, our models span a range of bolometric luminosities from $4 \times 10^3 \usk\mathrm{L_\odot}$ to $2.2 \times 10^6 \usk\mathrm{L_\odot}$.
    These values are typical of observations of high-mass star-forming clumps \citep[e.g.][]{Elia+2017_mnras471_100, Urquhart+18_mnras473_1059}, encompassing the luminosities associated with massive clusters in the MW.
    This validates our grid for the analysis of these objects.

    The final number of models per grid is $630$.
    We repeated the calculations for two additional grids, with $p=-1.2$ and $p=-1.8$ to investigate the effect of a different density distribution of gas on the ratios.
    Table~\ref{tab:grids} reports the entire parameter space that was covered by the grids in our experiments, for a total of $9450$ individual simulated clumps.

    By assuming a power-law distribution in density and temperature, our clump models simulate the effects of the envelope on the line ratios.
    We remark that common radiative transfer tools, like RADEX \citep{VanderTak+2007_aa468_627}, cannot account for density and temperature variations along the line of sight, leaving a potentially large source of uncertainty unknown.

    \subsection{Line ratio computation}\label{subsec:ratio_computation}

    We simulated the seven lines in Table~\ref{tab:lines_list}, and computed the following ratios of integrated intensities: \ratioESevenESix, \ratioEEightESeven, \ratioEEightESix, \ratioTwHFSevenTwHFSix, and \ratioThHEOneTHE.
    Two different methods are considered for investigating the ratios.
    We compute the line ratios averaged over the entire clump, mimicking single-dish observations of high-mass clumps in the Milky Way or high-resolution imaging of similar objects in nearby galaxies.
    The line ratios were also derived along each individual LOS across the clumps in the final simulated images ($101 \times 101$ pixels, Table~\ref{tab:clump_properties}).
    Calibrating the relation between average number density along the LOS and the line ratios is key for deriving accurate density profiles from observations and reconstructing the 3D structure of the fragments.
    As we will see in Sect.~\ref{sec:robustness_distribution_distance}, the two methods give consistent results as a function of average density.

    The integrated line fluxes from all pixels in each line map were summed, and then the ratio of these summed fluxes was computed to obtain the integrated line ratio for the entire clump.
    The maps of the LOS-averaged physical properties (number density and temperature) were produced by a weighted mean of the grid cells values along the line of sight.
    We used as weights the number density of methanol in the individual grid cells.
    This weighting scheme is naturally motivated by the fact that in the optically thin case, the line intensity is proportional to the number of molecules.
    Thus cells with more methanol molecules contribute more to the observed ratio.
    Furthermore, when both lines in a ratio are heavily optically thick, the ratio tends to a constant and its sensitivity to the volume density disappears.

    \subsection{Inferring the H$_2$ number density from the line ratios}\label{subsec:density_inference}

\begin{figure*}
        \centering
        \includegraphics[width=0.9\hsize]{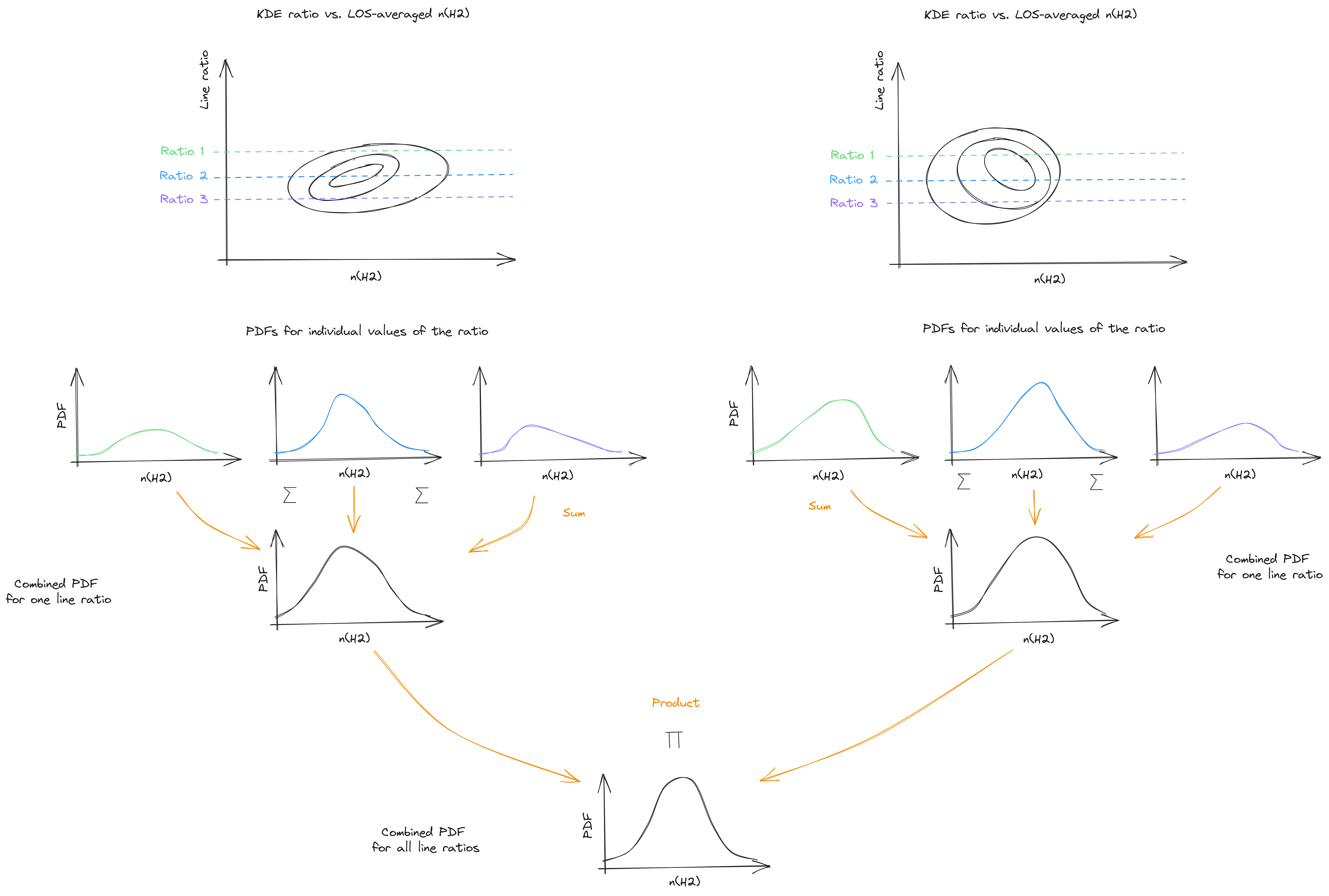}
        \caption{Schema of the procedure to derive the final density posterior from the combination of several line ratios. Two distinct ratios are considered in this example, the relations of which with $n(\mathrm{H_2})$ are shown in the top row. After generating a number of ratios, according to observations and uncertainties, the PDFs of individual ratios are summed, and then the resulting distribution for each line ratio is multiplied to obtain the final number density posterior.}
        \label{fig:density_posterior_from_pdf}
    \end{figure*}

    To derive the number density, given the methanol line ratios, we have used the kernel density estimate (KDE), a non-parametric method to estimate the probability density function.
    We have computed the KDE with the SKLearn package \citep{scikit-learn}, and we leveraged the Bayes theorem to extract the most probable value of $n_\mathrm{H_2}$ and its uncertainty.
    Each marginal probability is computed by first generating 1000 realizations of the two integrated line intensities to compute 1000 realizations of the ratio.
    These are used to extract the probability density function (PDF) for that ratio from the KDE, and then summed for each realization to obtain the final PDF:
    \begin{equation}
        \label{eq:marginalized_pdf}
        P_{jk}(n_\mathrm{H_2}|D) \propto \sum_{i=1}^{1000} P_{jk}(n_\mathrm{H_2}, R_{jk, i} | D),
    \end{equation}
    where $P_{jk}(n_\mathrm{H_2}, R_{jk, i}|D)$ is the ratio - number density PDF for the ratio $jk=I(J_K-J'_{K'})/I(J''_{K''}-J'''_{K'''})$, $P_{jk}(n_\mathrm{H_2}|D)$ is the marginalized PDF for the number density from ratio $jk$, and $R_{jk, i}$ is the $i$th realization of the ratio $jk$.
    The PDFs for the different ratios are combined by multiplication to obtain the final posterior, by considering the marginal PDFs for each ratio $jk$:
    \begin{equation}
        \label{eq:combined_pdf}
        P(n_\mathrm{H_2}|D) \propto \prod_{jk=1}^{5} P_{jk}(n_\mathrm{H_2} | D),
    \end{equation}
    because all the ratios provide independent information, and the model needs to reproduce all the available ones at the same time.
    A basic schema of the procedure is given in Fig.~\ref{fig:density_posterior_from_pdf}.
    Note that calibration uncertainties do not affect this combination, as they cancel out in each of the ratios.
    The tool derives from the posterior the highest-probability-density (HPD) interval for a given probability mass (i.e. the total probability within the interval), so that it is possible to estimate the uncertainty of the best-fit value, assumed to be the one corresponding to the maximum probability density.

    \section{Calibrating CH$_3$OH line ratios as density probes with density and temperature stratification}\label{sec:results}

    To investigate the robustness of methanol line ratios as density probes under varying assumptions, we developed a series of models, progressively adding layers of complexity and evaluating their impact on the general behaviour of the line ratios.
    The sets are designed to explore: 1. constant density and temperature (Appendix~\ref{subsec:uniform}), used as a benchmark with the results from \citep{Leurini+04_aa422_573}; 2. isothermal clumps with power-law density distributions (Appendix~\ref{subsec:power_law_density}), as an intermediate step to isolate the effect of the clump density structure; 3. internally-heated clumps with a temperature gradient (Appendix~\ref{subsec:fiducial_model}) and a constant abundance; 4. thermal sublimation from ices on dust grains (Appendix~\ref{subsec:hot_core}), and 5. different column densities and line width (Appendix~\ref{subsec:optical_depth}), which explore the effect of optical depth on the proposed diagnostics.

    For the radiative transfer we are only considering the E symmetry state, because the E and A states do not readily interconvert under typical interstellar conditions, and can thus be considered different species.
    In most of our experiments we assume a constant abundance of E-CH$_3$OH equal to $10^{-9}$ \citep{vanderTak+2000_aa361_327, Sabatini+21_aa652_71}.
    In fact, the low angular resolutions typical of clump surveys makes observations mostly sensitive to the cold methanol in the envelope.
    This abundance is intentionally chosen to be lower than what is commonly reported in star-forming regions for the total methanol abundance \citep{Giannetti+17_aa603_A33, Sabatini+21_aa652_71}.
    The choice is justified by the fact that we are considering only the E symmetry state, and that the abundances reported in literature are estimated assuming LTE, that typically tend to overestimate the total column density of methanol (\citealt{Leurini+04_aa422_573}, and Sect.~\ref{sec:tau_density_degeneracy}).
    Combined, the two effects account for nearly one order of magnitude.
    To investigate the effect of this assumption, we consider two additional sets: set 4, where we consider an abundance jump of a factor 100 in the inner core due to ice sublimation, and set 5 where we adopt abundance values of $10^{-8}$ and $10^{-10}$.

    We discuss below the results from the comparison between the observed line ratio and the number density of the emitting material, for different temperature conditions.
    For this we present the results derived from the analysis from each individual LOS, which represents a large ensemble of density and temperature conditions ideal to investigate the relevance and general trends of the proposed ratio.
    We also discuss these quantities derived from the clump-average, that are directly comparable with available survey observations.

    \begin{table}[tb]
        \caption{\label{tab:correlation_coefficients}Spearman correlation coefficients between the line ratios considered in this work, the LOS-averaged number density and temperature.}
        \small
        $$
        \begin{array}{ccc}
            \hline
            \hline
            \mathrm{Line\usk ratio}   & \rho_{<n_{\mathrm{H_2}}>} & \rho_{<T>}  \\
            \hline
            \ratioESevenESix       & 0.86                      & 0.06        \\
            \ratioEEightESix       & 0.88                      & 0.29        \\
            \ratioEEightESeven     & 0.85                      & 0.40        \\
            \ratioTwHFSevenTwHFSix & 0.09                      & 0.03        \\
            \ratioThHEOneTHE       & 0.65                      & 0.23        \\
            \hline
        \end{array}
        $$
        \tablefoot{The dependence on temperature is always a second-order effect which have no impact on the density inference, as can seen in Figure~\ref{fig:los_map_summary}.}
    \end{table}

    \subsection{Robustness of the ratios to density and temperature gradients}\label{sec:robustness_ratios_density_temperature}
\begin{figure*}
            \centering
            \begin{tikzpicture}[font=\sffamily]
                \node at (-7,0) {};
                \node at (7,0) {};
                \node[anchor=center] at (2.2,0) {\tiny Internally-heated clump (RT)};
                \node[anchor=center] at (-4.1,0) {\tiny Isothermal clump};
                \node[anchor=center] at (8.4,0) {\tiny Internally-heated clump (RT+ML)};
            \end{tikzpicture}\\
            \includegraphics[width=0.32\hsize]{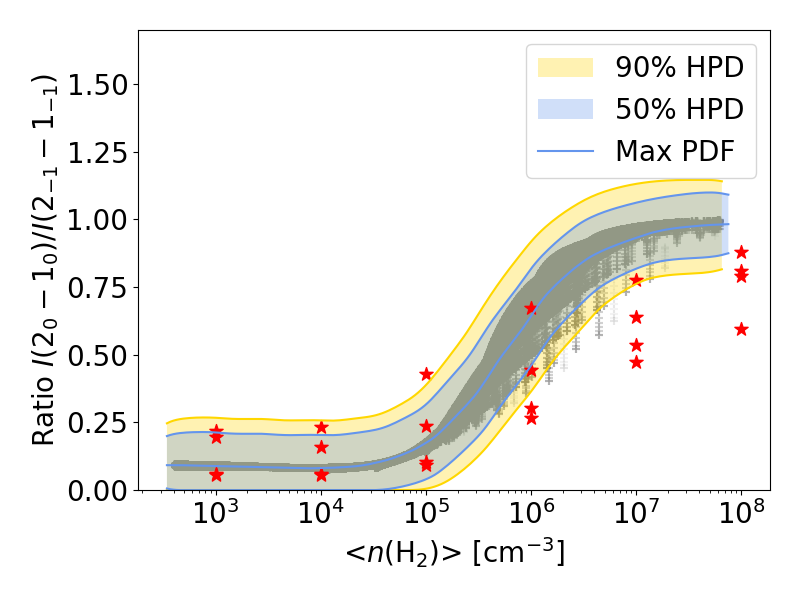}\hfill
            \includegraphics[width=0.32\hsize]{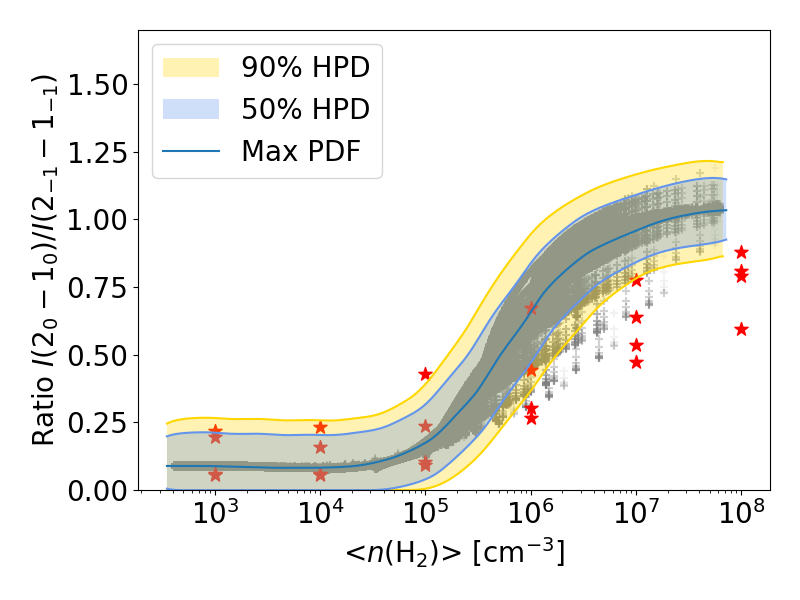}\hfill
            \includegraphics[width=0.32\hsize]{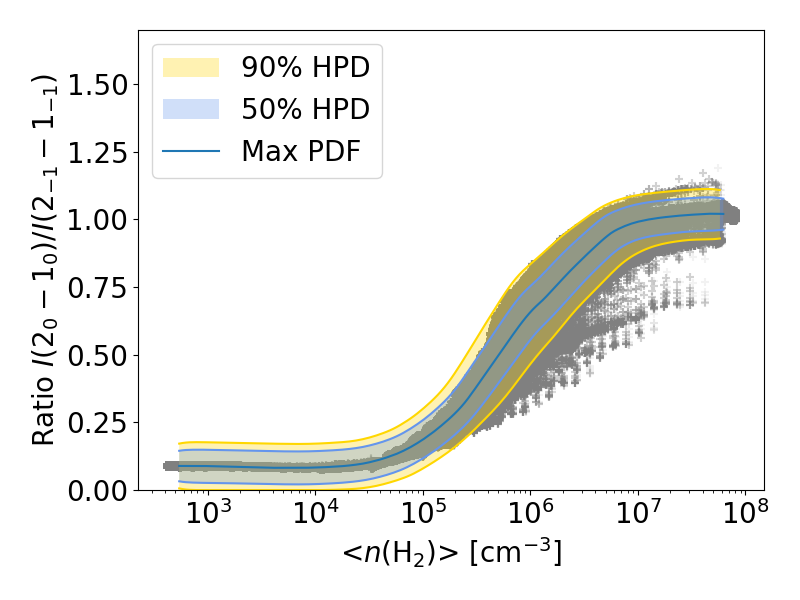}\\
            \includegraphics[width=0.32\hsize]{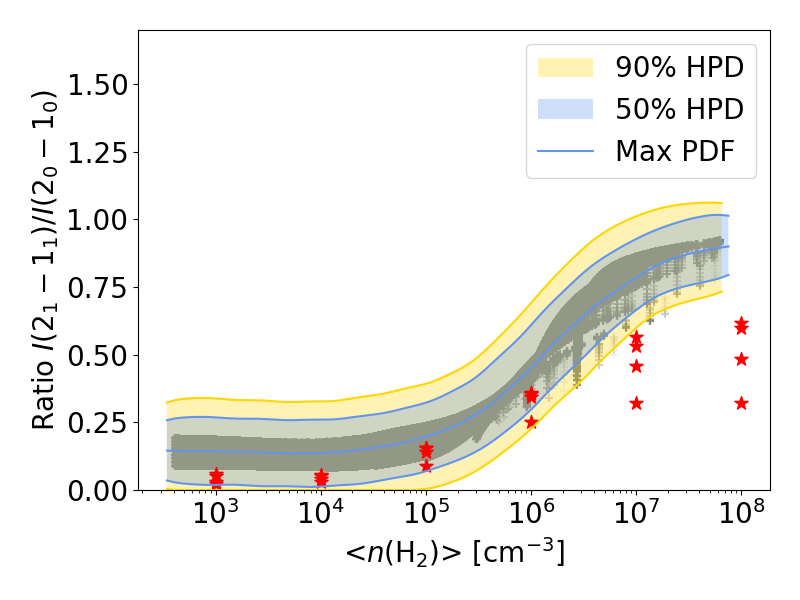}\hfill
            \includegraphics[width=0.32\hsize]{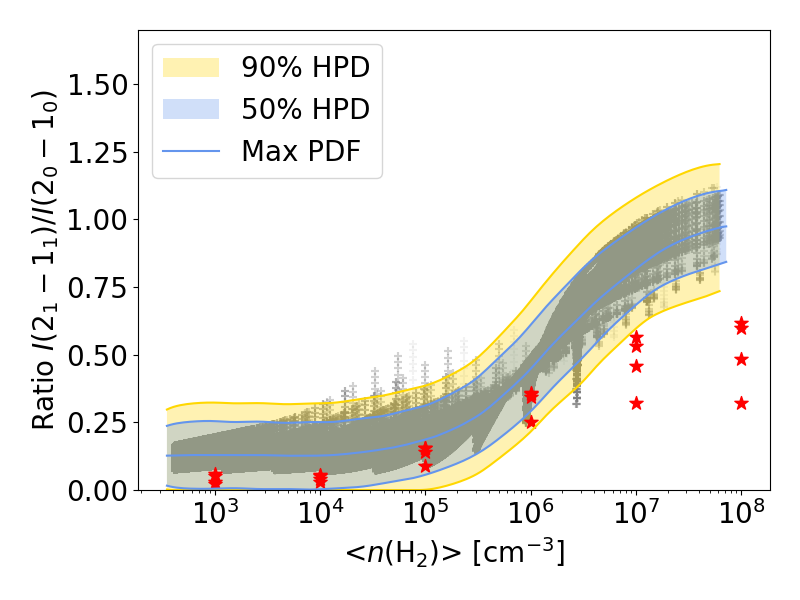}\hfill
            \includegraphics[width=0.32\hsize]{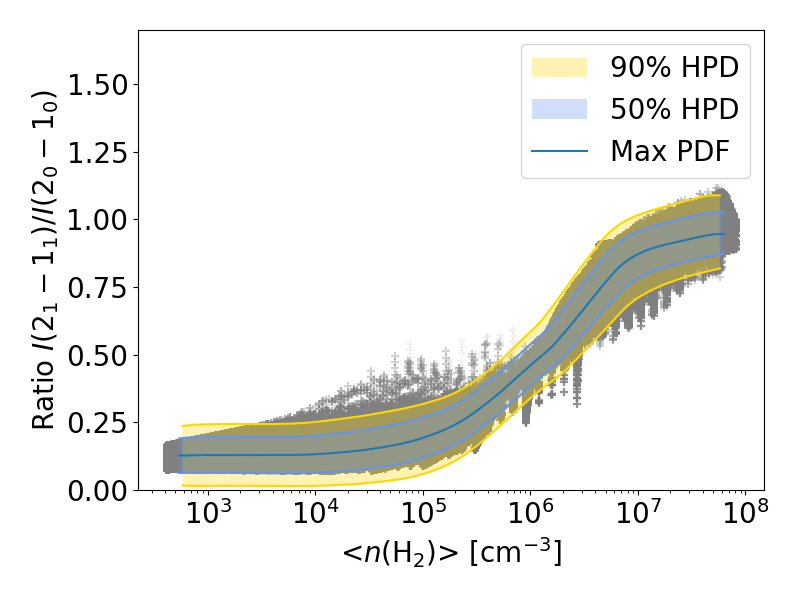}\\
            \includegraphics[width=0.32\hsize]{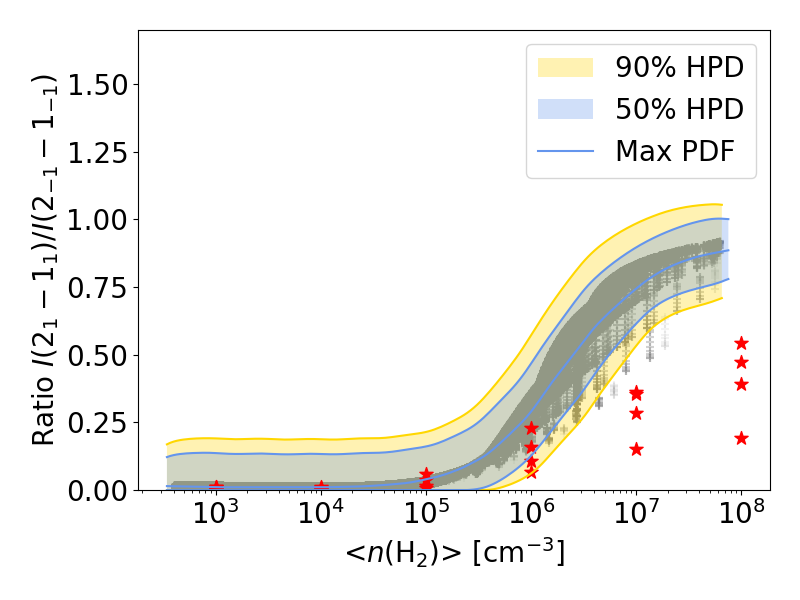}\hfill
            \includegraphics[width=0.32\hsize]{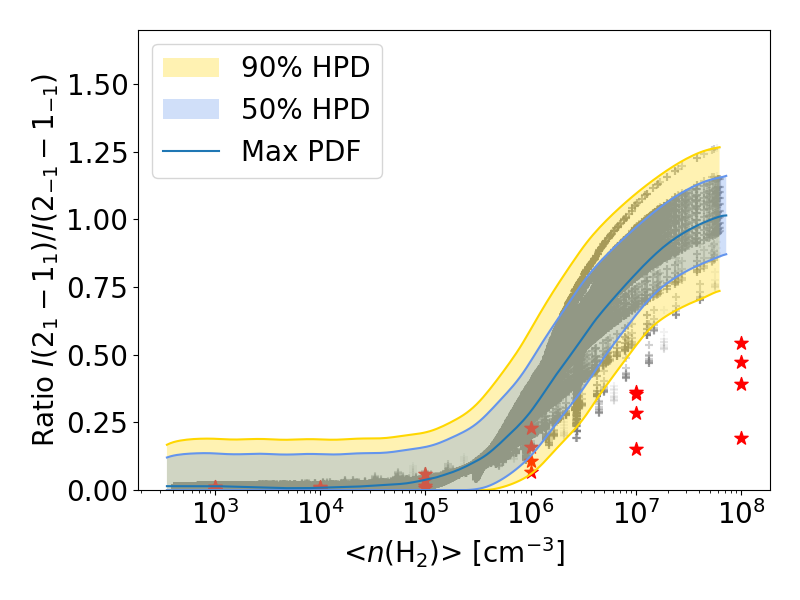}\hfill
            \includegraphics[width=0.32\hsize]{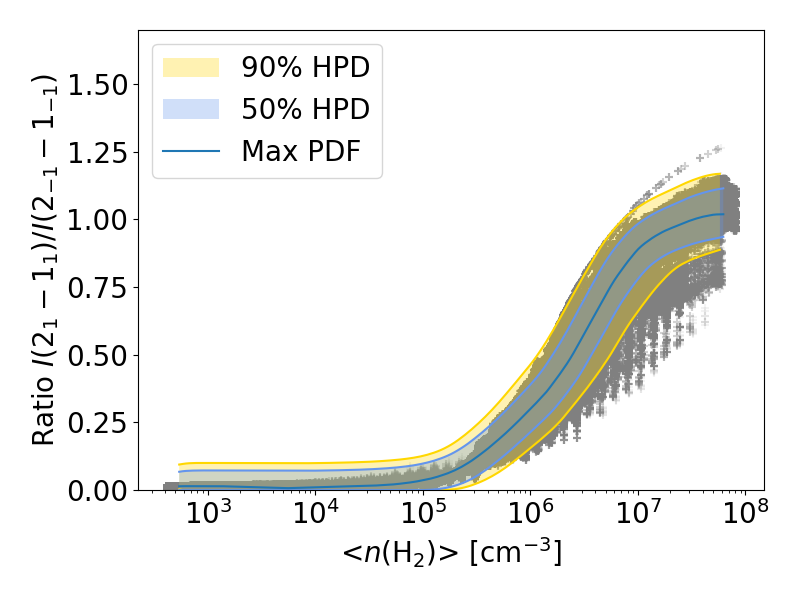}\\
            \includegraphics[width=0.32\hsize]{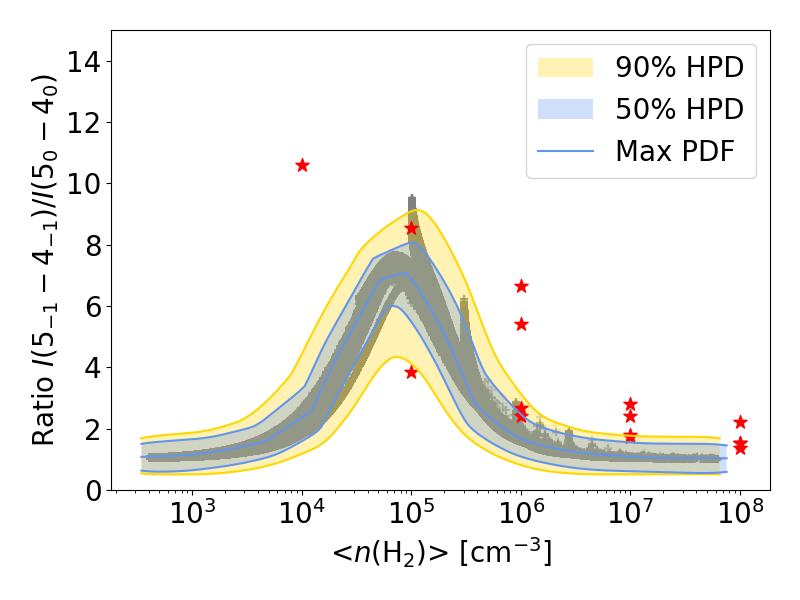}\hfill
            \includegraphics[width=0.32\hsize]{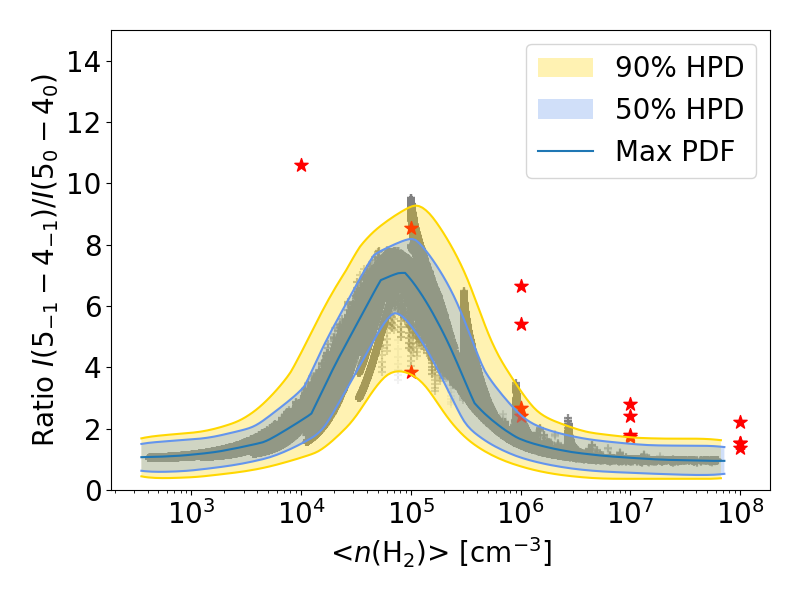}\hfill
            \includegraphics[width=0.32\hsize]{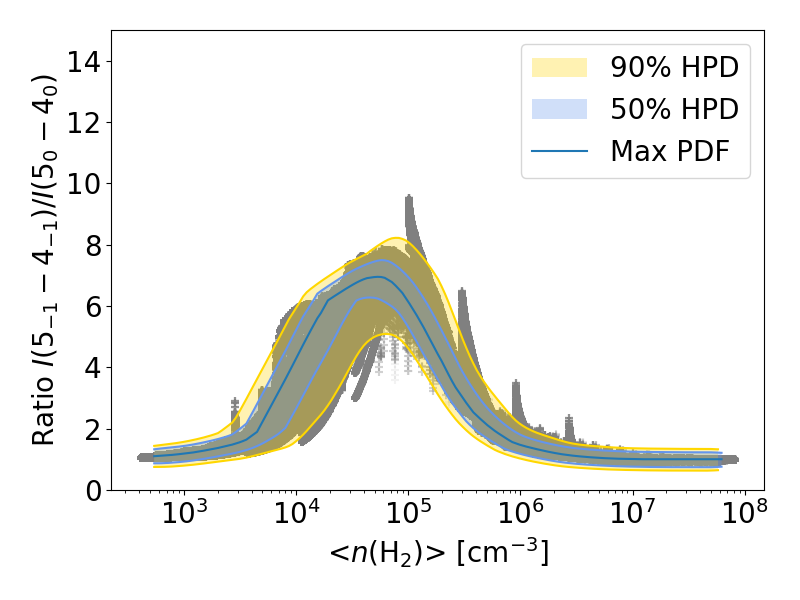}\\
            \includegraphics[width=0.32\hsize]{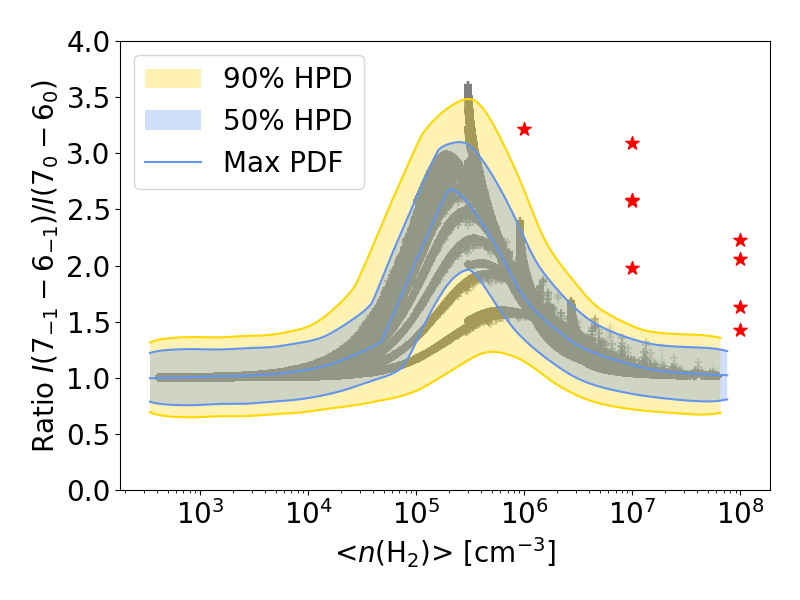}\hfill
            \includegraphics[width=0.32\hsize]{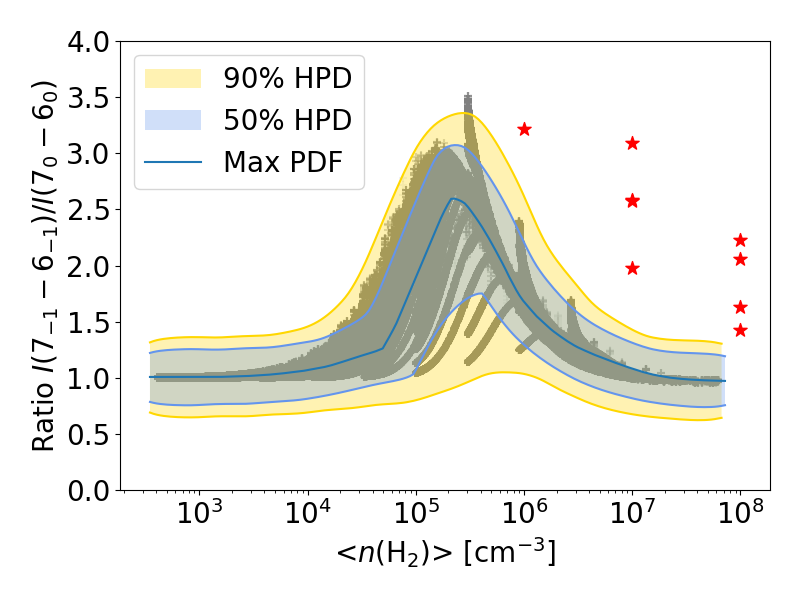}\hfill
            \includegraphics[width=0.32\hsize]{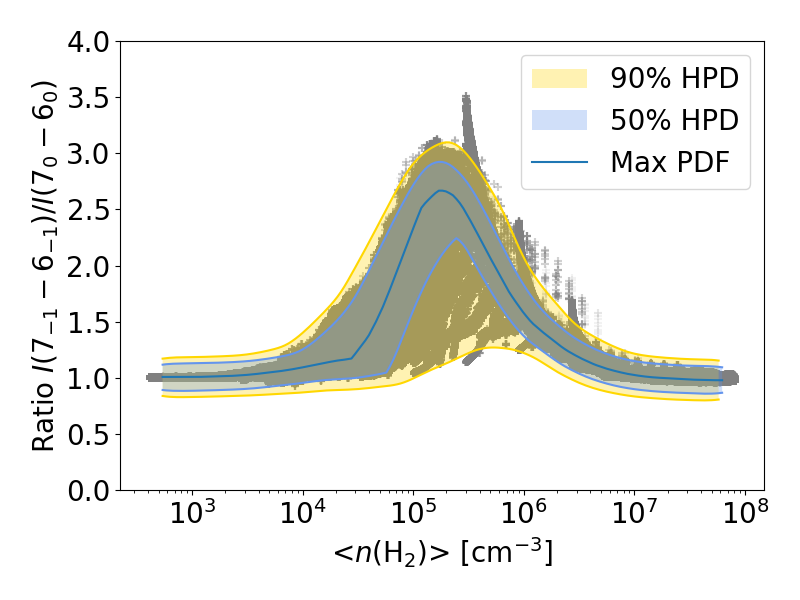}\\

            \caption{\label{fig:los_map_summary}Kernel density estimate of the probability density function for the line ratios as a function of average number density along the line-of-sight, for all the temperatures considered in the grid, under the assumption of an isothermal clump. The rows show the ratios \ratioESevenESix, \ratioEEightESeven, \ratioEEightESix, \ratioTwHFSevenTwHFSix, and \ratioThHEOneTHE, in order, from top to bottom. RADEX computations for an isothermal, uniform clump are shown as red stars; the parameters used are described in the text. Note that for the \ratioTwHFSevenTwHFSix, and \ratioThHEOneTHE\ ratios the RADEX points at low densities are above the plotting limits, reaching values around 25 in both cases.}
        \end{figure*}

        Figure~\ref{fig:los_map_summary} shows the line ratios as a function of the average number density of H$_2$ along the LOS for each pixel in the final simulated image as grey plus signs, comparing three cases.
        The left column shows the isothermal clump case, while the central column displays the internally heated clump case, characterised by a temperature gradient with $q=-0.5$.
        The left- and central columns also contain a comparison of our results with a simpler isothermal and uniform model carried out with RADEX, indicated by the red stars.
        Due to the sparsity of our RT grid, the sampling of the average number density vs. the line ratio space is not uniform, generating artifacts in the final number density posterior in specific regimes.
        ML algorithms are known to be extremely good in emulating complex computations, such as the radiative transfer, and can be used to address this issue.
        We therefore tuned such models to provide the individual line fluxes and line ratios on a four-times finer grid.
        The right column of Fig.~\ref{fig:los_map_summary} shows the results of this refinement, for our best-performing models.
        The details of the training are presented in Appendix~\ref{sec:ml_details}.

        To connect the line ratios to the corresponding average number density along the LOS for each individual pixel across all models in our grids, we use a KDE.
        The shaded areas in the figures representing two (smoothed) HPD intervals of the resulting PDF.
        This overcomes saturation effects in the distribution of the $\approx 6 \times 10^{5}$ points in each plot.
        Details on how this is computed are given in Appendix~\ref{app:kde}.

        Figure~\ref{fig:los_map_summary} shows that the proposed line ratios are consistently correlated with density, despite the presence of a density gradient.
        Different temperatures of the gas cause the scatter that is visible in the left column in the measured ratios for a specific value of the average density along the LOS.
        Comparing the left- and central columns of the figure reveals the impact of realistic temperature gradient in the clump.
        This gradient, however, does not significantly alter the trend observed for isothermal clumps.
        On the other hand, the RADEX results reveal the potential errors in the density inference caused by a oversimplified clump model.
        The right column of the figure demonstrates how the inclusion of the emulated data improves the sampling of the parameter space.
        The model errors have no qualitative impact on the shape of the relation, indicating that the model accurately emulates the RT computations.
        This qualitative check serves to verify that the trained models generalise well for all the predicted data points, confirming that the final inference is reliable.

        Examining in more detail the isothermal and internally heated cases shows the effect of the simulated heating.
        The primary effect of the embedded protocluster is a slight increase in the scatter of the relations.
        Additionally, it causes a $10-20\%$ higher maximum value for the \ratioEEightESeven\ and \ratioEEightESix\ ratios at very high densities ($<n_\mathrm{H_2}> \gtrsim 5\times 10^6\usk \mathrm{cm^{-3}}$) in the internally-heated case, where the material is mostly thermalised.
        Confirmation that the temperature only has a second order effect comes also from the Spearman correlation coefficients $\rho$.
        The coefficient are listed in Table~\ref{tab:correlation_coefficients}, and they are far larger for the average density than for the average temperature.
        Because of their non-monotonic behaviour, the ratios of lines in the $241.7\usk\mathrm{GHz}$ and $338.1\usk\mathrm{GHz}$ bands show a much weaker correlation with the average number density compared to those for the lines in the $96.7\usk\mathrm{GHz}$ band.

        Figure~\ref{fig:los_map_summary} also shows the importance of including the fragment envelope when calibrating the ratio values.
        We used RADEX to construct simple homogeneous and isothermal clump models, and compared them to our results.
        The RADEX points (red stars in the left- and central columns) represent extreme values of the ratios for typical high-mass clumps in the galactic disk.
        The line intensities were computed with two values of E-CH$_3$OH column densities, $10^{14}\usk\text{cm}^{-2}$ and $3 \times 10^{15}\usk\text{cm}^{-2}$.
        These values correspond to the minimum and maximum column density of the cold methanol component in \citet{Giannetti+17_aa603_A33}, after accounting for an equal abundance of the E and A variants of methanol, and the likely overestimate of the LTE column density \citep[][and Sect.~\ref{sec:tau_density_degeneracy}]{Leurini+04_aa422_573}.
        We varied the number density in the range $10^{3}- 10^{8}\usk\text{cm}^{-3}$ in steps of one order of magnitude, and assumed two kinetic temperatures, $10\usk\mathrm{K}$ or $30\usk\mathrm{K}$; the full width at half maximum (FWHM) of the line was fixed to $3.5\usk\mathrm{\kilo\metre\usk\second^{-1}}$, as in the majority of our models.
        Comparing the most probable value of the ratio for the PDFs and the values predicted by RADEX reveals that the calibration of the relations can be off by even two orders of magnitude for a given value of the line ratio.
        In fact, they should roughly encompass the PDFs obtained through SAK if the calibrations were equivalent, because the RADEX points represent extreme values of the ratios.
        This stresses the importance of more sophisticated modeling when trying to infer $n_\mathrm{H_2}$.

\begin{figure}
            \centering
            \includegraphics[width=0.95\hsize]{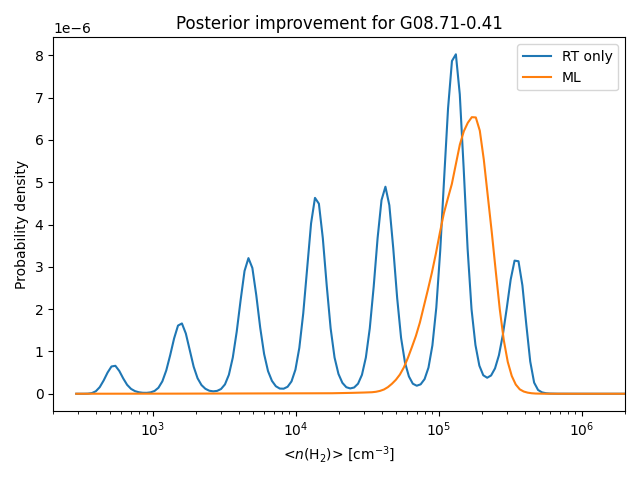}
            \caption{Improvement in the final number density posterior when adding ML-emulated data. In blue, a source displaying oscillations in the posterior is shown for the RT-only case; in orange, the same source with the number density posterior evaluated on the finer grid, generated by our final ML model. The oscillations caused by grid sparsity disappear.}
            \label{fig:ml_improvement_posterior}
        \end{figure}

        It is difficult to appreciate the improvement in the final posterior that ML emulation brings from Fig.~\ref{fig:los_map_summary} alone.
        We have mentioned above that our grid introduces an uneven sampling of densities when considering the individual LOSs.
        Figure~\ref{fig:ml_improvement_posterior} shows the resulting oscillatory pattern in the number density PDFs for one of the sources included in the sample of high-mass clumps used to investigate the evolution of number density (cf. Sect.~\ref{sec:poc}).
        The source has been chosen to be in the critical number density regime affected by this issue, and it shows the posterior before and after the dataset has been enriched with ML-emulated points.
        The finer grid produced through ML emulation removes artificial features from the relation, even using a kernel with half the bandwidth as for the RT-only case.

    \subsection{Regimes of density probed by the ratios}\label{sec:density_regimes}
        Figure~\ref{fig:los_map_summary} shows that ratios computed with lines in the $(2_K-1_K)$ band around $96.7\usk\mathrm{GHz}$ are sensitive to average densities in the range $\approx 5 \times 10^4 - 3 \times 10^7\usk\mathrm{cm^{-3}}$.
        This interval reflects the density at which collisional excitation starts to be important, and the density at which the lines are fully thermalised.
        In this case, the ratio starts to be correlated to the density at values comparable with the critical densities of the lines ($\approx 3 \times 10^4\usk\mathrm{cm^{-3}}$, see Table~\ref{tab:lines_list}).

        On the other hand, the ratios of lines in the $(5_K-4_K)$ and $(7_K-6_K)$ bands around $241.7\usk\mathrm{GHz}$ and $338.1\usk\mathrm{GHz}$ show a more complex behaviour.
        The density at which they thermalise (and quickly become optically thick) is comparable to the critical density, while dependence on the density starts around a few~$10^3\usk\mathrm{cm^{-3}}$ -- $10^4\usk\mathrm{cm^{-3}}$ for the internally-heated clump case.
        These lines have an Einstein-A coefficient $\approx 20-40$ times higher than the lines in the $(2_K-1_K)$ band.
        They have a significantly higher optical depth, and radiative trapping is more important, making them sensitive to densities much lower than their critical density.

        The \ratioTwHFSevenTwHFSix\ and \ratioThHEOneTHE\ ratios are less powerful than those computed with 3~mm lines for estimating number densities, because each value of the ratio has two solutions.
        However, they are valuable to extend the range of conditions where we can actually estimate this parameter to lower densities, not accessible through the lower-frequency lines.
        One example where this is useful is represented by entire molecular clouds, which have average densities typically in the range probed by these higher-frequency ratios.

    \subsection{Robustness of the ratios to gas distribution and distance}\label{sec:robustness_distribution_distance}

        The effect of the specific distribution of material in the clumps has been investigated by comparing the shape and location of the relations across the three grids with a different $p$-value.
        This comparison reveals that the relations between the line ratios and the average density along the LOS are unaffected by the exponent of the density power-law: whatever the distribution of material within the explored cases, a given ratio of the lines always corresponds to the same average density along the LOS (Fig.~\ref{fig:density_distribution_comparison}).

\begin{figure*}
            \centering
            \begin{tikzpicture}[font=\sffamily]
                \node at (-7,0) {};
                \node at (7,0) {};
                \node at (-0.5,0) {\tiny \textit{p}=-1.5};
                \node at (-6.0,0) {\tiny \textit{p}=-1.2};
                \node at (5.3,0) {\tiny \textit{p}=-1.8};
                \node at (-0.5,-0.5) {\tiny Integrated line ratios \ratioESevenESix};
            \end{tikzpicture}\\
            \includegraphics[width=0.25\hsize]{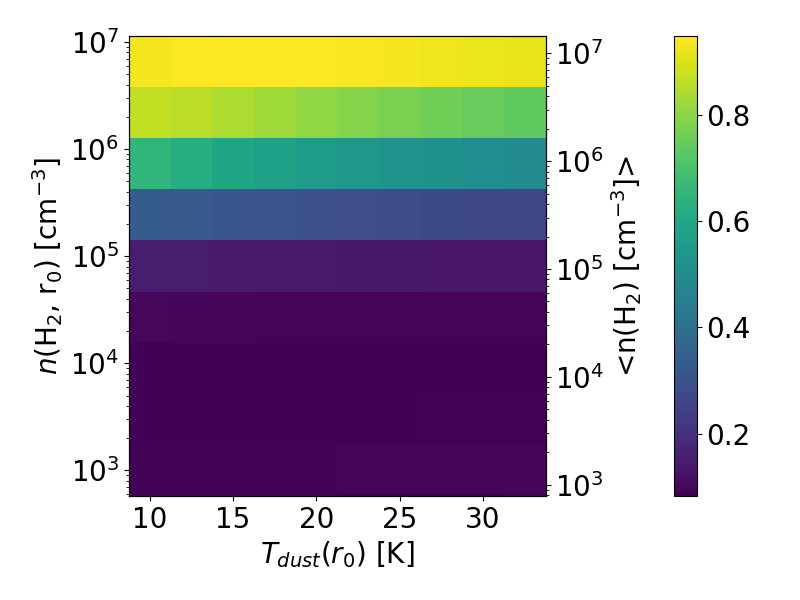}\hspace{0.05\hsize}
            \includegraphics[width=0.25\hsize]{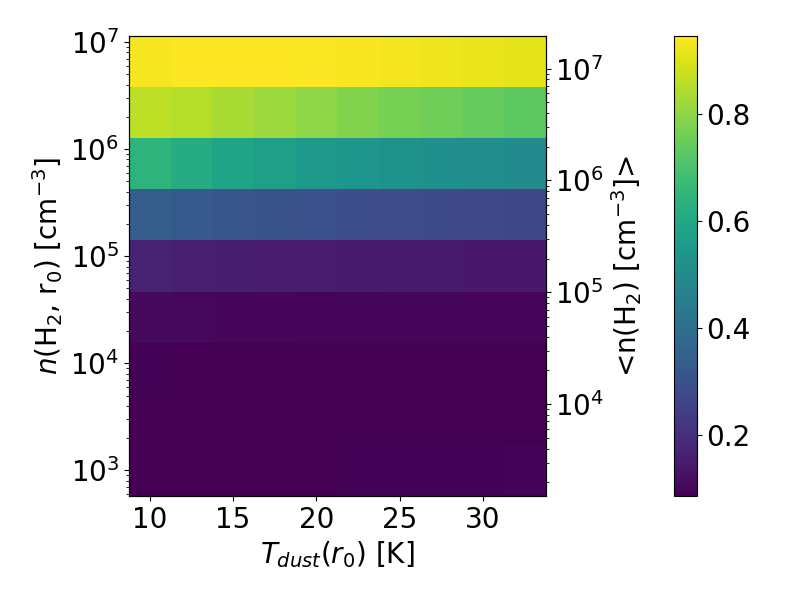}\hspace{0.05\hsize}
            \includegraphics[width=0.25\hsize]{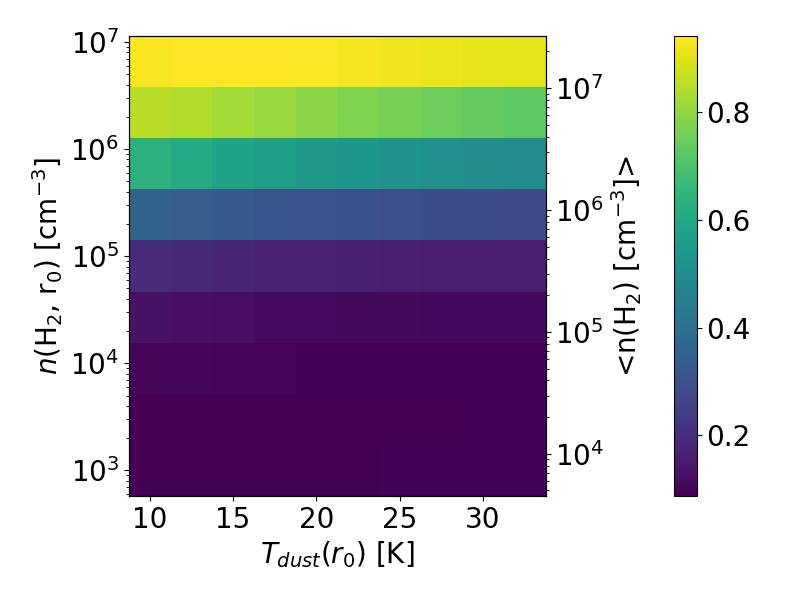}\\
            \begin{tikzpicture}[font=\sffamily]
                \node at (-7,0) {};
                \node at (7,0) {};
                \node at (-0.5,-0.2) {\tiny Integrated line ratios \ratioTwHFSevenTwHFSix};
            \end{tikzpicture}\\
            \includegraphics[width=0.25\hsize]{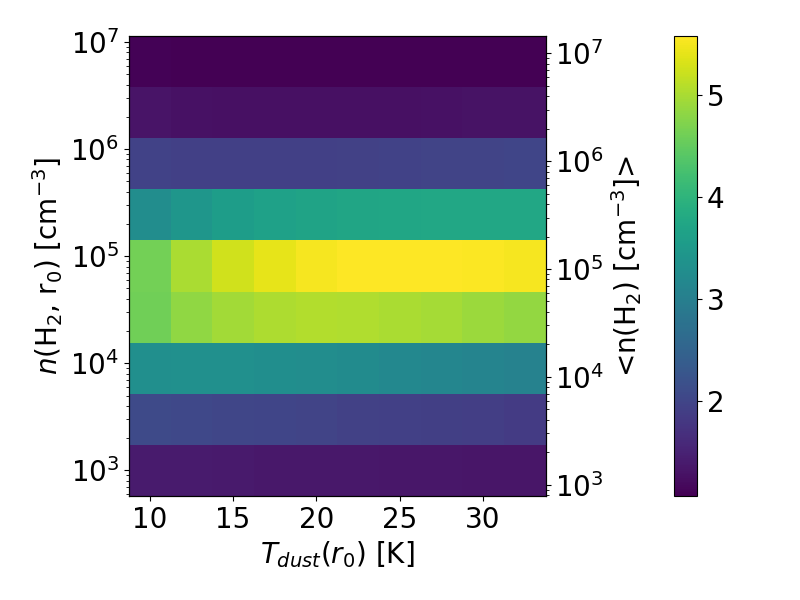}\hspace{0.05\hsize}
            \includegraphics[width=0.25\hsize]{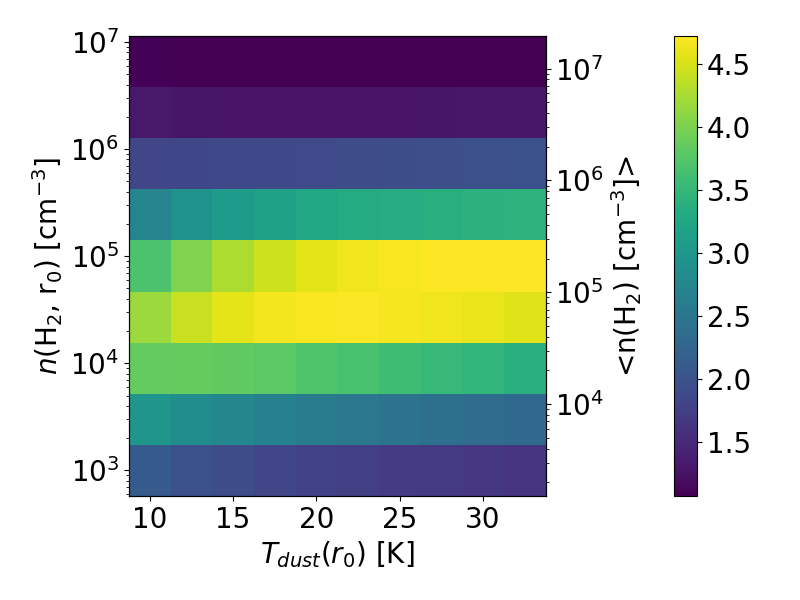}\hspace{0.05\hsize}
            \includegraphics[width=0.25\hsize]{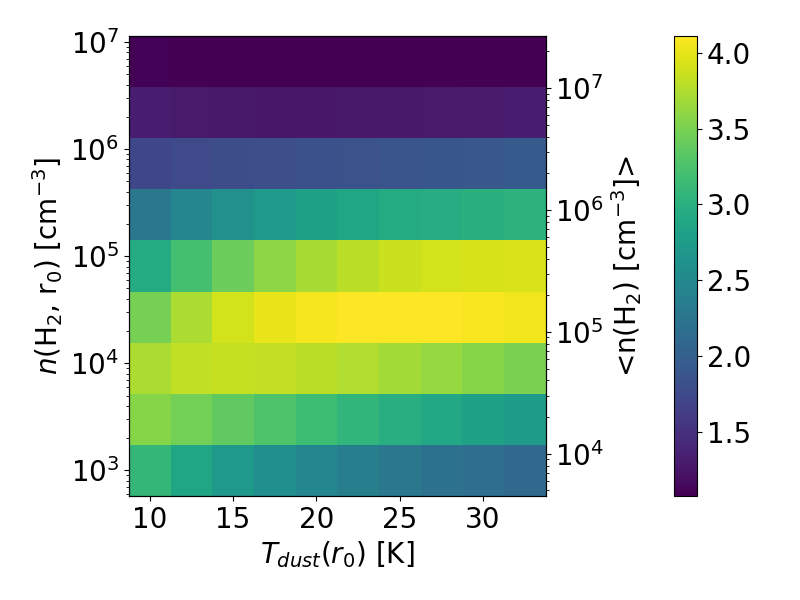}\\
            \begin{tikzpicture}[font=\sffamily]
                \node at (-7,0) {};
                \node at (7,0) {};
                \node at (-0.5,-0.2) {\tiny Integrated line ratios \ratioThHEOneTHE};
            \end{tikzpicture}\\
            \includegraphics[width=0.25\hsize]{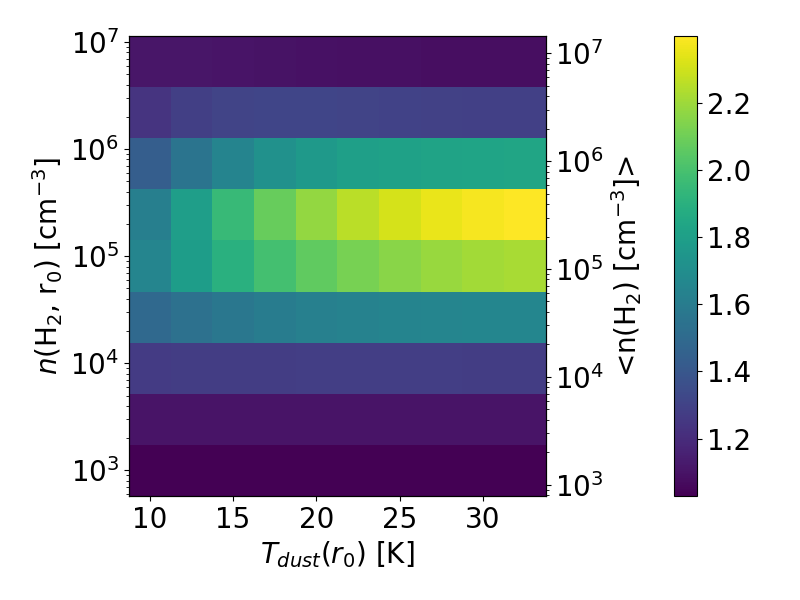}\hspace{0.05\hsize}
            \includegraphics[width=0.25\hsize]{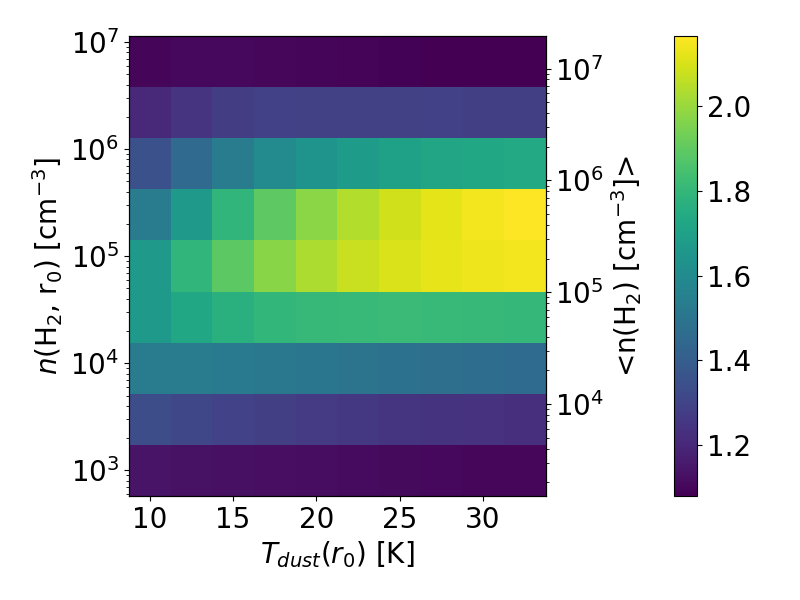}\hspace{0.05\hsize}
            \includegraphics[width=0.25\hsize]{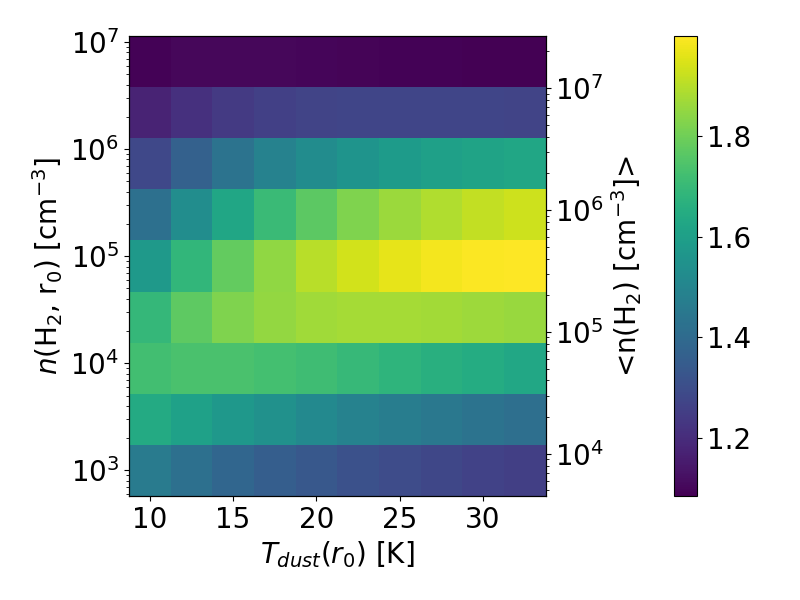}\\
            \caption{Comparison of the integrated line ratios for different density distributions, corresponding to the three grids considered in the case of internally-heated clumps. The columns of plots refer to the values of $p$ indicated at the top of each column. We only show the \ratioEEightESeven\ for the 3~mm lines, because all three have the same behaviour.}
            \label{fig:integrated_ratios_across_grids}
        \end{figure*}
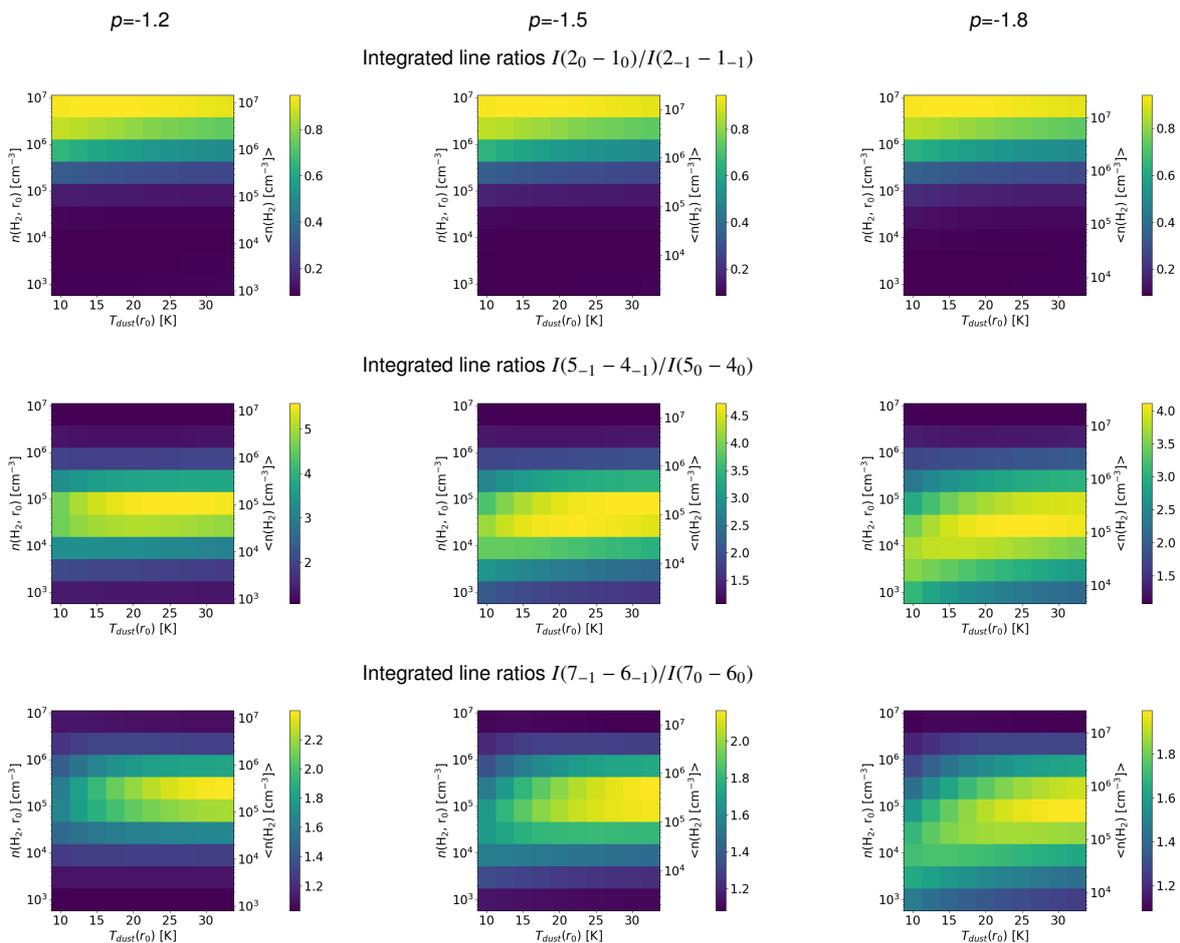

        Figure~\ref{fig:integrated_ratios_across_grids} further decouples the temperature and density dependence, by showing the line ratio values derived by integrating over the entire clump as function of $n_{\mathrm{H_2},0}$ and T$_0$, for different exponent $p$.
        In addition to the minor role of temperature, Fig.~\ref{fig:integrated_ratios_across_grids} shows that the ratio values depends only on the overall average density of the clumps, as indicated in the right axes of the panels of Fig.~\ref{fig:integrated_ratios_across_grids}, confirming that the results are independent of the fine details of the material distribution.

        The values and ranges of the ratios for a given average density can be compared when averaging along the LOS or over the entire clump by contrasting Figures~\ref{fig:los_map_summary} and \ref{fig:integrated_ratios_across_grids}.
        They are found to be broadly consistent, indicating that the ratios present only a marginal dependence on distance, because they are computed on widely different (a factor of $100$) spatial scales in the two cases.

    \subsection{Effects of optical depth and a hot-core-like abundance profile}\label{sec:hotcores_tau}
\begin{figure*}
        \centering
        \includegraphics[width=0.28\hsize]{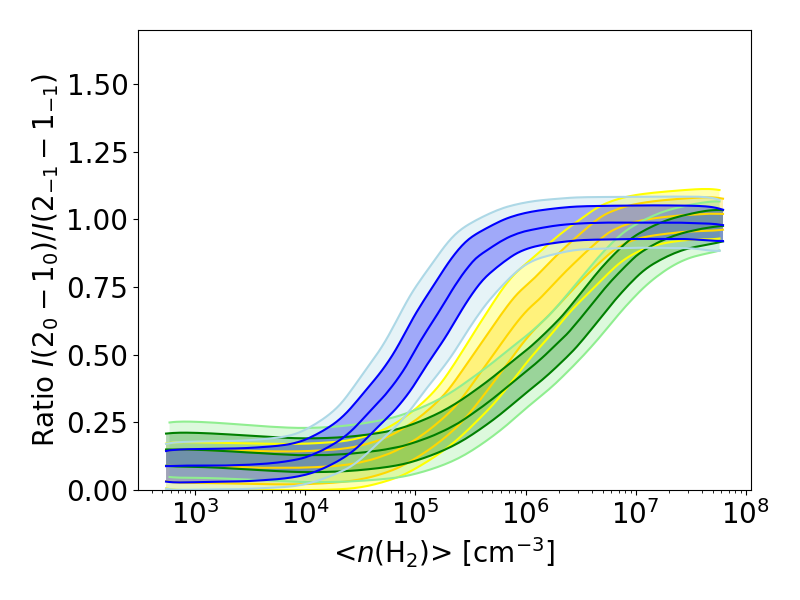}\hspace{0.05\hsize}
        \includegraphics[width=0.28\hsize]{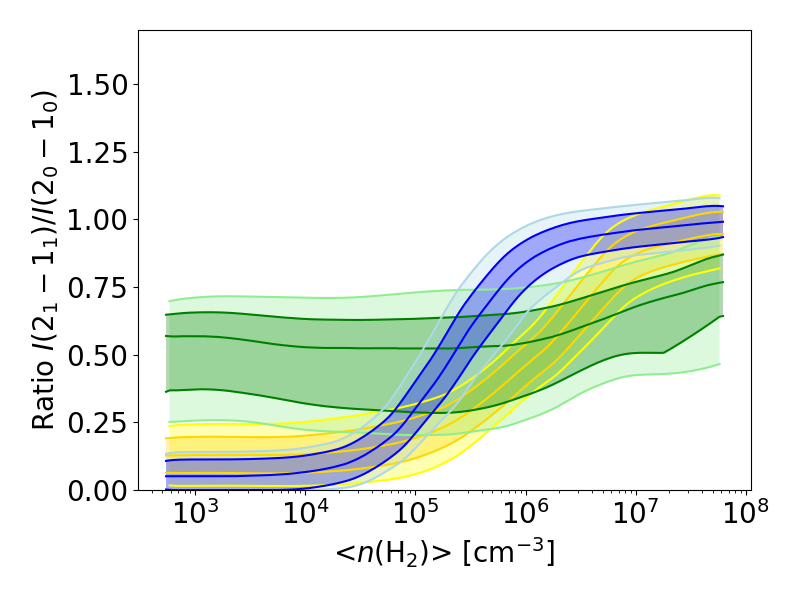}\hspace{0.05\hsize}
        \includegraphics[width=0.28\hsize]{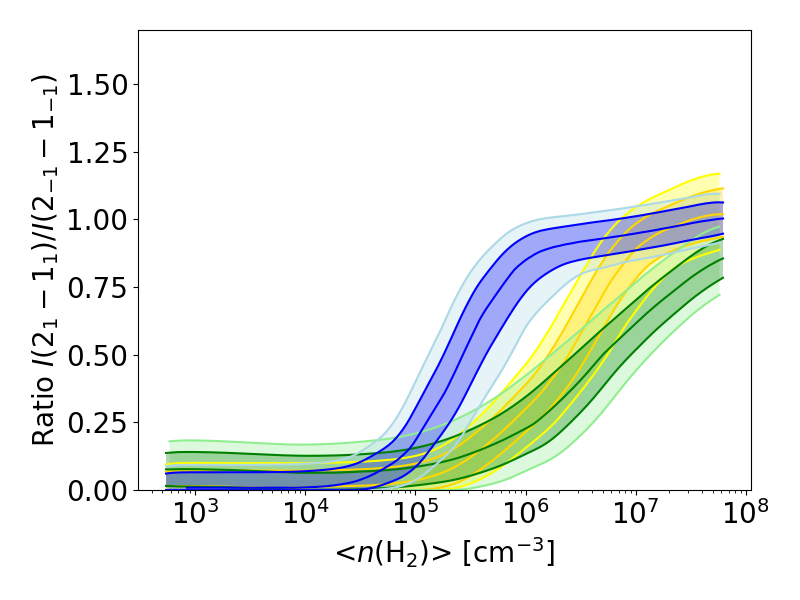}\\
        \includegraphics[width=0.28\hsize]{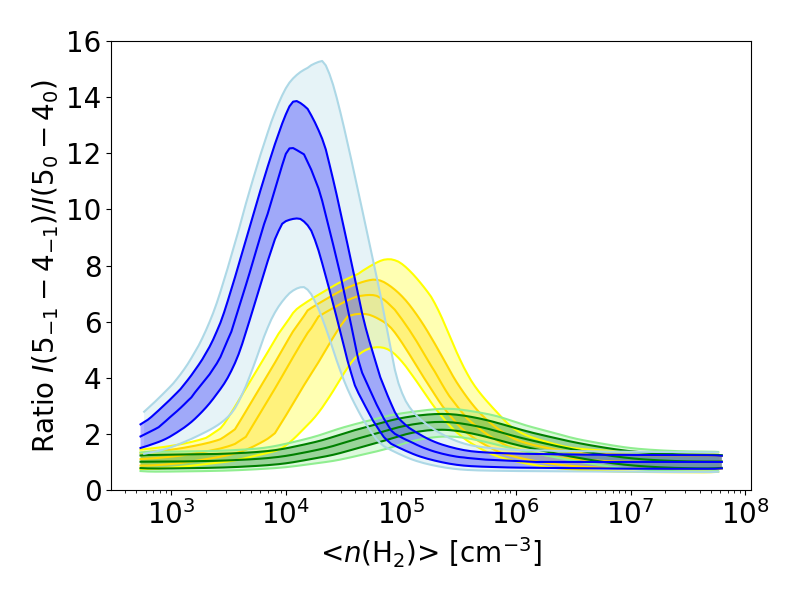}\hspace{0.05\hsize}
        \includegraphics[width=0.28\hsize]{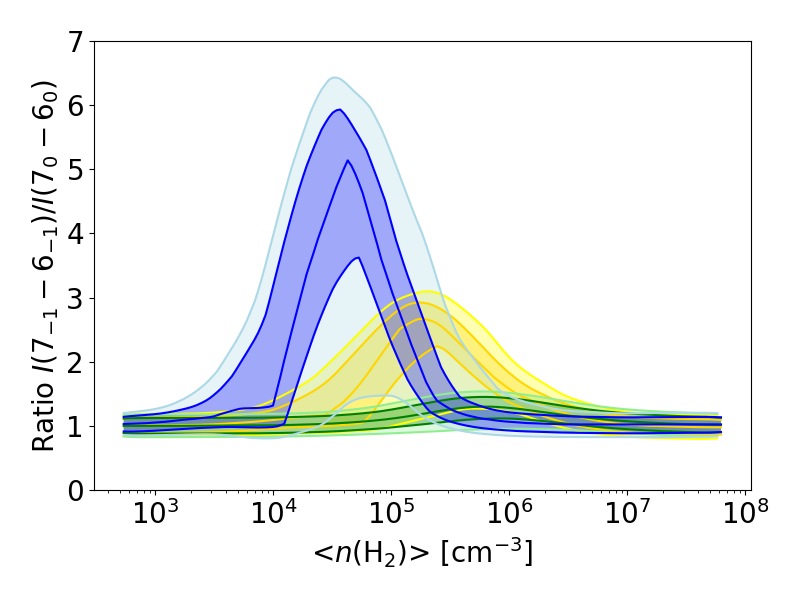}
        \caption{\label{fig:abundance_shift}Comparison between the internally-heated clump with the fiducial E-CH$_3$OH methanol abundance of $10^{-9}$ (yellow shaded areas), the low-abundance ($\mathrm{E-CH_3OH} / \mathrm{H_2} = 10^{-10}$, 1/10 fiducial), low-optical depth case (green shaded areas), and for \ratioESevenESix, \ratioEEightESeven, and the high-abundance ($\mathrm{E-CH_3OH} / \mathrm{H_2} = 10^{-8}$, 10 times fiducial), high-optical depth case (blue shaded areas) for the \ratioEEightESix, \ratioTwHFSevenTwHFSix, and \ratioThHEOneTHE, respectively. The light-shaded areas correspond to the $90\%$ HPD, while the dark shaded areas correspond to the $50\%$ HPD.}
    \end{figure*}

        In Appendix~\ref{subsec:optical_depth} we quantify the impact of different optical depths by scaling the abundance up and down by one order of magnitude.
        As shown in Fig.~\ref{fig:abundance_shift}, which compares models with E-CH$_3$OH abundances of $10^{-10}$ and $10^{-8}$, a 100-fold change in abundance results in a density shift of approximately a factor of 15 in the line ratio relations.
        Moreover, the dependence on the density of the \ratioEEightESeven\ ratio is mostly lost for the lowest methanol abundance ($10^{-10}$), and thus the lowest optical depths of the lines.
        It is interesting to note that this ratio has the largest dependence on the temperature (cf. Table~\ref{tab:correlation_coefficients}).

        Figure~\ref{fig:abundance_shift} allows also to check if the behaviour of the line ratios is consistent with the effect of radiative trapping discussed in Sect.~\ref{sec:density_regimes}.
        Radiative trapping decreases the effective radiative de-excitation rate \citep{Shirley2015_pasp127_299}, thus making the lines more efficient in tracing low-density gas \citep{Kauffmann+2017_aa605_5}.
        Thus, a larger abundance should shift the relation to lower densities by increasing the optical depth of the lines, and an opposite behaviour is expected for lower abundances.
        Another consequence is that the ratio should become thermalised, and thus insensitive to density, at lower average density along the LOS.
        Indeed a lower optical depth, corresponding to the lower methanol abundance, shifts a given value of the line ratios to higher average densities while a higher abundance, and thus higher optical depth, has the contrary effect.

        We finally note that the higher-frequency ratios show different maxima values, depending on the abundance assumed. From Fig.~\ref{fig:abundance_shift} one can see that the maximum value of the \ratioTwHFSevenTwHFSix\ ratio varies from $\approx 2.5$ in the case of the lowest methanol abundance ($10^{-10}$) to $\approx 14$ in the case of the highest methanol abundance ($10^{-8}$); similarly, the \ratioThHEOneTHE\ ratio varies from $\approx 1.3$ to $\approx 5$.
        This increase is again an effect of radiative trapping.
        By making the lines easier to excite at lower density, the emitting region becomes larger, particularly so for the lower excitation line.
        The outer layers are also progressively colder due to the power-law behaviour of the temperature.
        Therefore, the combination of optical depths, density, and temperature creates a region where only the upper level of the lower-excitation line is efficiently populated by the collisions, resulting in brighter line fluxes, compared to the higher-excitation one.
        Where this region is located determines the maximum value of the ratio: the larger the emitting region, the higher the expected ratio due to the lower gas temperatures traced.
        The higher values of the ratio maximum are indeed observed at lower densities, corresponding to outer and colder layers in our models.

        Methanol is a species that also efficiently traces the hot gas present close to protostars.
        This molecule forms in large quantities on the dust grains via successive hydrogenation of CO, and observations show that CH$_3$OH is an ubiquitous and important constituent of interstellar ices, with abundances that can reach values close to $10^{-6}$ (e.g. \citealt[][]{Grim_1991_aa243_473, Pontoppidan+2003_aa404_17, Boogert+2008_apj678_985, Boogert+2011_apj729_92}, and \citealt{Boogert+2015_araa53_541} for a review).
        When temperatures exceed $\sim90\usk\mathrm{K}$ CH$_3$OH is evaporated from grains, increasing its gas-phase abundance by around two orders of magnitude \citep[e.g.][]{vanderTak+2000_aa361_327, Maret+2005_aa442_527M, Giannetti+17_aa603_A33, Sabatini+21_aa652_71}.
        Therefore, we simulated evaporation of methanol from grains by adopting a step-function profile for the abundance, with a jump of a factor of 100 in the regions exceeding $90\usk\mathrm{K}$ (see Appendix\usk\ref{subsec:hot_core}).
        The 3~mm clump-integrated ratios are not affected by the presence of a hot core, while those at higher frequencies are reduced by $\approx 10\%$ due to optical depth issues in the innermost layers.
        Because of the small area affected by the increased abundance, only a minimal fraction of the ratios computed along the LOS is influenced, showing values consistent with a constant abundance of $10^{-8}$.

    \subsection{The optical depth-number density degeneracy}\label{sec:tau_density_degeneracy}

    Our results point to the fact that the considered line ratios are, in general, robust indicators of number density, with the largest uncertainties coming from the abundance of methanol itself.
    As mentioned earlier, changing the abundance of E-CH$_3$OH by a factor of 10 up or down from the fiducial value, respectively decreases or increases the estimated density by a factor $\approx 2-5$.
    This introduces a degeneracy between the column density or opacity of the lines and the number density, that is comparable to or larger than the uncertainty due to the spread in the modelled points.

    This degeneracy can be broken relatively easily.
    One possibility would be to account for the line opacity before computing the ratios, estimating the corrections from the intensity of the corresponding $^{13}$CH$_3$OH isotopologue lines, assuming that the isotopic ratio is known.
    This has also the potential of expanding the sensitivity to somewhat larger density values, if we use the low-abundance, low-opacity case for the inference, for which the lines trace denser gas and the thermalisation happens at higher densities (cf. green-shaded area in Fig.~\ref{fig:abundance_shift}).

    \begin{figure}
        \centering
        \includegraphics[width=0.75\columnwidth]{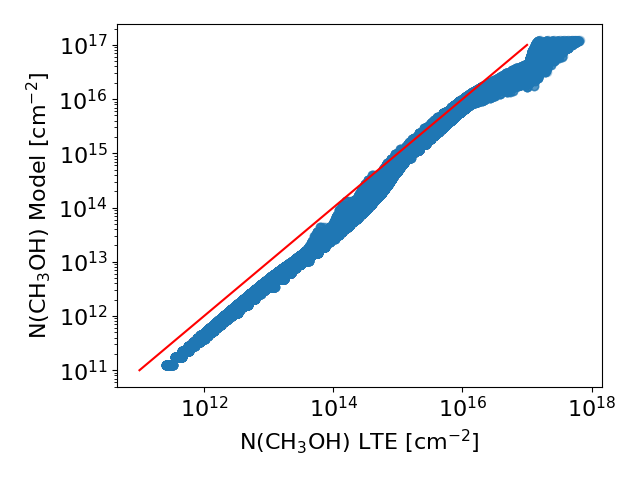}
        \caption{Comparison of the column density, as derived from a simple rotational diagram and computed from the model grid, on a pixel-by-pixel basis. The red line shows the one-to-one locus.}
        \label{fig:lte_approx}
    \end{figure}

    Knowing a priori the column density and the abundance of methanol would also be a useful constraint.
    \citet{Leurini+04_aa422_573} show that LTE and LVG analyses for the cold methanol component yield similar results in terms of column density, even though it underestimates the excitation temperature.
    In their analysis the authors find that the CH$_3$OH column density is slightly overestimated (by a factor $\approx 1.3$) by the LTE assumption.
    If this systematic error can be characterised and accounted for, it is straightforward to have a rough estimate of the fractional abundance in the source, assuming that the H$_2$ column density can be estimated from dust continuum emission.
    This allows to then select the most appropriate grid of models.

    Figure~\ref{fig:lte_approx} shows the comparison of the column densities of methanol from our models and those obtained via a simple population diagram analysis applied to the $96.7\usk\mathrm{GHz}$ lines, that assumes LTE \citep{GoldsmithLanger1999_apj517_209}.
    The figure demonstrates that the two column densities for each pixel are very well correlated, with a tendency towards overestimation for the LTE approximation.
    The median difference between the LTE and model column density of E-CH$_3$OH is approximately a factor of three.
    Of course this result depends on the number density of the material being probed, but Fig.~\ref{fig:lte_approx} shows that this holds for all the models we created, with number densities along the LOS in excess of $\sim \mathrm{few} \times 10^2\usk\text{cm}^{-3}$ (cf., e.g., Fig.~\ref{fig:los_map_summary}).
    This shows that such an approximation is valid for clumps and cores; averages over entire clouds are not likely to be dominated by heavily optically thick emission.

    Alternatively, one could produce a model finely tuned to the source properties (radius, mass, temperature, distance, line width, etc.), and compare directly the synthetic and observed line intensities.
    In this case, the tool would not be universal, but tailored grids would have to be produced, and the results would be more model-dependent.
    Another possibility is to introduce a prior, according to the estimate of the density from the dust, given an estimate of the source size, but this would make it impossible to estimate the source dimension along the LOS from the number- and column densities of H$_2$.
    The volume density from the dust already assumes a dimension of the source along the LOS and the final estimate would not be independent of this assumption.
    A final possibility is to select the grid based on the ratio vs. methanol column density values. This would bypass the uncertainties connected to the dust. Of course the beam size must be taken into account in this case, when estimating the molecular column density for a given data set.

    \section{Analytic approximation of the ratio -- number density relations}\label{sec:analytic_representation}
    To facilitate adoption of these results without having to run the SAK inference code locally, we provide analytic formulas to derive the average number density along the LOS from the ratios, based on the internally-heated model.
     This model reflects the most typical scenario for Galactic clumps \citep[e.g.][]{Urquhart+18_mnras473_1059}.
     Notably, even isothermal sources follow similar relations between line ratios and $n(\mathrm{H_2})$, because these are relatively insensitive to variations in temperature.
    The only case where significant adjustments have to be made is the observation of hot-cores with high-spatial resolution, for which the lines have much larger optical depths.

    These expressions were obtained by fitting a curve to the maximum of the PDF in Fig.~\ref{fig:los_map_summary}.
    To represent the relation between ratios and number density we used the function:
    \begin{equation}
        log_{10}(n_\mathrm{H_2}) = \left(arctanh\left(\frac{R - a}{s}\right) + b\right)^\nu\label{eq:analytical_relation},
    \end{equation}
    where $R$ is the measured line ratio.
    The best-fit parameters for Eq.~\ref{eq:analytical_relation} and the applicability ranges of these approximations are given in Table~\ref{tab:analytical_relations}.
    The fit was performed through the "curve\_fit" function of \texttt{scipy.optimize} \citep{2020SciPy-NMeth}, using the limits in density listed in Table~\ref{tab:analytical_relations}

    Not accounting for observational uncertainties, this is accurate to a factor of two at the $1\sigma$ level, for densities $\gtrsim 5\times10^4\usk\mathrm{cm^{-3}}$ when using the ratios of lines in the $(2_K-1_K)$ band.
    The use of these relations for ratios computed with lines in the $(5_K-4_K)$ and $(7_K-6_K)$ bands requires a priori knowledge of whether we are in a high- ($\gtrsim 10^5\usk\mathrm{cm^{-3}}$) or low-density regime ($2\times10^3\usk\mathrm{cm^{-3}} \lesssim n_\mathrm{H_2} \lesssim 10^5\usk\mathrm{cm^{-3}}$) to avoid degeneracy.
    The formal uncertainties are within a factor of two for the $(5_K-4_K)$-band ratio for both density regimes, and for the high-density regime of the one computed with the lines in the $(7_K-6_K)$ band.
    The latter ratio in the low-density regime is associated with a larger uncertainty of a factor $\approx 4$.

    \begin{table*}
        \centering
        \small
        \caption{Best-fit parameters for the ratio - number density relations from Eq.~\ref{eq:analytical_relation}, and applicability ranges for these approximations.}
        \label{tab:analytical_relations}
        \begin{tabular}{cccccc}
            \hline
            \hline
            Ratio                  & Validity range                & $a$   & $s$   & $b$   & $\nu$\\
                                   & $[\mathrm{cm^{-3}}]$          &       &       &       &       \\
            \hline
            \ratioESevenESix       & $5\times10^4 - 2\times10^7$   & 0.526 & 0.505 & 6.086 & 0.971\\
            \ratioEEightESeven     & $10^5 - 3\times10^7$          & 0.565 & 0.466 & 5.826 & 1.044\\
            \ratioEEightESix       & $10^5 - 10^7$                 & 0.520 & 0.532 & 6.728 & 0.976\\
            \multirow{2}*{\ratioTwHFSevenTwHFSix}
                                   & $2\times10^3 - 7.1\times10^4$ & 4.788 & 3.584 & 6.727 & 0.780\\
                                   & $7.1\times10^4 - 4\times10^6$ & 5.292 & -4.175& 7.320 & 0.832\\
            \multirow{2}*{\ratioThHEOneTHE}
                                   & $10^4 - 1.9\times10^5$        & 4.442 & 3.397 & 8.104 & 0.823\\
                                   & $1.9\times10^5 - 10^7$        & 2.247 & -1.229& 6.483 & 0.925\\
            \hline
        \end{tabular}
    \end{table*}

    \section{Application to galactic massive clumps}\label{sec:poc}
    We use our newly-derived calibration for the methanol line ratios to investigate the average number density in a sample of massive star-forming clumps in different evolutionary phases. Indeed,
    despite some indications that  the density progressively increases up to the development of a compact H\textsc{ii} region, it is still uncertain how
     the average number density of high-mass clumps evolve with time \citep{Elia+2017_mnras471_100, Giannetti+17_aa603_A33, Tang+18_aa611_6, Elia+2021_mnras504_2742, Urquhart+2022_mnras510_3389}.

    \subsection{Observations and data processing}
    We selected sources in the TOP100 sample \citep{Giannetti+2014_aa570_65} of the ATLASGAL survey \citep{Schuller+2009_aa504_415}. This sample contains around $110$ massive clumps spanning the entire evolutionary sequence, selected to be the brightest per class in the entire ATLASGAL survey.
    The TOP100 has been extensively followed up with spectroscopic observations, covering more than $120~\mathrm{GHz}$ of bandwidth in unbiased surveys in the 3~mm, 1.4~mm and 0.9~mm windows, and with observations of specific lines of interest \citep[e.g.][]{Tang+18_aa611_6, Navarete+19_aa622_135, Lee+22_aa664_80}. Therefore,
    the sample is extremely well characterised \citep[e.g.][]{Csengeri+16_aa586_149, Koenig+17_aa599_A139, Giannetti+17_aa603_A33, Wienen+21_aa649_21}, and is representative of the Galactic protocluster population.

    Based on their IR- and radio-continuum properties \citep{Koenig+17_aa599_A139},
    the TOP100 sample has been classified into four evolutionary stages.
    The four classes, ordered from the least- to the more evolved, are:
    \begin{itemize}
        \item Quiescent: sources with no compact emission at $70\usk\mathrm{\mu m}$ in Hi-GAL data;
        \item Protostellar: clumps associated with a compact source at $70\usk\mathrm{\mu m}$. These sources are either not detected in MIPSGAL at $24\usk\mathrm{\mu m}$ \citep{Carey+2009_pasp121_76} or with an integrated flux lower than 2.6 Jy, corresponding to a B3 star at 4 kpc \citep{Heyer+2016_aa588_29};
        \item MYSO: objects associated with a mid-IR source with a flux in excess of 2.6 Jy in MIPSGAL or MSX data (if the former is saturated);
        \item HII: compact H\textsc{ii} regions revealed by radio-continuum emission at 5- or 9~GHz. We considered the  the CORNISH \citep{Hoare+2012_pasp124_939, Purcell+2013_apjs205_1} and RMS surveys \citep{Urquhart+2007_aa461_11, Urquhart+2009aa501_539}, as well as targeted observations towards methanol masers \citep{Walsh+1998_mnras301_640}. Note that at this stage dispersal is just beginning and the clump envelope is still very massive.
    \end{itemize}
    This evolutionary sequence has been confirmed in terms of increasing temperatures, luminosity-to-mass ratios and incidence of hot-cores \citep{Koenig+17_aa599_A139, Giannetti+17_aa603_A33, Wienen+21_aa649_21}, decreasing CO depletion \citep{Giannetti+2014_aa570_65}, as well as chemical composition in general \citep{Sabatini+21_aa652_71}.
    Similar results in terms of temperatures, luminosities, and luminosity-to-mass ratios were obtained on a larger scale using the entire ATLASGAL sample \citep{Urquhart+18_mnras473_1059, Urquhart+2022_mnras510_3389}.

    For our experiment we focus on the sub-sample for which we have the lines listed in Table~\ref{tab:lines_list} (31 sources: 3 Quiescent, 10 Protostellar, 11 MYSO, and 7 HII).
    This is ideal, because we have all the lines we need and a very reliable determination of the evolutionary stage of the clumps, so that we can investigate if and how the average density changes with time.
    Although it is true that the sub-sample is small, an Anderson-Darling test confirms, considering the sources with $300\usk\mathrm{M_\odot} < M < 10^4\usk\mathrm{M_\odot}$, that the clumps for each evolutionary phase are drawn from the same population in terms of mass, diameter and distance (p-values 0.16, 0.1, and 0.25, respectively) as the others, and therefore there is no bias in the samples we use.

    Details of the observations and data reduction are given in \citet{Csengeri+16_aa586_149, Giannetti+2014_aa570_65, Giannetti+17_aa603_A33} and \citet{Sabatini+21_aa652_71}.
    Lines were fitted with a Gaussian curve, to derive their peak brightness temperature, their integrated emission, and the rms noise of the spectrum.
    These parameters of the line emission is reported in Tables \ref{tab:poc_lines_3mm} and \ref{tab:poc_lines_hf}.

    \begin{table*}
        \centering
        \small
        \caption{Sources included in the proof-of-concept, classification, distance, mass, and number density results.}
        \label{tab:poc_results}
        \begin{tabular}{llrlll}
            \hline
            \hline
            Source & Classification & best fit($n\mathrm{H_2}$) & 67\%  HPD & Mass & Distance \\
             &  & $[10^{5}\,\mathrm{cm^{-3}}]$ & $[10^{5}\,\mathrm{cm^{-3}}]$ & $[10^2\,\mathrm{M_\odot}]$ & [kpc] \\
            \hline
            AGAL008.706-00.414 & Protostellar & 1.7 & [1.1, 2.3] & 16.0 & 4.8 \\
            AGAL010.444-00.017 & Protostellar & 5.1 & [4.0, 6.5] & 16.0 & 8.6 \\
            AGAL010.472+00.027 & HII & 28.8 & [25.9, 36.1] & 100.0 & 8.6 \\
            AGAL010.624-00.384 & HII & 14.2 & [11.8, 19.5] & 37.0 & 5.0 \\
            AGAL012.804-00.199 & HII & 7.9 & [6.4, 10.6] & 18.0 & 2.4 \\
            AGAL013.178+00.059 & Protostellar & 2.8 & [2.3, 3.7] & 3.6 & 2.4 \\
            AGAL013.658-00.599 & MYSOs & 5.1 & [4.1, 6.4] & 5.6 & 4.5 \\
            AGAL014.114-00.574 & Protostellar & 2.2 & [1.7, 3.0] & 3.5 & 2.6 \\
            AGAL014.194-00.194 & Protostellar & 3.0 & [2.4, 3.8] & 8.2 & 3.9 \\
            AGAL014.492-00.139 & Quiescent & 1.9 & [1.5, 2.7] & 19.0 & 3.9 \\
            AGAL014.632-00.577 & MYSOs & 2.8 & [2.2, 3.5] & 2.5 & 1.8 \\
            AGAL015.029-00.669 & MYSOs & 5.1 & [4.3, 6.8] & 11.0 & 2.0 \\
            AGAL018.606-00.074 & Protostellar & 2.1 & [1.6, 2.9] & 8.7 & 4.3 \\
            AGAL018.734-00.226 & Protostellar & 4.2 & [3.4, 5.3] & 79.0 & 12.5 \\
            AGAL018.888-00.474 & Protostellar & 2.3 & [1.9, 3.2] & 28.0 & 4.7 \\
            AGAL019.882-00.534 & MYSOs & 4.2 & [3.5, 5.3] & 7.9 & 3.7 \\
            AGAL022.376+00.447 & Protostellar & 1.9 & [1.4, 2.6] & 6.2 & 4.0 \\
            AGAL023.206-00.377 & Protostellar & 5.8 & [4.6, 7.7] & 12.0 & 4.6 \\
            AGAL028.564-00.236 & Quiescent & 1.8 & [1.3, 2.6] & 54.0 & 5.5 \\
            AGAL028.861+00.066 & MYSOs & 5.8 & [4.7, 8.1] & 10.0 & 7.4 \\
            AGAL030.848-00.081 & Quiescent & 2.2 & [1.6, 3.1] & 12.0 & 4.9 \\
            AGAL031.412+00.307 & HII & 30.7 & [27.4, 38.6] & 30.0 & 4.9 \\
            AGAL034.258+00.154 & HII & 23.7 & [20.2, 29.7] & 8.0 & 1.6 \\
            AGAL034.401+00.226 & HII & 2.8 & [2.2, 3.5] & 2.7 & 1.6 \\
            AGAL034.411+00.234 & MYSOs & 5.8 & [4.9, 7.5] & 2.1 & 1.6 \\
            AGAL034.821+00.351 & MYSOs & 2.2 & [1.6, 3.0] & 1.1 & 1.6 \\
            AGAL035.197-00.742 & MYSOs & 4.2 & [3.7, 5.3] & 4.6 & 2.2 \\
            AGAL037.554+00.201 & MYSOs & 5.4 & [4.1, 7.2] & 12.0 & 6.7 \\
            AGAL049.489-00.389 & HII & 25.3 & [21.0, 31.3] & 110.0 & 5.4 \\
            AGAL053.141+00.069 & MYSOs & 3.2 & [2.4, 4.0] & 1.0 & 1.6 \\
            AGAL059.782+00.066 & MYSOs & 4.5 & [3.6, 5.6] & 2.5 & 2.2 \\
            \hline
        \end{tabular}
        \tablefoot{We indicate the highest-probability density interval containing the 67\% of the probability mass as HPD 67\%.}
    \end{table*}

    \subsection{Compression of clump-scale material}
    One thousand realisations of the intensities are generated according to the fit results, and the ratios are computed.
    When a line is not detected, the intensity is generated following a Gaussian distribution with mean and standard deviation equal to the measured RMS noise.
    From these ratios the average number density along the LOS is extracted, with its uncertainty, according to the procedure delineated in Section~\ref{subsec:density_inference}.
    The different ratios are combined as summarised in Figure~\ref{fig:density_posterior_from_pdf} to obtain the final posterior for the density that best represents all of the data.
    The best-fit results and the uncertainties are listed in Table~\ref{tab:poc_results}, together with the classification, the mass and distance of the sources.
    Despite the fact that the observations of the ($7_K-6_K$) band have a $\sim35\%$ smaller beam ($\sim19\arcsec$) than the ($2_K-1_K$) and ($5_K-4_K$) bands ($\sim28\arcsec$), each of the ratios is computed between lines with identical beams.
    When combining all the ratios for the fit, there might be a small discrepancy due to the fact that lower-density material enters the larger beams of the ($2_K-1_K$) and ($5_K-4_K$) bands, lowering somewhat the average density along the LOS with respect to what is traced by the \ratioThHEOneTHE\ line ratio in the ($7_K-6_K$) band.
    The effect of the difference in beam sizes is expected to be small.
    This is mainly because the \ratioThHEOneTHE\ ratio has a lower predictive power for density.
    Additionally, the material at the edge of the beam has a limited influence on the overall average due to the lower beam response in the integrated flux and lower column density.
    Indeed, a run without the \ratioThHEOneTHE\ ratio shows that results are unaffected for the $\sim85\%$ of sources, with the remaining $\sim15\%$ showing variations of around $10\%$ of the best-fit density value.

\begin{figure}
        \centering
        \includegraphics[width=0.8\columnwidth]{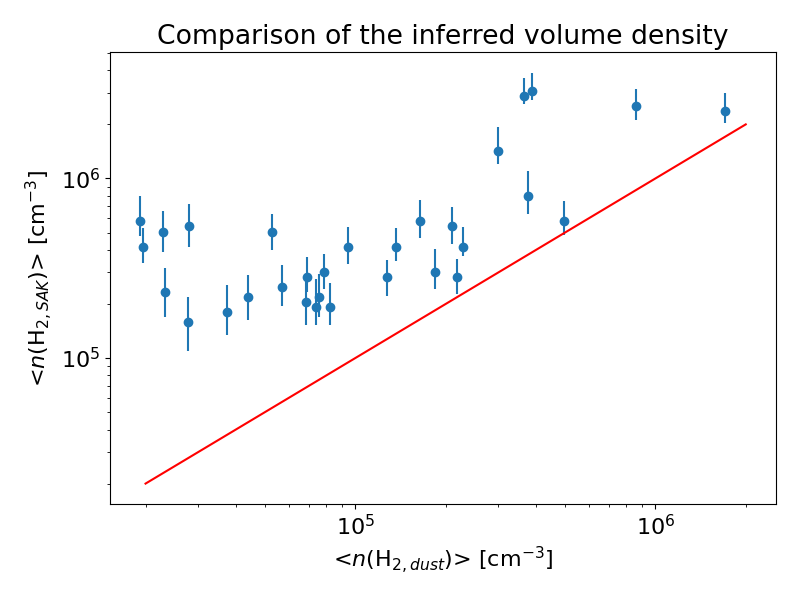}
        \caption{Comparison of the number density derived from the dust continuum emission, assuming spherical geometry, and that derived with the method proposed in this work. The red line indicates the 1:1 correspondence.}
        \label{fig:vol_density_comparison}
    \end{figure}

    Figure~\ref{fig:vol_density_comparison} shows the comparison of the number density derived from the continuum emission, as reported in \citet{Giannetti+17_aa603_A33}, assuming a spherical geometry, and those derived with SAK.
    The former estimate of the number density is obtained dividing the peak column density by the diameter of the source, as derived from the circularization of the FWHM of the ATLASGAL continuum emission.
    The uncertainties are reported only for SAK, because they are extremely difficult to quantify accurately when deriving the density from the dust continuum emission, using the assumption of spherical geometry.
    This estimate is affected by the unknown intrinsic morphology of the fragments, gas-to-dust ratio \citep{Giannetti+17_aa606_12}, the dust properties \citep[cf. ][]{Giannetti+2014_aa570_65}, temperature distribution and uncertainty \citep{Elia+2017_mnras471_100, Elia+2021_mnras504_2742}.
    The uncertainty is at least a factor of three, considering only the unknown size along the LOS.
    The figure shows that the CH$_3$OH-based estimates are always higher than those derived from dust continuum, with a discrepancy within a factor $\sim 2 - 5$.
    Above $n(\mathrm{H_2})\approx 10^5\usk\mathrm{cm^{-3}}$, the two measures become more correlated, with a relatively constant difference of a factor of $\sim 2$.

    Combining the measurement of column- and number densities, it is possible to have an independent estimate of the extension of the object along the LOS.
    Our approach therefore potentially provides a detailed view of the 3D structure of molecular condensations, allowing for pixel-by-pixel analysis and yielding insights into both individual objects and broader statistical samples.
    We stress that the correlation between the ratios and the density along the LOS (Fig.~\ref{fig:los_map_summary}) and the fact that it does not depend on the detailed distribution of the material (Sect.~\ref{subsec:fiducial_model} and Fig.~\ref{fig:density_distribution_comparison}) indicate that we can use our results also if the clumps are not strictly spherical.
    Thus, the systematic discrepancy in the number densities derived using the dust continuum and the independent ones from SAK is possibly suggesting that the effective radius is not a good representation of the size of the clump along the LOS.
    The higher density measured with SAK indicates that the size along the LOS is shorter than the size derived in the plane-of-the-sky.
    If we consider the typical aspect ratio of $1.5-2$ of ATLASGAL clumps \citep{Urquhart+2014_mnras443_1555}, the size along the LOS would actually be comparable to the projected minor axis.
    The discrepancy increases for sources with an average density $n(\mathrm{H_2})<10^5\usk\mathrm{cm^{-3}}$, as estimated from the dust continuum.
    In these clumps the density shows a median difference of a factor $\sim5$ compared to the CH$_3$OH-based estimates.
    Such a large discrepancy suggests that these young sources, less affected by feedback from YSOs, may even be more elongated compared to the other clumps.

\begin{figure*}
        \centering
        \includegraphics[width=0.39\hsize]{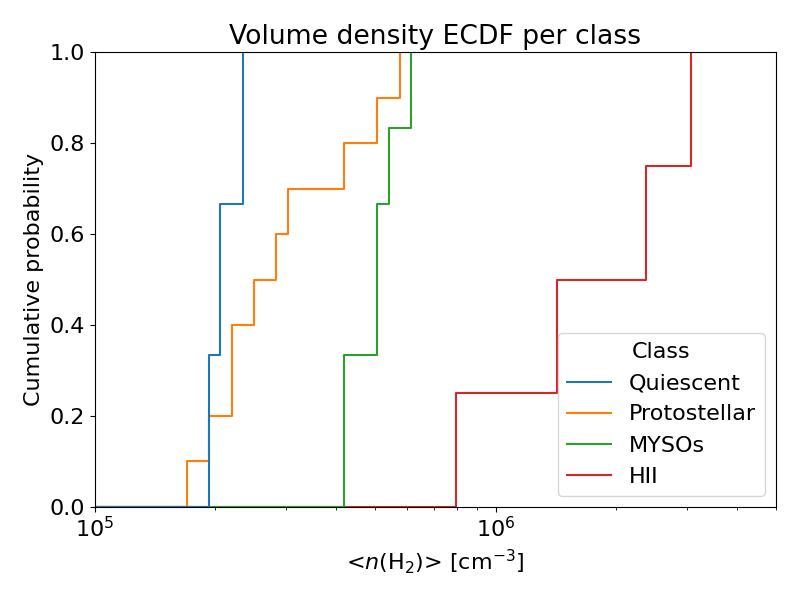}\hspace{0.05\hsize}
        \includegraphics[width=0.39\hsize]{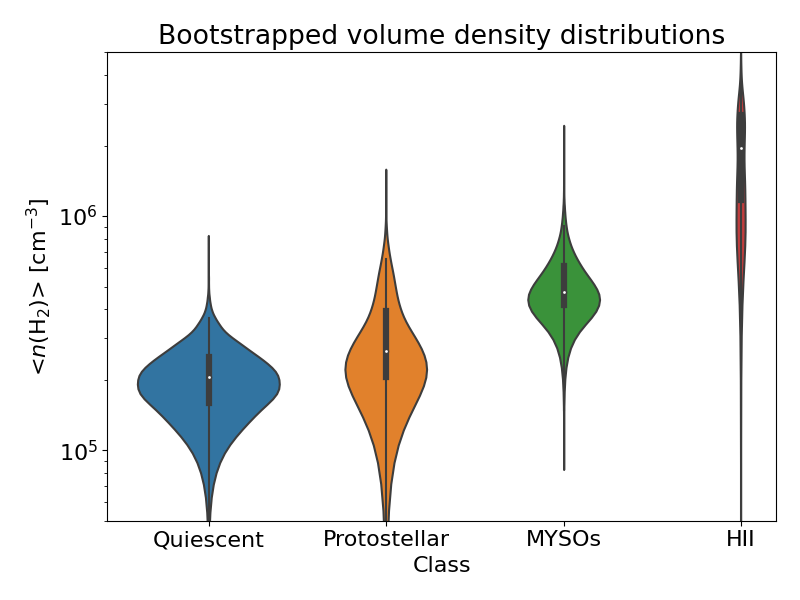}
        \caption{\label{fig:density_evolution}Left: Empirical cumulative distribution functions of the best-fit density for each evolutionary class in the TOP100 sample. Right: Violin plot of the bootstrapped density distribution per class; 10000 samples have been drawn for each source from the inferred number density posterior. The combined PDF, determined via a KDE smoothing, is shown around the quartiles and whiskers of a box plot.}
    \end{figure*}

    Figure~\ref{fig:density_evolution} summarises the temporal evolution of the number density.
    The number density of H$_2$ increases with evolutionary stage, indicating that the clump contracts and concentrates with time, under the action of gravity.
    Over the course of the early evolution, lasting around $5 \times 10^5\usk\mathrm{yr}$ \citep[e.g.][]{Urquhart+18_mnras473_1059, Sabatini+21_aa652_71}, there is a noticeable increase in the median density within the star-forming clumps.
    The typical density rises from around $2 \times 10^5\usk\mathrm{cm^{-3}}$ to approximately $1.5 \times 10^6\usk\mathrm{cm^{-3}}$, showing an acceleration in the later stages.
    Using again an Anderson-Darling test confirms that the four samples are drawn from different population, having a p-value lower than $0.001$.
    This result remains valid even after removing the H\textsc{ii} regions from the samples, or comparing only Protostellar clumps and those hosting a MYSO (p-value $\sim 0.01$ in both cases), confirming that the density evolves with time, even though by a limited amount in the first stages of high-mass star formation.
    Of course, larger samples are needed to better quantify typical properties per stage, and firmly characterise the amount of compression among them on pc scales, but this demonstrates the power of the proposed method on scales where such measurements are notoriously difficult.

    A similar acceleration is also observed in the evolution of the luminosity-to-mass ratio, as found by \citet{Urquhart+2022_mnras510_3389} from the analysis of the full ATLASGAL sample.
    Such an acceleration is typical of free-fall, and, in general, of scenarios with accretion rates that increase over time.
    If one computes the free-fall time of an early clump with a number density of $10^5\usk\mathrm{cm^{-3}}$ and compares it to the statistical lifetimes of the various phases, taken from \citet{Sabatini+21_aa652_71}, one finds that the former is around five times shorter than the latter.
    Considering that ATLASGAL clumps are already relatively compact, a fluctuation with a number density of $10^4\usk\mathrm{cm^{-3}}$ would have a free-fall time within a factor of two of the chemical and statistical lifetimes.
    This, together with the moderate and accelerating compression from cold, quiescent sources to those hosting a very compact H\textsc{ii} region, is an important point to test in theories of high-mass star- and cluster formation.
    In fact, depending on whether the collapse proceeds inside-out or outside-in the evolution of the clump average density could proceed differently.
    On the one hand, in a clump that starts collapsing from the outside, the inner Jeans-stable layers are crushed by the outer ones falling on them, progressively increasing the average density \citep{VazquezSemadeni+2019_mnras490_3061}.
    On the other hand, if the clump is in near equilibrium, has a fixed mass, and the collapse starts in the inner layers \citep{Shu1977_apj214_488}, the average density is constant, or could decrease due to the material deposited onto the protostars.

    \section{Summary and conclusions}\label{sec:conclusions}
    This study has developed a straightforward framework for constructing and post-processing simplified yet realistically accurate models of molecular gas condensations.
    The framework provides an easy-to-use, accurate method for estimating number densities of molecular hydrogen.
    This method relies on methanol, a widespread, abundant species and some of its most common and accessible emission bands, around $96.7$, $241.7$, and $338.2\usk\mathrm{GHz}$ (see Table~\ref{tab:lines_list}).
    We aim to improve the accuracy of deriving physical parameters of molecular clumps by moving away from the standard, yet uncertain, method of estimating the number density from column density or mass measurements based on size assumptions from plane-of-sky projections.
    Our approach helps balance the process by reducing the heavy reliance on temperature and spherical geometry assumptions in these calculations.

    The results demonstrate that the selected line ratios, especially those at three millimetres, are robust against variations in temperature and density distributions along the LOS.
    The lines in the $(2_K-1_K)$ band, in fact, increase steadily with density in the range $\sim 5\times 10^4- 3\times 10^7\usk\mathrm{cm^{-3}}$.
    This interval can be expanded to lower densities if all transitions are available, by combining the five line ratios analysed here.
    In this scenario, the minimum density that can be derived is as low as a $\mathrm{few} \times 10^3$, typical of entire molecular clouds.
    A comparison with uniform and isothermal clumps simulated with RADEX show that for a given value of the ratios the discrepancy with the average number density along the LOS can reach two orders of magnitude, showing the importance of including the clump envelope in the modeling, in the form of density gradients.

    In Sect.~\ref{sec:analytic_representation} we provide an analytical approximation of the relation that we find between each line ratio and the average number density along the LOS.
    The limits of validity and typical model uncertainties (not accounting for observational ones) are also provided.
    While the analytical approximations provide a less precise estimate of the uncertainty compared to the full SAK code, they can be used to reconstruct a first estimate of the density structure from line ratio maps, without local code execution.

    The main limitation of the method arises from the impact of molecular abundance on optical depth, which can shift the relationship.
    An increase in abundance by a factor of 100 results in an average decrease in the inferred density of a factor $\approx 15$.
    Such a large jump is typical of regions affected by strong shocks, as those associated with outflows, or large temperatures causing desorption from grains, as in hot cores, but it remains relatively constant in the cool envelope \citep[e.g.][]{Kristensen+2010_aa516_57, Gerner+14_aa563_97, Giannetti+17_aa603_A33, Sabatini+21_aa652_71}.
    In addition, under typical clump or core conditions, and more in general, for sufficiently high densities, the abundance of CH$_3$OH can be estimated via a simple LTE analysis, allowing the selection or up-weighting of the best-fitting grid.
    For a higher level of precision, optical depths can be estimated, and corrected for, when computing ratios by observing the corresponding lines of $^{13}$CH$_3$OH.
    In this case, using a run with low abundance to interpret the results would be preferable, because the impact of the optical depth would be already accounted for.

    We applied our method to the ATLASGAL TOP100 dataset to investigate the evolution of number density in high-mass clumps, selectively focusing on sources where both IRAM 30m telescope- and Atacama Pathfinder Experiment (APEX) data were at our disposal.
    These clumps showed a spectrum of average densities at the dust emission peak spanning from approximately $2 \times 10^5$ to around $\approx 2.5 \times 10^6\usk\mathrm{cm^{-3}}$.
    As the cluster-forming objects evolve, the density exhibited a remarkable, gradual ascent, which gained momentum with the progression of their life cycle, as depicted in Figure~\ref{fig:density_evolution}.
    This behaviour agrees with the theoretical expectations of an outside-in collapse, where the clumps are initially relatively uniform and slowly increase their average density under the compression of the infalling envelope.
    The timescales of the whole cluster-formation process as measured through statistical considerations \citep{Urquhart+18_mnras473_1059} or the chemical evolution \citep{Sabatini+21_aa652_71} are also roughly consistent with the free-fall time of the early clump, in agreement with the acceleration of the compression with evolution.

    Comparing the densities from SAK and those made under the assumption of a spherical geometry (Fig.~\ref{fig:vol_density_comparison}), we find indications that clumps might be elongated structures, with a size along the LOS comparable to the projected minor axis of the clumps, given their typical aspect ratio of $1.5-2$.
    This behaviour is even more pronounced in the earliest stages of evolution.

    Thanks to the robustness of specific ratios (e.g., \ratioESevenESix, \ratioEEightESeven, \ratioEEightESix, \ratioTwHFSevenTwHFSix, and \ratioThHEOneTHE) to temperature variations and the distribution of material along the LOS, we have calibrated a reliable relationship between these ratios and the LOS-averaged number density.
    This provides a straightforward method to independently measure the number density of molecular gas in various astrophysical contexts, from high-resolution observations in external galaxies to clumps, cores, and disks in the Milky Way.

    The primary advantage of SAK is its ability to reliably estimate the average number density along the LOS using both quick analytical relations and a more accurate numerical method based on multiple line ratios. This approach eliminates the need for creating a grid of LVG models and remains robust to variations in temperature and density gradients.
    Additionally, SAK facilitates the creation of LVG grids for different use cases, leveraging distributed systems for speed. It also simplifies user experience through effortless installation with containerization and zero-code configuration.
    Our framework can be easily extended to handle more complex models. We plan to include this feature to enable postprocessing of fully-fledged astrochemical simulations of clumps and cores.

    \begin{acknowledgements}
        We thank the referee for their constructive comments that greatly helped in the improvement of this work.
        This publication is based on data acquired with the Atacama Pathfinder Experiment (APEX). APEX is a collaboration between the Max-Planck-Institut fur Radioastronomie, the European Southern Observatory, and the Onsala Space Observatory.
        This work also make use of observations carried out with the IRAM 30 m Telescope, under project number 181-10.
        We acknowledge financial support PRIN-INAF 2019 From yOuNg Star clusters to planETary systems (ONSET)  (P.I. Leurini)
    \end{acknowledgements}

    \bibliographystyle{aa}
    \bibliography{references/master_biblio}

    \begin{appendix}
        \onecolumn
        \section{SAK Detailed infrastructure and automation}\label{app:automation}

        In the following, we provide additional information on how SAK is built, on what it can do, and on its internal workings.
        We first describe the Extract-Transform-Load (ETL) pipeline, that contains all of the SAK logic and code.
        Subsequently, we discuss how the application is organised as a whole.

        The Python ETL pipeline consists of three different layers:
        \begin{enumerate}
            \item staging (stg);
            \item model (mdl);
            \item presentation (prs),
        \end{enumerate}
        that correspond to the "classic" extraction, transformation, and loading of data.
        The pipeline is defined and orchestrated in a dedicated Python file, that executes the process for a grid of parameters.
        It is configured with external files in YAML format, a global one, and one for each of the ETL layers.
        The available parameters are documented in a markdown file in the code repository\footnote{\url{https://www.ict.inaf.it/gitlab/andrea.giannetti/swiss_army_knife_stable/documentation/documentation.md}}.

\begin{figure*}[!hb]
            \centering
            \includegraphics[width=0.67\hsize]{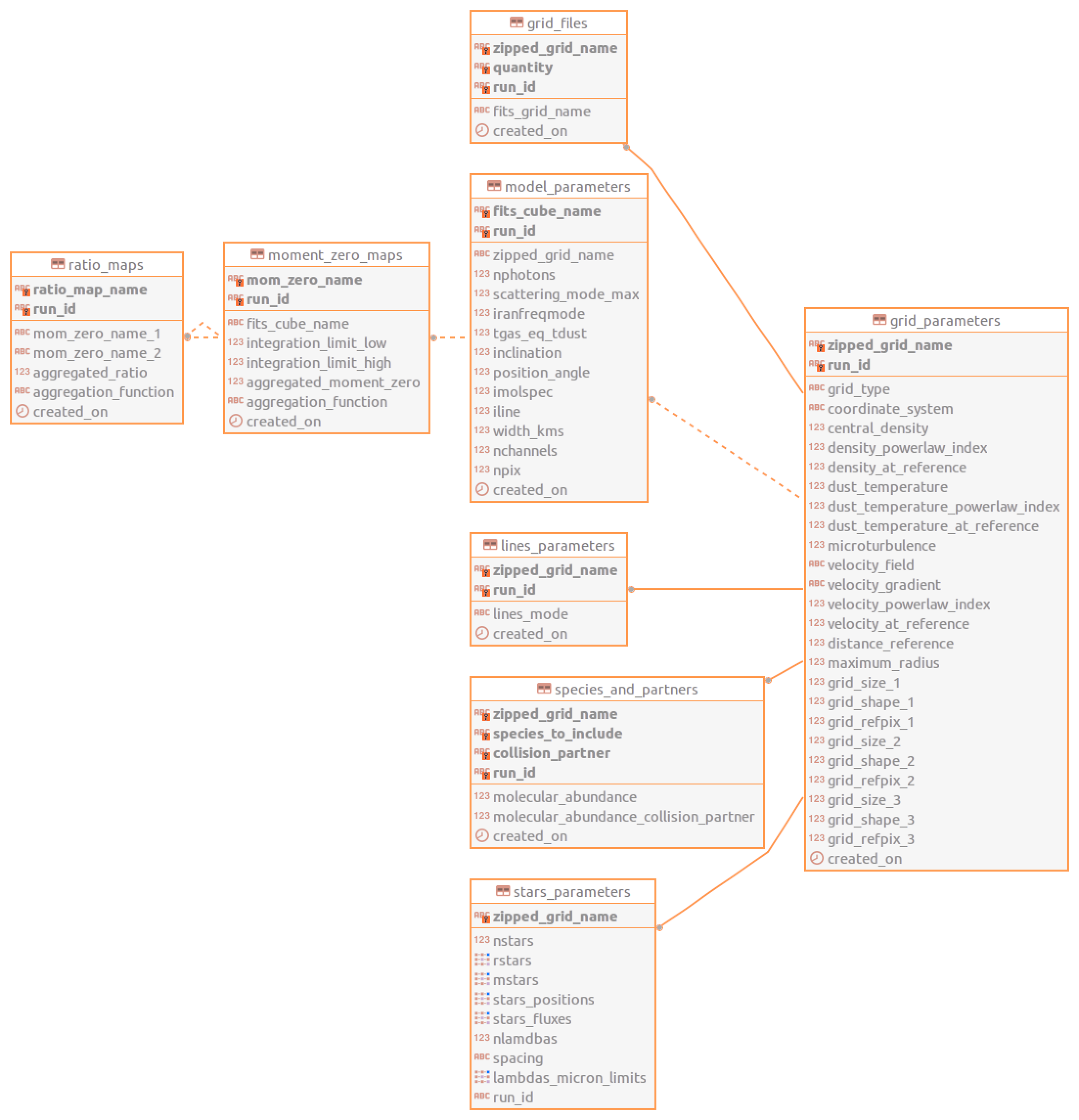}
            \caption{\label{fig:sak_erd}Entity-relation diagram created by the SAK tool. A description of the individual tables and columns can be found in the repository documentation.}
        \end{figure*}

        In the staging layer, a PostgresSQL database (DB) for storing and structuring all the data is initialised, i.e. the tables are created (see Fig.~\ref{fig:sak_erd}), and the RADMC 3D input files are generated to reflect the model physical grid of parameters.
        The complexity of the models created depends on the parameters specified in the staging configuration file.
        The postprocessing input files are modified automatically to reflect the desired configuration.
        For example, if the ``star'' section is defined, then one or more heating source(s) are placed in the grid, and the temperature distribution is recomputed based on their energy output, via the RADMC 3D ``mctherm'' command.
        Similarly, a step-function-like profile for the abundance can be adopted instead of a constant one, to simulate the evaporation of molecular species from grains in hot-cores.
        In this case, the user has to define a threshold temperature and the factor by which the abundance of a species is increased above this threshold.
        Molecular data can retrieved automatically from the LAMDA database and stored in the execution directory.
        It is also possible to use custom molecular levels, transition, and collisional coefficients data, by specifying that the molecular data file is cached, and by providing it to the system.
        Once all the input files for RADMC 3D are created, they are compressed and stored in an archive.
        The name of the compressed archive is stored into the database for future reference.
        All of the grid parameters are saved into the DB.

        In the model layer RADMC 3D is executed to compute the synthetic spectrum including one or more line of one or more molecular species (and optionally the temperature distribution is computed and updated in the compressed archive of the input files), the output is created and converted to fits format.
        The postprocessing computation mode (LTE, LVG, optically-thin LVG, as offered by RADMC 3D) can be selected directly from the input files, for example to investigate the impact of non-LTE excitation.
        The cube name and the model parameters are saved into the database.

        Finally, in the presentation layer, the cubes are collapsed to compute the moment-zero maps, the ratio maps are derived, and the ratios themselves are aggregated over the entire map.
        All the relevant information is stored in the database in the appropriate table (cf. Fig.~\ref{fig:sak_erd}).

        The pipeline has been containerised both with Docker \citep{merkel2014docker} and Apptainer \citep[formerly Singularity; ][]{Kurtzer+2017_singularity, gregory_m_kurtzer_2021_4667718}, to facilitate the execution of the system by different people on different machines.
        The application is divided into two services: the database, responsible for storing and structuring the data, and the ETL service, which encompasses the code itself.
        These two services are orchestrated by docker compose, a tool for defining and running multi-container applications\footnote{https://docs.docker.com/compose/}.
        The docker-compose.yaml file is available in the repository.
        The Dockerfile for the ETL container is provided, and the image can be retrieved from the repository.
        The Docker image can be imported in Apptainer for the execution on remote, distributed systems that do not support Docker, due to security issues.
        Indeed, a key security consideration is that Docker typically requires root access to function, which can be a limitation in some environments.
        We provide a script to facilitate this conversion step.
        The Apptainer definition for the database is also provided, so that this container can be created and run before executing the code itself.

        \FloatBarrier
        \section{Kernel density estimates computation}\label{app:kde}
        To obtain the KDE of the ratio-number density relations, an Epanechnikov kernel (inverted parabola) was used.
        Bandwidths for the kernel were selected independently for average density and line ratio.
        The average densities along the LOS are not uniformly-spaced in the diagram.
        Instead, they tend to cluster around the characteristic value of the density at 0.5~pc.
        Due to the model spacing in this parameter, using small bandwidth values can lead to larger values of the PDF at these specific density values (cf. Sect.~\ref{sec:robustness_ratios_density_temperature}).
        This could artificially introduce a bias in the density inference, and bandwidths were chosen to reduce this effect.
        When we did not consider ML-data augmentation, we adopted a bandwidth of $0.4$ in logarithmic space for the number density axis for all ratios; for the line ratio axes we used $0.2$ for \ratioESevenESix, \ratioEEightESeven, and \ratioEEightESix\ ratios, and $0.8$ and $0.4$ for \ratioTwHFSevenTwHFSix\ and \ratioThHEOneTHE\ ratios, respectively.
        Adding the data emulated though ML we could reduce these bandwidths by a factor of two, improving the final uncertainties in the inferred density.
        In general, larger values of the bandwidth reflect the lower sensitivity of specific ratios to the number density, according to the Spearman correlation coefficients listed in Table~\ref{tab:correlation_coefficients}.
        In fact, using a larger kernel for less sensitive ratios allows to reduce their weight on the final result, because the final posterior is wider.
        To reduce local variations and oscillations in figures showing the number density along the LOS (Figures~\ref{fig:los_map_summary}, \ref{fig:abundance_shift}, \ref{fig:density_distribution_comparison}, \ref{fig:hot_core_map_ratio}, \ref{fig:microturbulence_map_ratio}), the curve representing the maximum location of the PDFs, and the corresponding uncertainty, were smoothed using a moving average of 20 points.

        \section{Detailed description of individual experiments}\label{app:experiments}

        In the following the individual experiments and their results are described, as well as how they fit with one another.

        \subsection{Experiment 1 - Constant density and temperature}\label{subsec:uniform}

        Our first experiment was designed to be simple, essentially replicating the setup of \citet{Leurini+04_aa422_573}.
        We updated the collisional coefficients and narrowed the temperature range to be more representative of average values in high-mass clumps \citep{Urquhart+18_mnras473_1059, Elia+2021_mnras504_2742}.
        The gas was distributed in a spherical fashion, with constant density and temperature.
        The usual grid of 630 clumps is built, similar to the one for the fiducial models described in Sect.~\ref{subsec:model_details}, but with $p=q=0$ in Equations~\ref{eq:pl_density_profile} and \ref{eq:pl_temperature_profile}.
        This experiment essentially acted as a first, quick, benchmark for the SAK architecture.

        \begin{figure*}[!h]
            \centering
            \begin{tikzpicture}[font=\sffamily]
                \node at (0,2.3) {\tiny Integrated line ratio \ratioESevenESix};
                \node at (0,0) {\includegraphics[width=0.25\hsize]{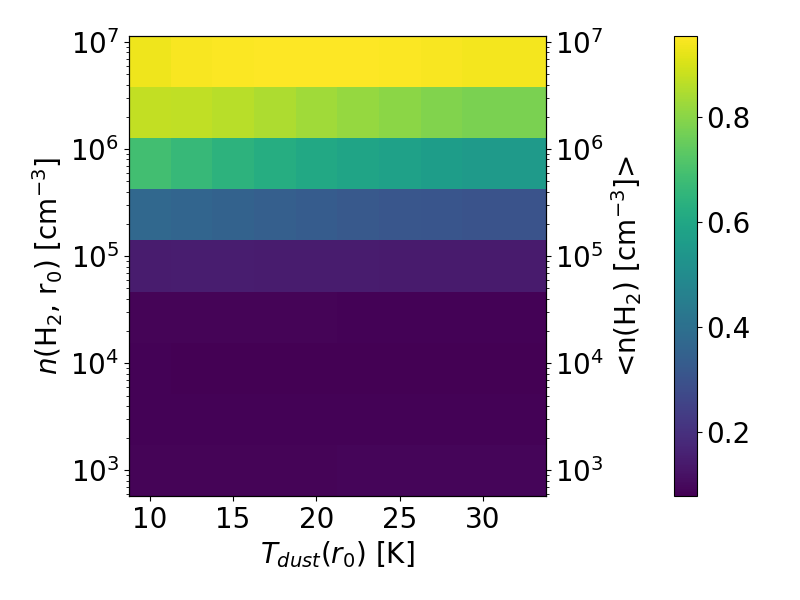}};
            \end{tikzpicture}\hspace{0.05\hsize}
            \begin{tikzpicture}[font=\sffamily]
                \node at (0,2.3) {\tiny Integrated line ratio \ratioEEightESeven};
                \node at (0,0) {\includegraphics[width=0.25\hsize]{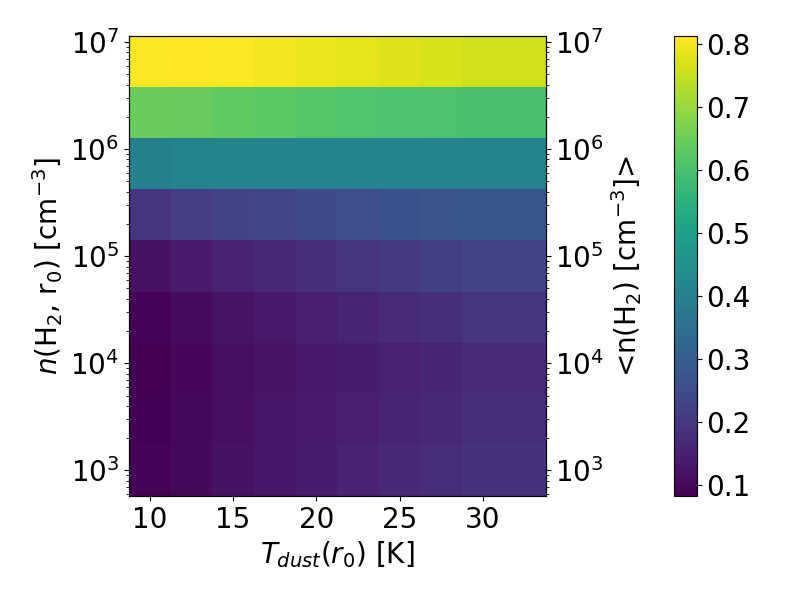}};
            \end{tikzpicture}\hspace{0.05\hsize}
            \begin{tikzpicture}[font=\sffamily]
                \node at (0,2.3) {\tiny Integrated line ratio \ratioEEightESix};
                \node at (0,0) {\includegraphics[width=0.25\hsize]{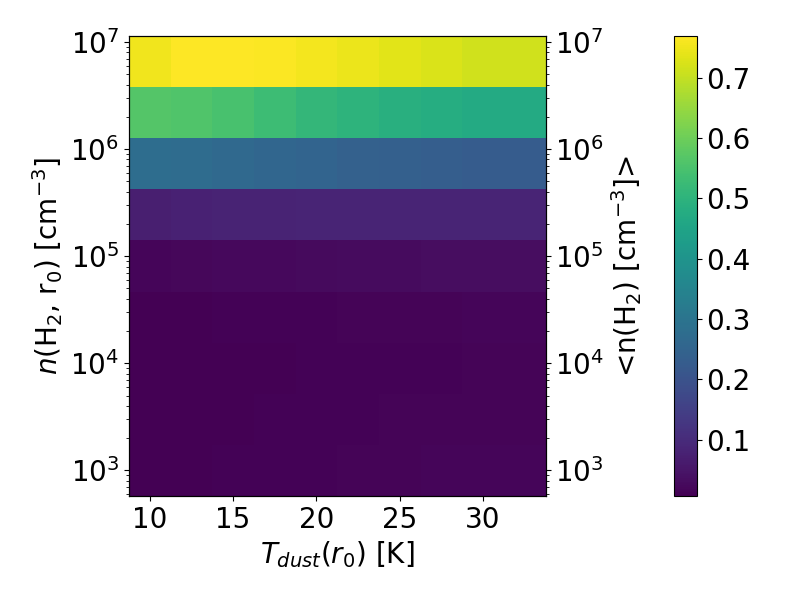}};
            \end{tikzpicture}

            \begin{tikzpicture}[font=\sffamily]
                \node at (0,2.3) {\tiny Integrated line ratio \ratioTwHFSevenTwHFSix};
                \node at (0,0) {\includegraphics[width=0.25\hsize]{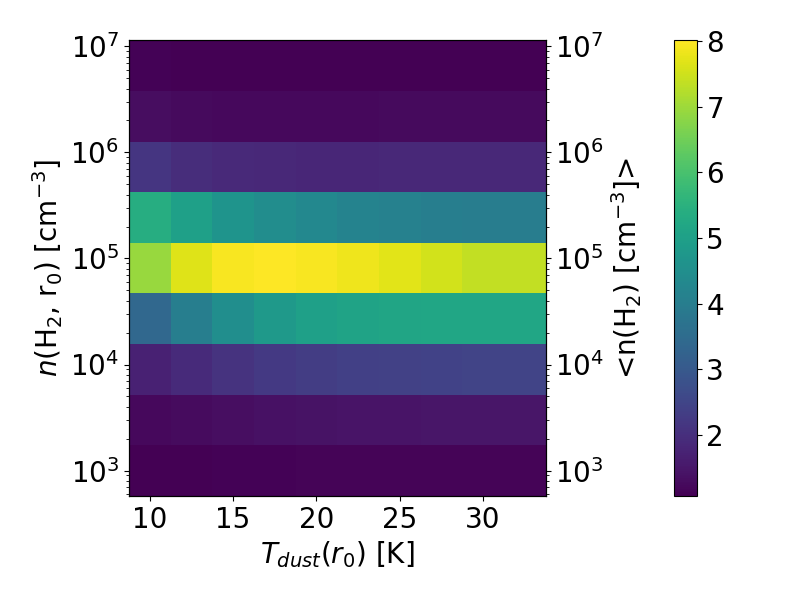}};
            \end{tikzpicture}
            \hspace{0.05\hsize}
            \begin{tikzpicture}[font=\sffamily]
                \node at (0,2.3) {\tiny Integrated line ratio \ratioThHEOneTHE};
                \node at (0,0) {\includegraphics[width=0.25\hsize]{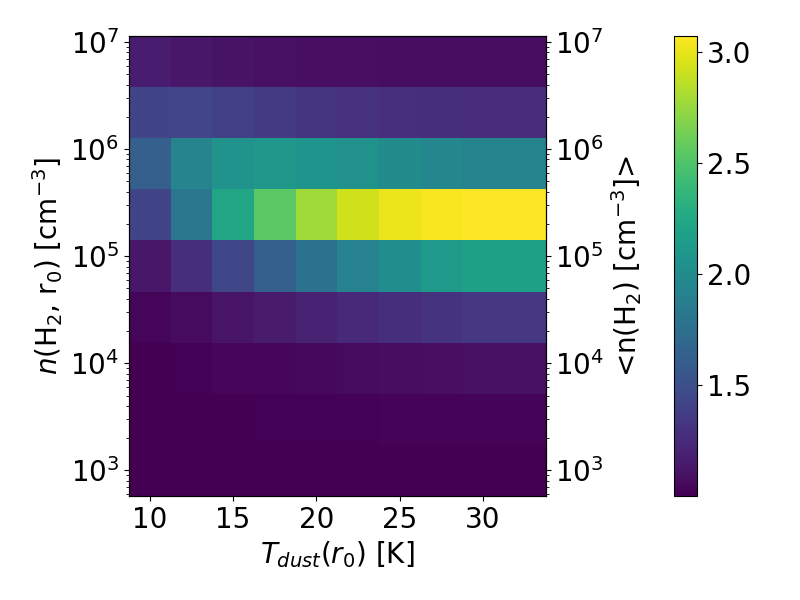}};
            \end{tikzpicture}
            \caption{\label{fig:uniform_aggregated_ratio}Integrated line ratios over the entire clump as a function of temperature and characteristic number density for the isothermal, constant-density case, for lines at around $96.7\usk\mathrm{GHz}$ (top row, for line ratios \ratioESevenESix, \ratioEEightESeven, and \ratioEEightESix), those at around $241.7\usk\mathrm{GHz}$ (bottom left, ratio \ratioTwHFSevenTwHFSix), and around $338.1\usk\mathrm{GHz}$ (bottom right, ratio \ratioThHEOneTHE).}
        \end{figure*}

        The results (Fig.~\ref{fig:uniform_aggregated_ratio}) are in qualitative agreement with those reported in \citet{Leurini+04_aa422_573}.
        In particular, we compared directly the ratios \ratioESevenESix\ and \ratioTwHFSevenTwHFSix\ for average column densities of methanol of the order of $10^{15}-10^{16}$ cm$^{-2}$.
        This shows that the experiment architecture is solid, and confirms that methanol lines can be used to estimate the number density of H$_2$, in principle with very little dependence on $T$.
        Moreover, it shows that the specific value of the collisional coefficients has limited influence on the general behaviour of the ratios.

        Being such a widespread molecule, CH$_3$OH could become the tracer of choice for measuring local values of the number density, both in our Galaxy and in external, nearby galaxies.
        However, the real world is more complex, and the method needs to be calibrated for ease of use, and its robustness to density variations along the line-of-sight, as well as to other model parameters, has to be evaluated.

        \subsection{Experiment 2 - Isothermal clumps with power-law density distributions}\label{subsec:power_law_density}

        The interstellar medium is not uniform, and a probe working only for constant densities would lose its usefulness in real-world applications.
        The values of the line ratios should depend the average value of $n_\mathrm{H_2}$ along the line-of-sight, ideally not changing exceedingly for different distributions of the gas.
        To verify whether this is the case, we constructed three grids of models, similar to those described in Sect.~\ref{subsec:model_details}, but with a constant temperature, i.e. using $q=0$ in Eq.~\ref{eq:pl_temperature_profile}.

        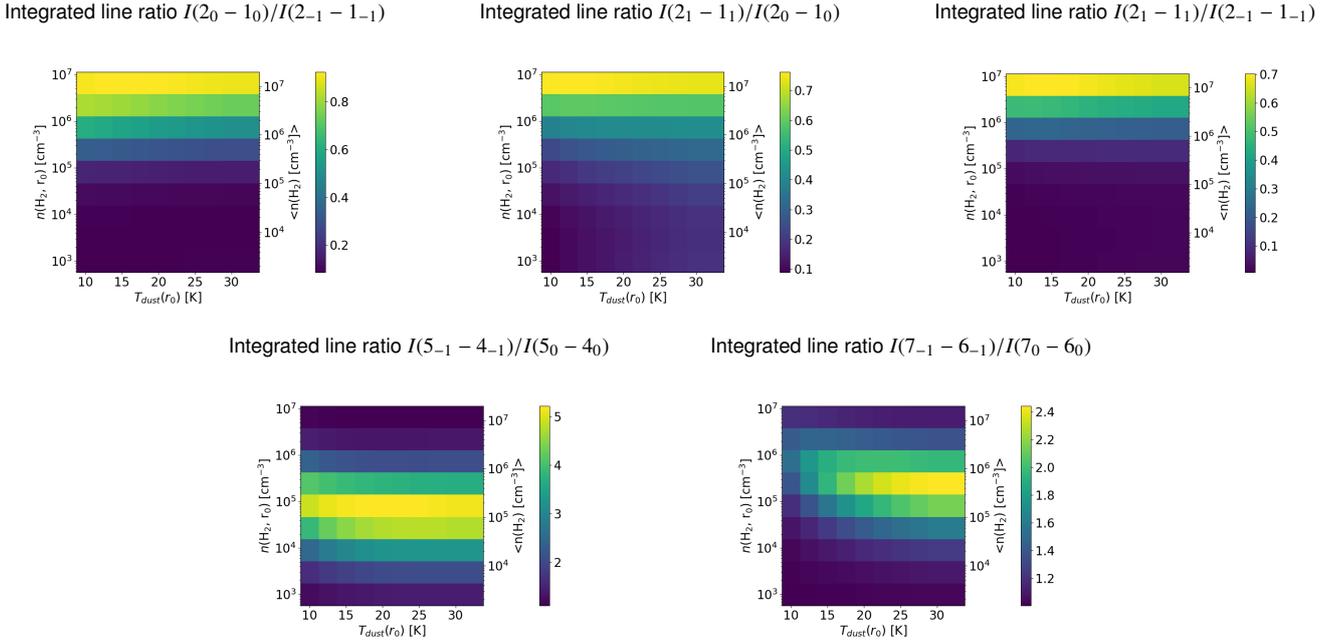
\begin{figure*}[!h]
            \centering
            \begin{tikzpicture}[font=\sffamily]
                \node at (0,2.3) {\tiny Integrated line ratio \ratioESevenESix};
                \node at (0,0) {\includegraphics[width=0.25\hsize]{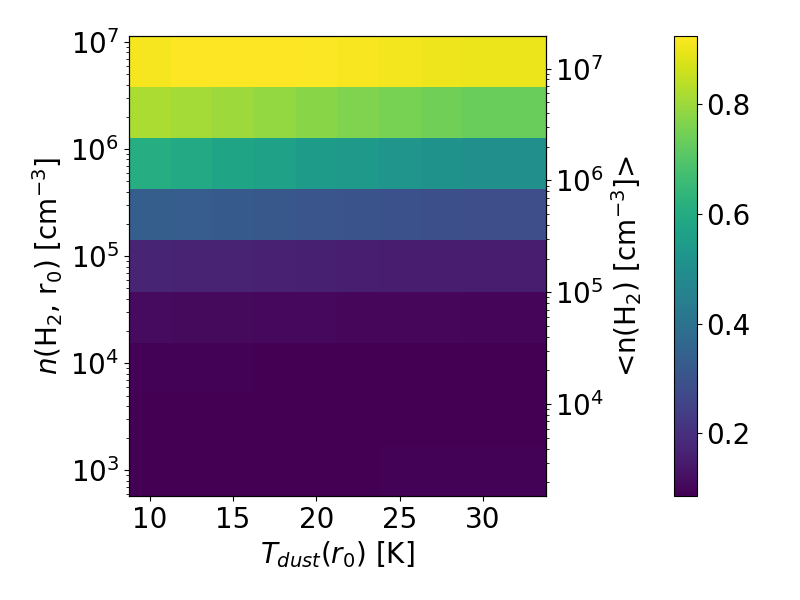}};
            \end{tikzpicture}\hspace{0.05\hsize}
            \begin{tikzpicture}[font=\sffamily]
                \node at (0,2.3) {\tiny Integrated line ratio \ratioEEightESeven};
                \node at (0,0) {\includegraphics[width=0.25\hsize]{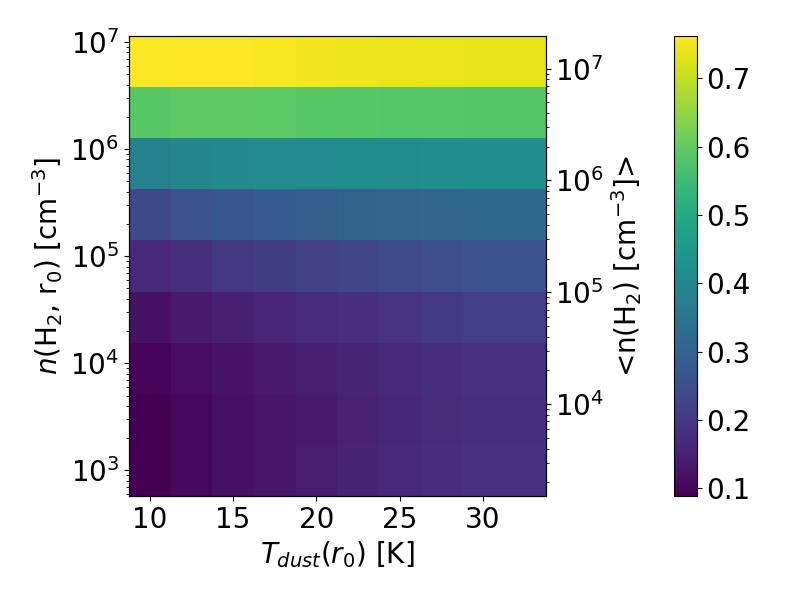}};
            \end{tikzpicture}\hspace{0.05\hsize}
            \begin{tikzpicture}[font=\sffamily]
                \node at (0,2.3) {\tiny Integrated line ratio \ratioEEightESix};
                \node at (0,0) {\includegraphics[width=0.25\hsize]{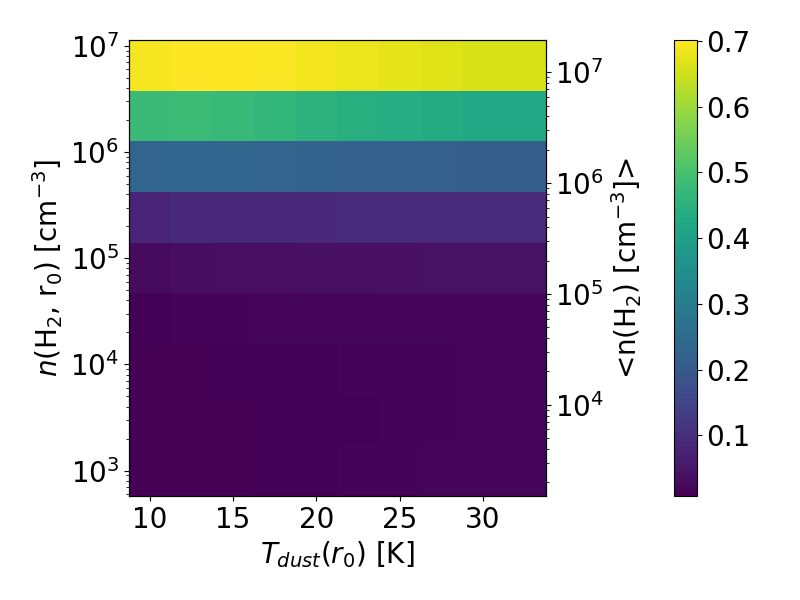}};
            \end{tikzpicture}

            \begin{tikzpicture}[font=\sffamily]
                \node at (0,2.3) {\tiny Integrated line ratio \ratioTwHFSevenTwHFSix};
                \node at (0,0) {\includegraphics[width=0.25\hsize]{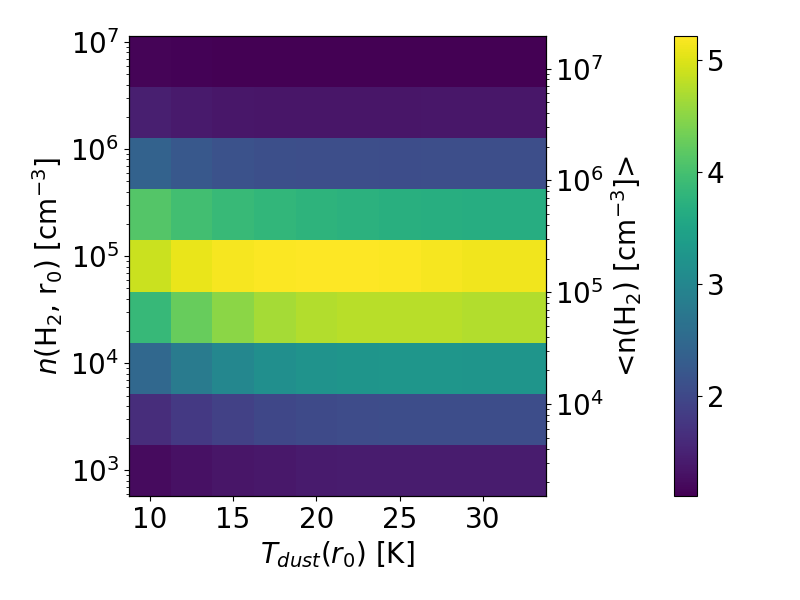}};
            \end{tikzpicture}
            \hspace{0.05\hsize}
            \begin{tikzpicture}[font=\sffamily]
                \node at (0,2.3) {\tiny Integrated line ratio \ratioThHEOneTHE};
                \node at (0,0) {\includegraphics[width=0.25\hsize]{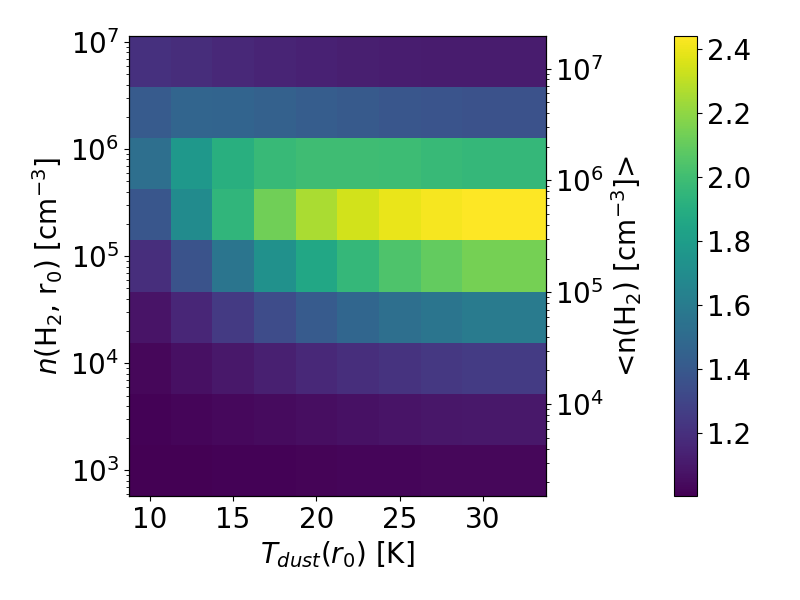}};
            \end{tikzpicture}
            \caption{\label{fig:isothermal_aggregated_ratio}Integrated line ratios over the entire clump as a function of temperature and number density for the isothermal case with a density distribution $\propto r^{-1.5}$, for lines at around $96.7\usk\mathrm{GHz}$ (top row, for line ratios \ratioESevenESix, \ratioEEightESeven, and \ratioEEightESix), those at around $241.7\usk\mathrm{GHz}$ (bottom left, ratio \ratioTwHFSevenTwHFSix), and around $338.1\usk\mathrm{GHz}$ (bottom right, ratio \ratioThHEOneTHE).}
        \end{figure*}

        Again, to have an overview of the ratio behaviour in the clump (relevant for the observations of unresolved clumps or clouds, too), we show in Fig.~\ref{fig:isothermal_aggregated_ratio} the aggregated ratio as a function of temperature and characteristic density at 0.5~pc.
        The $96.7\usk\mathrm{GHz}$ line ratios have a monotonic behaviour, increasing for larger characteristic densities, which is proportional to the average density of the clump.
        The ratios also show very weak dependence on the temperature of the gas.

        Compared to the $96.7\usk\mathrm{GHz}$ lines, higher-frequency lines have higher rates of radiative de-excitation.
        Consequently, for the same characteristic number density and temperature, they reach higher optical depths.
        They also include lines with energies above the ground state that can be significant compared to the gas temperature.
        This causes a more complex behaviour of the ratios, that first increase and then decrease approaching one when both the lines become heavily optically thick.
        Relying solely on these lines to infer $n_\mathrm{H_2}$ in typical clump density regimes is, thus, less effective.
        This is because, excluding the peak ratio value, each ratio can correspond to two distinct density values, leading to ambiguity.
        However, if only those lines are available, the use of a prior may help in breaking the degeneracy.
        Such ratios are also sensitive to lower densities than those at $96.7\usk\mathrm{GHz}$ and can be very useful when tracing gas on larger scales, for example in other galaxies, or to complement what is seen at $96.7\usk\mathrm{GHz}$, for the external layers of the clump.

        The extremely interesting result here is that the values of the ratios are essentially insensitive to the density distribution for different values of $p$, for the lines at $96.7\usk\mathrm{GHz}$, which are confirmed to be the best tracers of number density.
        So integrated values of this ratio can be used to infer average densities $\gtrsim10^{5} \mathrm{cm}^{-3}$.
        Combined $241.7\usk\mathrm{GHz}$ lines, this can be pushed down to $\sim\mathrm{few} \times 10^{3} \mathrm{cm}^{-3}$.

        We can obtain more from these line ratios than the clump-averaged density.
        Because of how the system is built, we don't just have access to the clump-scale ratios, but we also have the pixel-by-pixel maps of the ratios and of the LOS-average of the physical properties.
        So it is possible to test whether the relation holds for individual pixels, i.e., at scales much smaller than the clump itself, and independently of the detailed geometry.

        We followed the procedure described in Sect.~\ref{subsec:model_details}, and the results are shown in Figure~\ref{fig:los_map_summary}.
        All $96.7\usk\mathrm{GHz}$ line ratios show a tight correlation with the average density along the LOS.
        As in the case of integrated ratios, no appreciable difference is seen among models with different density distributions, indicating that the relationship is very robust, at least for isothermal clumps.
        Again, the information from the $241.7\usk\mathrm{GHz}$ and $338.1\usk\mathrm{GHz}$ bands can be used to increase the applicability regime of the $96.7\usk\mathrm{GHz}$ ratios to $\sim\mathrm{few} \times 10^{3} \mathrm{cm}^{-3}$.

        \subsection{Experiment 3 - Internally-heated clumps}\label{subsec:fiducial_model}

        Observations and theoretical models both indicate that active SF significantly alters clump temperatures on large scales, resulting in power-law like distributions \citep[e.g.][]{vanderTak+2000_apj537_283, Lin+21_aa658_128}.
        A steep temperature profile like the one we used ($q=-0.5$) maximises the potential variations in the line ratios.
        Because in Sect.~\ref{subsec:power_law_density} we considered isothermal clumps, i.e. with a power law exponent of 0, we encompass two extreme cases for the temperature distribution, maximising the effects on the ratios, exactly like we did for the density profiles.

        Figure~\ref{fig:integrated_ratios_across_grids} shows the integrated ratios over the clump for the three different grids.
        The ratios show little dependence on the density distribution slope, especially at $96.7\usk\mathrm{GHz}$, and have a similar dependence on gas temperature.
        For higher frequency lines, the ratios distribution becomes broader along the number density axis for steeper density gradients.
        This makes sense, because the material inside the scaling radius is denser for steeper gradients, and therefore the ratios are higher, on average.

        When one looks at the ratio as a function of the average value of the number density along the LOS (Fig.~\ref{fig:los_map_summary}), though, the only difference compared to the isothermal case is a marginally larger dispersion of the points, which increases with frequency.
        This larger dispersion is virtually invisible in the density plots, because only a small fraction of the points shows this behaviour, and the kernel is relatively large.
        Temperature gradients are the responsible for this effect, but their importance remain of second order.
        That said, the constraints on the number density can be tightened if we also compare line ratios sensitive to the temperature.
        While this is important, we want to maintain the method as simple as possible, and adding such information would make our results much more model-dependent.

\begin{figure*}[!h]
                \centering
                \includegraphics[width=0.28\hsize]{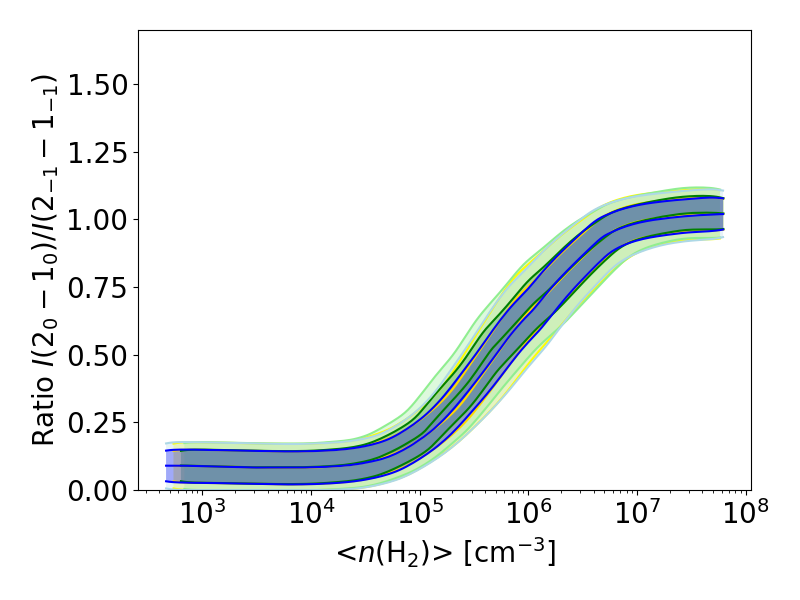}\hspace{0.05\hsize}
                \includegraphics[width=0.28\hsize]{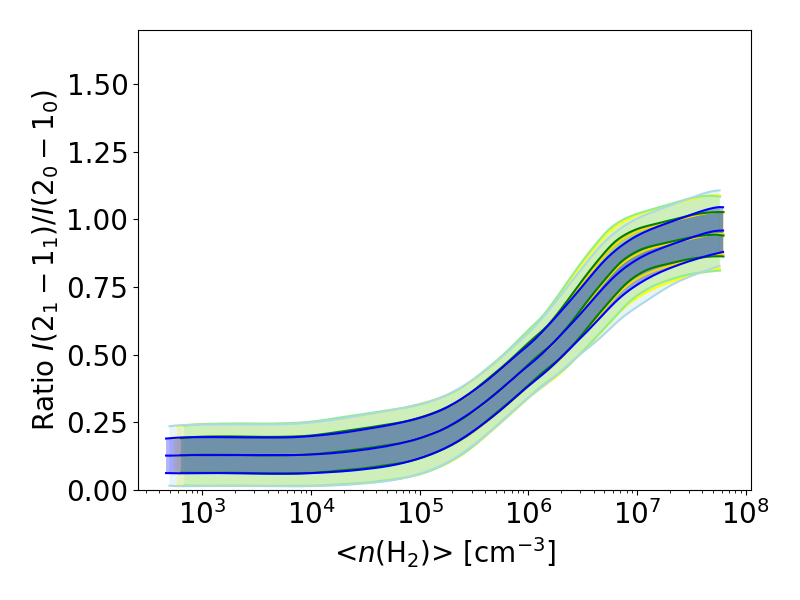}\hspace{0.05\hsize}
                \includegraphics[width=0.28\hsize]{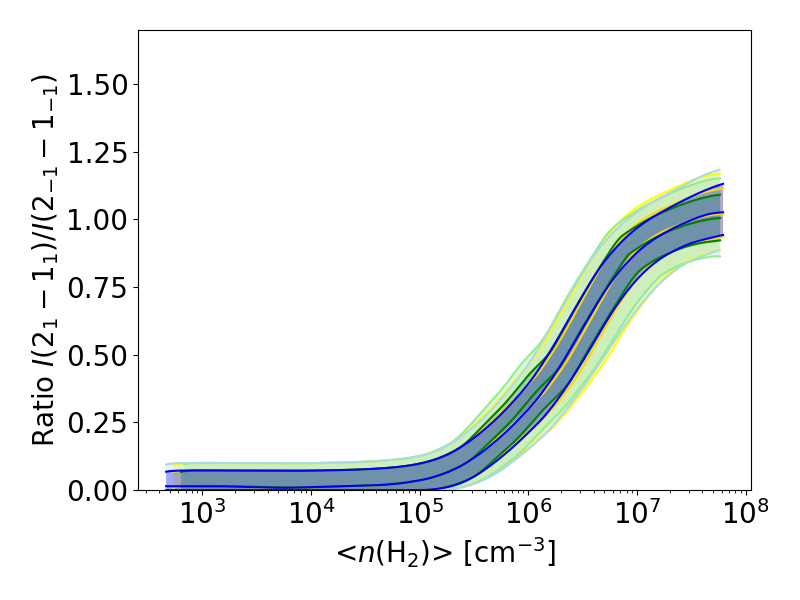}\hspace{0.05\hsize}
                \includegraphics[width=0.28\hsize]{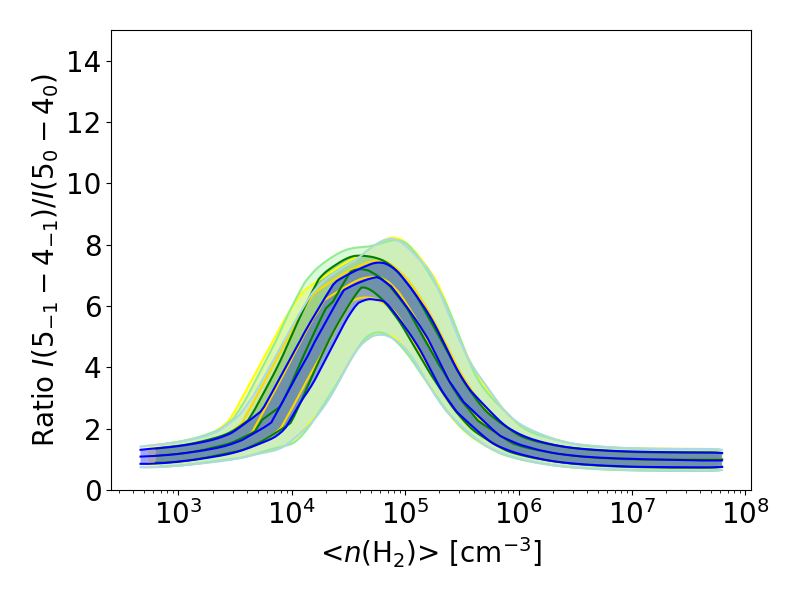}\hspace{0.05\hsize}
                \includegraphics[width=0.28\hsize]{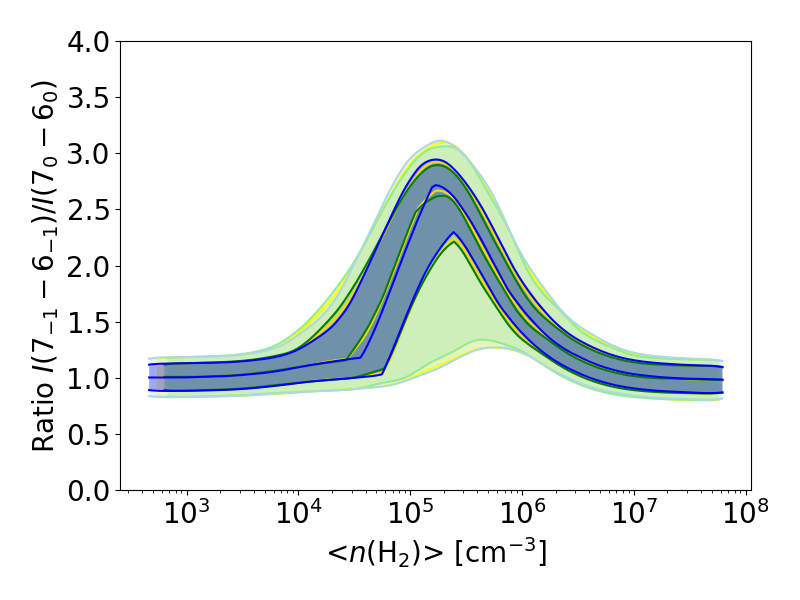}
                \caption{\label{fig:density_distribution_comparison}Comparison between the relations when using a different power-law distribution for density. The green-shaded region corresponds to $p=-1.2$, the yellow one to $p=-1.5$, while the blue area refers to $p=-1.8$.}
            \end{figure*}

        Different grids also show no appreciable differences in the LOS-averaged plots (Fig.~\ref{fig:density_distribution_comparison}).
        The larger average values of the integrated ratio for steeper grids are therefore driven by the larger average density of the clumps, rather than by the distribution of the material along the LOS.
        This confirms that the indicators that we propose are indeed sensitive to the number density, irrespective of how the material is distributed.

        \subsection{Experiment 4 - Simulating evaporation from grains}\label{subsec:hot_core}

        Our understanding of methanol chemistry remains incomplete.
        In fact, accurately predicting the abundance of CH$_3$OH, and related molecules like H$_2$CO, is a known weakness of current astrochemical models \citep{Gerner+14_aa563_97, Sabatini+21_aa652_71}.
        Despite this shortcoming, the qualitative behaviour can be reproduced.
        We know that CH$_3$OH is formed onto dust grains, and that its abundance is boosted in the gas phase by around two orders of magnitude in hot cores, following evaporation from the grains, above $T\sim90\usk\kelvin$ \citep[e.g.][]{Giannetti+17_aa603_A33, Sabatini+21_aa652_71}.
        We approximate this behaviour with a step function abundance profile. The abundance is increased by a factor of 100 above the threshold in temperature. Again, the same three grids of models with different density distributions are created.

\begin{figure*}[!h]
            \centering
            \includegraphics[width=0.28\hsize]{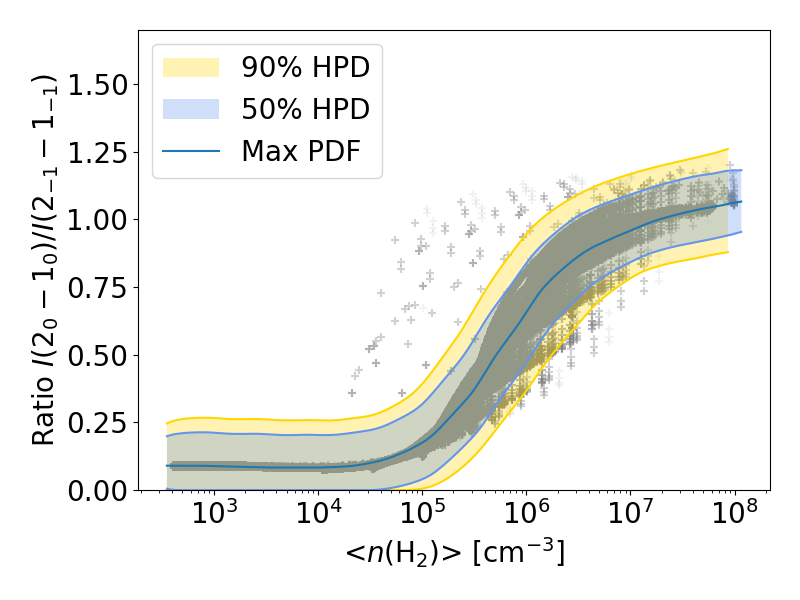}\hspace{0.05\hsize}
            \includegraphics[width=0.28\hsize]{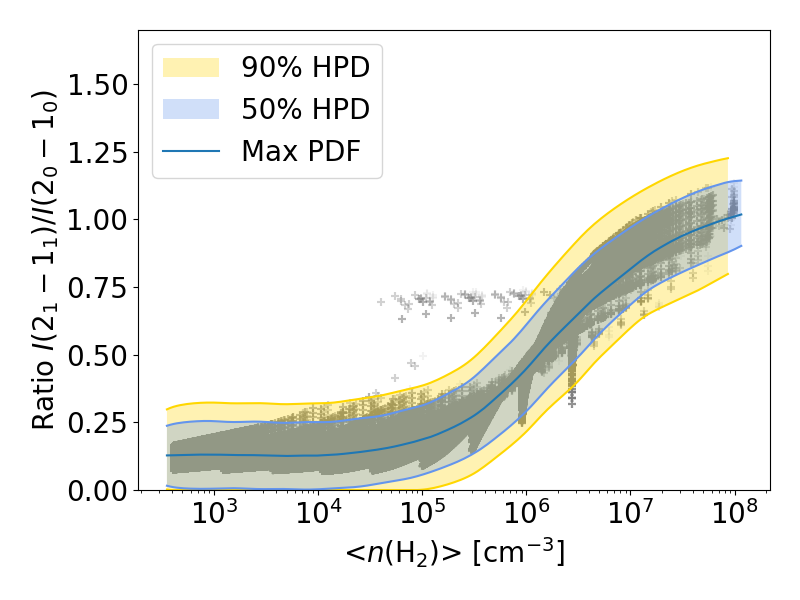}\hspace{0.05\hsize}
            \includegraphics[width=0.28\hsize]{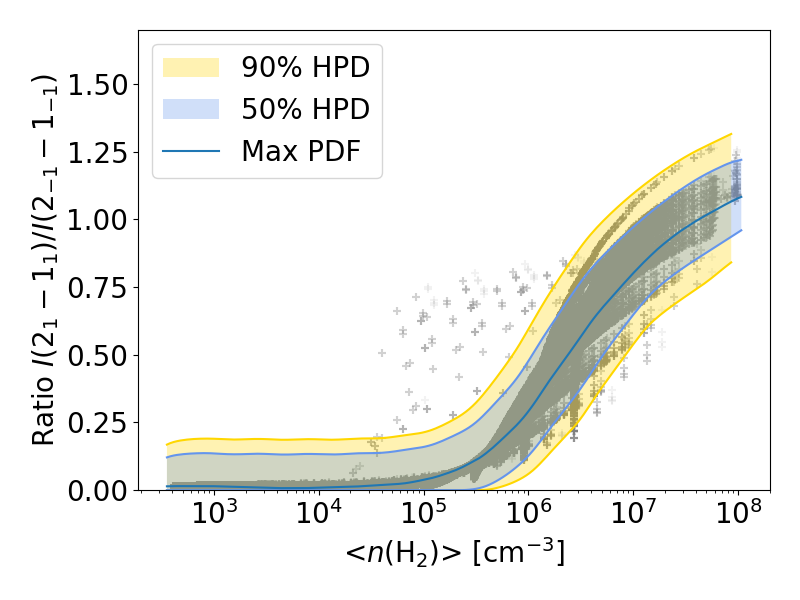}\\
            \includegraphics[width=0.28\hsize]{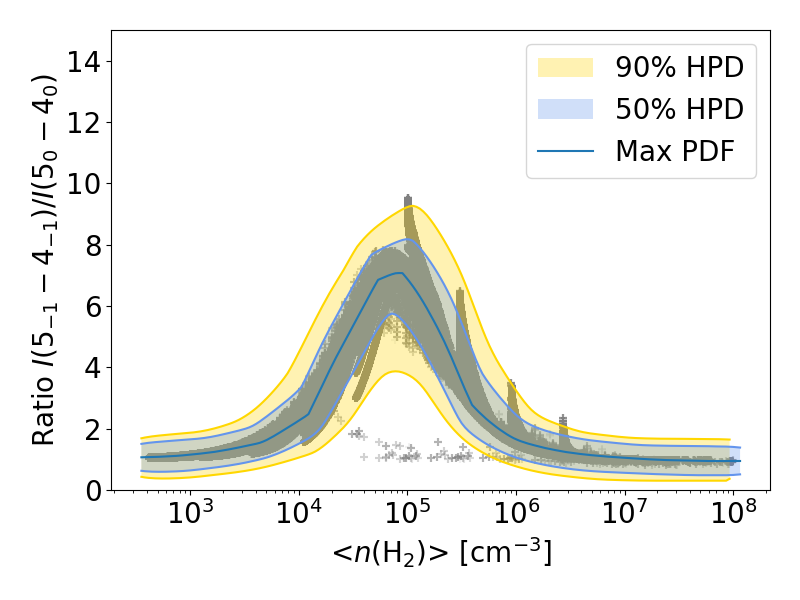}\hspace{0.05\hsize}
            \includegraphics[width=0.28\hsize]{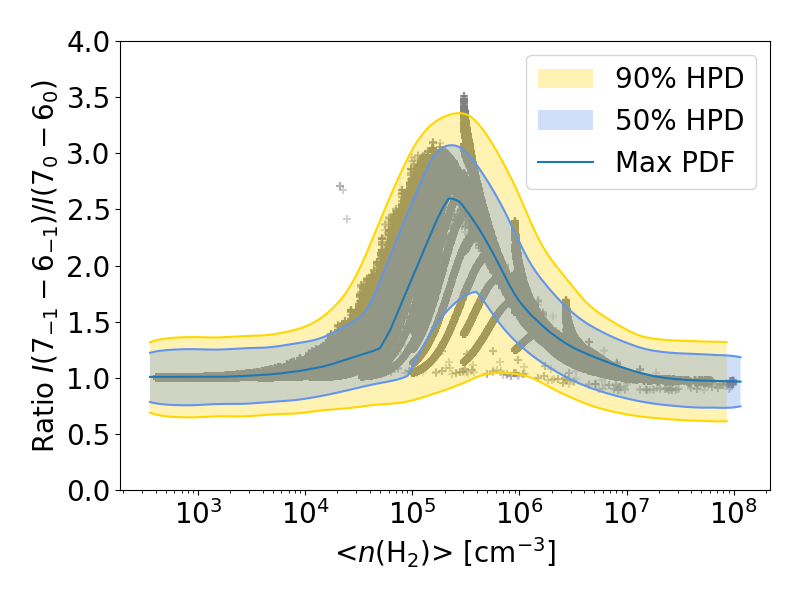}
            \caption{\label{fig:hot_core_map_ratio}Kernel density estimate of the probability density function for the line ratios as a function of temperature and average number density along the line-of-sight, for the internally-heated clump case with a density distribution $\propto r^{-1.5}$ and a hot-core-like abundance profile, for lines at around $96.7\usk\mathrm{GHz}$ (top row, for line ratios \ratioESevenESix, \ratioEEightESeven, and \ratioEEightESix), those at around $241.7\usk\mathrm{GHz}$ (bottom left, ratio \ratioTwHFSevenTwHFSix), and around $338.1\usk\mathrm{GHz}$ (bottom right, ratio \ratioThHEOneTHE).}
        \end{figure*}

        The results from this experiment are particularly interesting, and suggest the direction for additional tests.
        Specifically, the central pixels, where the LOS passes through the hot-core-like region, exhibit higher ratios.
        This increase is not due to higher densities, but rather to the increased opacity of the lines.
        For very optically thick lines the ratio would be close to one for the line ratios considered (Fig.~\ref{fig:hot_core_map_ratio}). While this seems to make density inference more complicated, these points are separable in the plot by measuring the average abundance along the LOS. The optical depth also affects different bands differently, because of the larger spontaneous emission rates at higher $J$s, as mentioned earlier.
        In addition to this possibility of breaking the degeneracy, only very few points end up outside the relationship found for the constant abundance case. The net effect is to increase somewhat the potential uncertainty on the inferred number density, because the probability would be more spread out. The behaviour of the ratios for high abundances highlights the sensitivity of any single ratio to the line opacity, and therefore to the average abundance of the species, as well as its line width.

        \subsection{Experiment 5 - Optical depth}\label{subsec:optical_depth}

        In this final check we wanted to go deeper on the impact of the line opacities on the ratio-number density relation. We constructed grids where the constant abundance is shifted one order of magnitude up and down from the fiducial value of $10^{-9}$, one where the microturbulence is doubled, and two with a hot-core abundance profile, where, again, the abundance is scaled up and down by a factor of 10.

\begin{figure*}[!h]
            \centering
            \includegraphics[width=0.28\hsize]{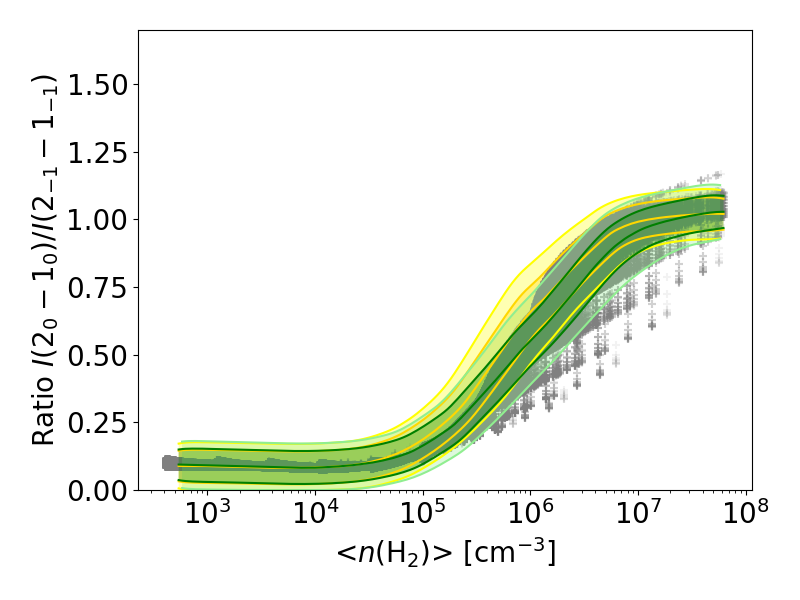}\hspace{0.05\hsize}
            \includegraphics[width=0.28\hsize]{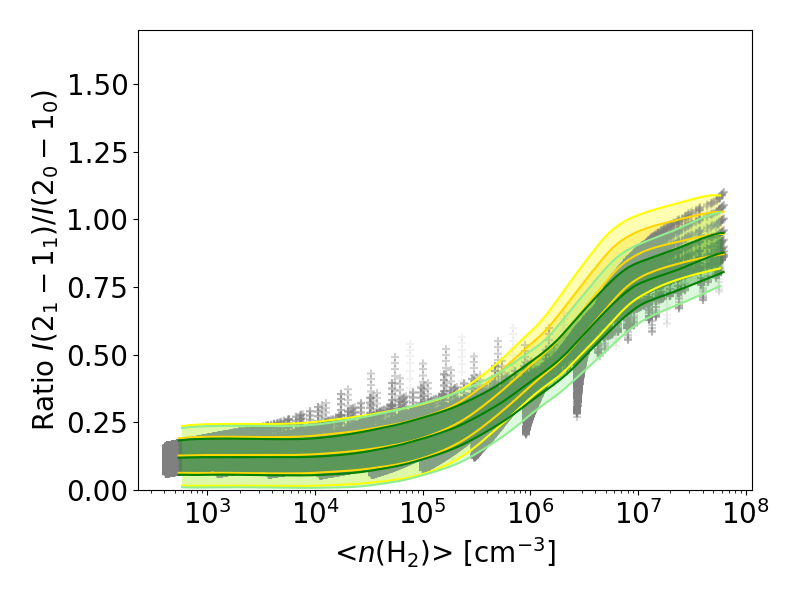}\hspace{0.05\hsize}
            \includegraphics[width=0.28\hsize]{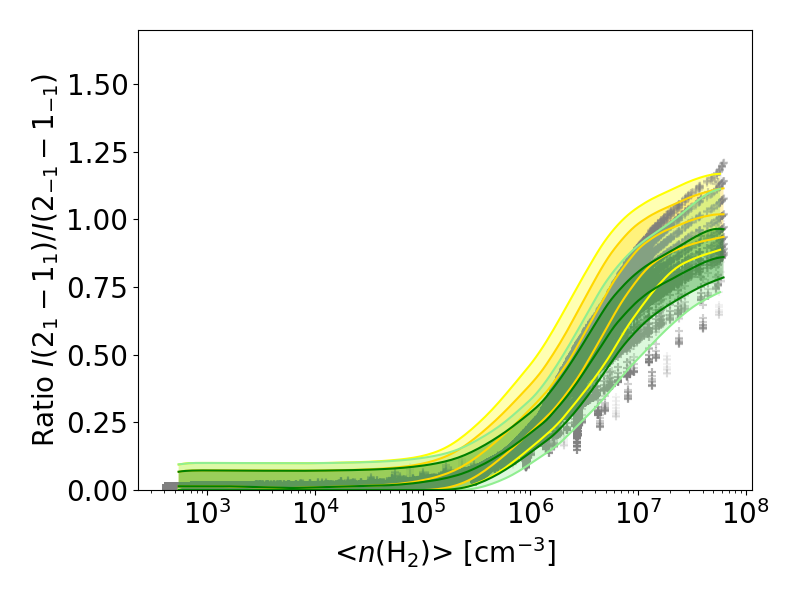}\\
            \includegraphics[width=0.28\hsize]{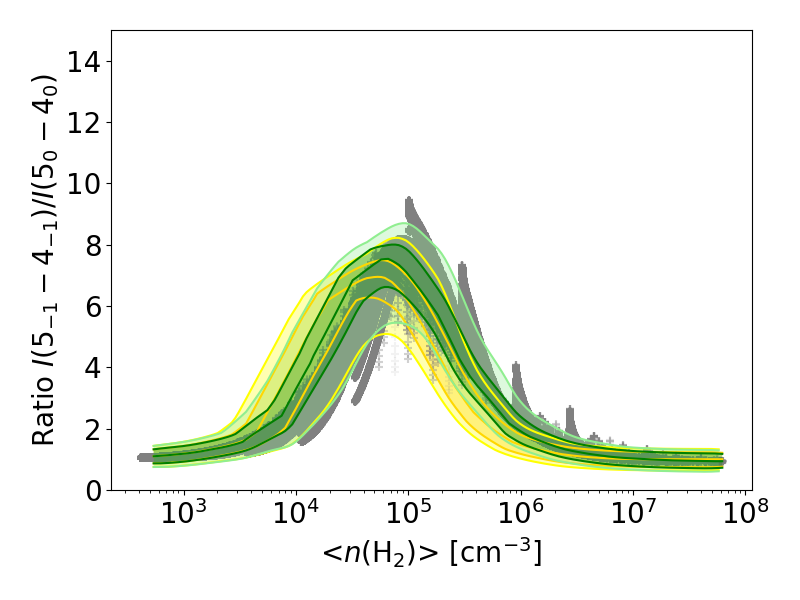}\hspace{0.05\hsize}
            \includegraphics[width=0.28\hsize]{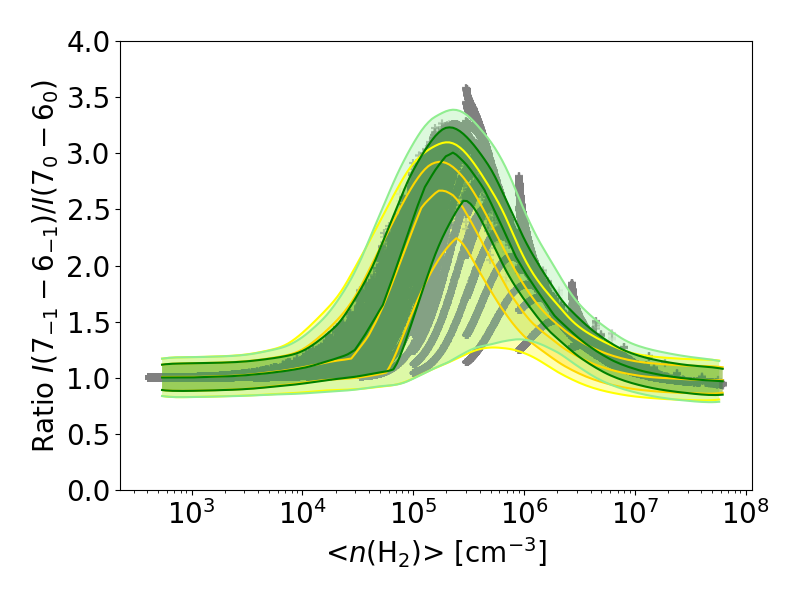}
            \caption{\label{fig:microturbulence_map_ratio}Kernel density estimate of the probability density function for the line ratios as a function of temperature and average number density along the line-of-sight, for the internally-heated clump case with a density distribution $\propto r^{-1.5}$, and a doubled microturbulence level, for lines at around $96.7\usk\mathrm{GHz}$ (top row, for line ratios \ratioESevenESix, \ratioEEightESeven, and \ratioEEightESix), those at around $241.7\usk\mathrm{GHz}$ (bottom left, ratio \ratioTwHFSevenTwHFSix), and around $338.1\usk\mathrm{GHz}$ (bottom right, ratio \ratioThHEOneTHE). The results for the fiducial values of microturbulence are shown in yellow, for reference.}
        \end{figure*}

        Doubling the microturbulence level has a limited effect, only slightly delaying the ratio saturation at the constant value associated with high optical depths.
        This effect is noticeable at high number- and column densities (cf. Fig.~\ref{fig:microturbulence_map_ratio}).
        The \ratioEEightESix\ is the most affected, with a discrepancy in the maximum location of slightly more than a factor of 2 at densities $\approx 10^7\usk \mathrm{cm^{-3}}$.
        A larger effect can be seen when the constant abundance is rescaled, as discussed in Sect.~\ref{sec:results}.

        In all cases, larger abundances make larger values of the ratio correspond to lower number densities, because lines are efficiently excited over a larger area of the clump, corresponding to lower densities.
        Also in this case, roughly estimating the average LOS fractional abundance allows to work with an appropriate set of models (cf. Sect.~\ref{subsec:density_inference}).
        We finally note that very large line ratios between high-frequency lines can only be reached if methanol abundance is high.
        This effect can be used, if we are in the right density regime, to add constraints to the abundance itself.

        Assuming a hot-core-like abundance profile has the same consequences observed for the fiducial value of the abundance, discussed in Sections~\ref{sec:results}, and \ref{subsec:hot_core}.
        We therefore do not report any results for this case.

        \section{Construction of the ML-emulation algorithms}\label{sec:ml_details}

            In this Section we provide details on the training procedure of the ML algorithms.
            We describe how we prepared the data for training, how we trained the algorithms, what performances we reached and which algorithm was selected.

            \subsection{Data preprocessing}
                Initially, we preprocessed the data by applying a base-10 logarithm transformation.
                This transformation was applied to several parameters: the average number density and temperature along the line-of-sight, the molecular column density, the reference number density and temperature, the standard deviation of the molecular number density along the LOS, and finally, either the moment zero or the line ratio (depending on the model).
                We chose the logarithmic transformation because these parameters span several orders of magnitude, which can hinder the algorithm's learning efficiency.
                All these properties must always be larger than zero, another advantage of this transformation.

                To augment the training data, we employed a binning strategy in the target variable to add columns evaluating the similarity of the input features.
                First, we divided the input data into 50 bins based on the value of the target variable.
                Within each bin, we calculated the average input features.
                Subsequently, for each data sample, we computed the cosine similarity between its input features and the average input features of its corresponding target bin.
                This approach is based on the assumption that similar input features should lead to similar output values.
                We finally applied a Quantile transformation with uniform output to the data, prior to training.

            \subsection{Model training and selection}
                The dataset was divided into training, development, and validation sub-samples.
                To ensure an accurate evaluation of the model's ability to generalize to new characteristic densities, especially in the context of recreating moment zero maps for varying densities, we deliberately avoided random selection for the development and validation sets.
                We left out all the models with a specific characteristic number density at 0.5~pc, so that the performance of creating a moment zero map for a never-seen value of this parameter could be estimated accurately.
                The development and validation data have been arbitrarily chosen to be those with characteristic number densities of $2.7
                \times 10^4 \usk\mathrm{cm^{-3}}$ and $2.187
                \times 10^6 \usk\mathrm{cm^{-3}}$, respectively.
                Our baseline was created using an AutoSklearnRegressor from the autosklearn package \citep{feurer-neurips15a, Feurer+22_jmlr23_261}, using the moment zero as the target.
                This ensemble of models reaches an MSE of $10^{-2} - 10^{-1}$ for the prediction of the logarithm of the moment zero, depending on the line.

                Given the limited dataset size (approximately $6\times10^5$ data points per grid), we opted to focus on decision-tree ensemble methods, such as Random Forest and XGBoost, rather than exploring deep neural network architectures.
                Decision-tree ensembles offered a faster hyperparameter tuning cycle and are known to provide robust results with datasets of this size.
                A grid search with a threefold cross-validation was carried out for a random forest regressor and an XGBoost \citep{xgboost} to optimize the hyperparameters.
                For these kind of models, we also applied sample weights, based on the frequency on the target variable, computed by a KDE smoothing on the training set, using an Epanechnikov kernel with a bandwidth of $0.1$ in the moment-zero log-space.
                Performances are finally evaluated on the validation dataset, which also helps to verify that the model is correctly capable of generating new images from scratch.
                The kind of model with the best performances is then selected on the basis of our validation dataset.
                The best performances were obtained with the XGBoost, reaching an MSE of $10^{-4} - 10^{-2}$, for our fiducial model and the same target as the baseline model.
                This model improved the accuracy of our emulated data between one and two orders of magnitude compared to the baseline model, depending on the specific line considered.
                The best-fit hyperparameters, as well as the scripts to reproduce the results are provided in the code repository (see Sect.~\ref{app:automation}).

                We then repeated the same training procedure, but this time using the line ratios as the target variable.
                This approach was adopted to prevent the accumulation of uncertainties associated with predicting each line individually before calculating their ratio.
                The noise in the emulated ratios is further reduced, particularly for ratios calculated using lines in the ($5_K-4_K$) and ($7_K-6_K$) bands at $241.7\usk\mathrm{GHz}$ and $338.2\usk\mathrm{GHz}$.
                For inferring the densities, we therefore rely on these models, which are the most precise.
                Those predicting the moment zero can still be useful to fine-tune calculations for specific sources; because the KDE contours are very similar between the two kinds of models, the description of the results holds also in this case.
        \section{Line fitting results}
            Tables~\ref{tab:poc_lines_3mm} and \ref{tab:poc_lines_hf} present the line properties of the lines in the $(2_K-1_K)$ and $(5_K-4_K)$ methanol bands, used in Sect.~\ref{sec:poc} to compute the line ratios and infer the density for the TOP100 sources.

            \begin{table*}[!h]
            \centering
            \small
            \caption{Line properties of the lines in the $(2_K-1_K)$ methanol band.}
            \label{tab:poc_lines_3mm}
            \begin{tabular}{lllll}
                \hline
                \hline
                Source & $T_{MB,(2_{-1}-1_{-1})}$ & $T_{MB,(2_{0}-1_{0})}$ & $T_{MB,(2_{1}-1_{1})}$ & $FWHM_{(2_K-1_K)}$ \\
                 & [K] & [K] & [K] & [km s$^{-1}$] \\
                \hline
                AGAL008.706-00.414 & $1.32 \pm 0.07$ & $0.3 \pm 0.07$ & \dots & $4.11 \pm 0.02$ \\
                AGAL010.444-00.017 & $0.61 \pm 0.06$ & $0.32 \pm 0.06$ & \dots & $5.54 \pm 0.05$ \\
                AGAL010.472+00.027 & $2.53 \pm 0.13$ & $1.89 \pm 0.13$ & $1.37 \pm 0.13$ & $7.29 \pm 0.02$ \\
                AGAL010.624-00.384 & $2.08 \pm 0.06$ & $1.25 \pm 0.06$ & $0.8 \pm 0.06$ & $7.79 \pm 0.02$ \\
                AGAL012.804-00.199 & $1.28 \pm 0.06$ & $0.61 \pm 0.06$ & $0.34 \pm 0.06$ & $6.19 \pm 0.03$ \\
                AGAL013.178+00.059 & $2.25 \pm 0.13$ & $0.83 \pm 0.13$ & \dots & $4.54 \pm 0.01$ \\
                AGAL013.658-00.599 & $1.61 \pm 0.15$ & $0.75 \pm 0.15$ & \dots & $4.98 \pm 0.03$ \\
                AGAL014.114-00.574 & $1.11 \pm 0.05$ & $0.34 \pm 0.05$ & \dots & $3.06 \pm 0.02$ \\
                AGAL014.194-00.194 & $2.66 \pm 0.21$ & $1.01 \pm 0.21$ & \dots & $5.57 \pm 0.02$ \\
                AGAL014.492-00.139 & $1.15 \pm 0.1$ & $0.35 \pm 0.1$ & \dots & $5.2 \pm 0.04$ \\
                AGAL014.632-00.577 & $1.86 \pm 0.04$ & $0.58 \pm 0.04$ & $0.21 \pm 0.04$ & $2.84 \pm 0.01$ \\
                AGAL015.029-00.669 & $0.84 \pm 0.05$ & $0.4 \pm 0.05$ & $0.2 \pm 0.05$ & $4.55 \pm 0.03$ \\
                AGAL018.606-00.074 & $1.01 \pm 0.06$ & $0.3 \pm 0.06$ & \dots & $4.58 \pm 0.02$ \\
                AGAL018.734-00.226 & $1.77 \pm 0.07$ & $0.69 \pm 0.07$ & $0.25 \pm 0.07$ & $5.36 \pm 0.02$ \\
                AGAL018.888-00.474 & $2.23 \pm 0.16$ & $0.62 \pm 0.16$ & \dots & $5.86 \pm 0.03$ \\
                AGAL019.882-00.534 & $1.68 \pm 0.08$ & $0.62 \pm 0.08$ & $0.22 \pm 0.08$ & $4.11 \pm 0.01$ \\
                AGAL022.376+00.447 & $1.86 \pm 0.12$ & $0.46 \pm 0.12$ & \dots & $4.52 \pm 0.02$ \\
                AGAL023.206-00.377 & $1.56 \pm 0.15$ & $0.78 \pm 0.15$ & \dots & $6.38 \pm 0.02$ \\
                AGAL028.564-00.236 & $1.13 \pm 0.09$ & $0.29 \pm 0.09$ & \dots & $6.42 \pm 0.04$ \\
                AGAL028.861+00.066 & $0.9 \pm 0.05$ & $0.4 \pm 0.05$ & $0.2 \pm 0.05$ & $3.91 \pm 0.02$ \\
                AGAL030.848-00.081 & $0.52 \pm 0.05$ & $0.15 \pm 0.05$ & \dots & $6.51 \pm 0.05$ \\
                AGAL031.412+00.307 & $1.29 \pm 0.09$ & $0.95 \pm 0.09$ & $0.75 \pm 0.09$ & $7.53 \pm 0.04$ \\
                AGAL034.258+00.154 & $2.53 \pm 0.1$ & $1.74 \pm 0.1$ & $1.16 \pm 0.1$ & $6.37 \pm 0.01$ \\
                AGAL034.401+00.226 & $2.53 \pm 0.11$ & $0.84 \pm 0.11$ & $0.22 \pm 0.11$ & $5.47 \pm 0.01$ \\
                AGAL034.411+00.234 & $3.27 \pm 0.25$ & $1.33 \pm 0.25$ & $0.59 \pm 0.25$ & $4.93 \pm 0.01$ \\
                AGAL034.821+00.351 & $0.62 \pm 0.03$ & $0.16 \pm 0.03$ & \dots & $3.59 \pm 0.02$ \\
                AGAL035.197-00.742 & $2.41 \pm 0.11$ & $0.87 \pm 0.11$ & $0.36 \pm 0.11$ & $4.46 \pm 0.01$ \\
                AGAL037.554+00.201 & $0.96 \pm 0.07$ & $0.44 \pm 0.07$ & $0.17 \pm 0.07$ & $5.11 \pm 0.03$ \\
                AGAL049.489-00.389 & $4.64 \pm 0.21$ & $3.55 \pm 0.21$ & $2.26 \pm 0.21$ & $8.11 \pm 0.01$ \\
                AGAL053.141+00.069 & $1.05 \pm 0.1$ & $0.34 \pm 0.1$ & \dots & $3.58 \pm 0.01$ \\
                AGAL059.782+00.066 & $1.08 \pm 0.04$ & $0.39 \pm 0.04$ & $0.17 \pm 0.04$ & $2.51 \pm 0.01$ \\
                \hline
            \end{tabular}
            \tablefoot{Only one FWHM value is listed because the fit is performed forcing all lines to have the same width. The main-beam temperature of the lines is indicated as $T_{MB,(J_K-J'_{K'})}$. A non-detection is indicated with three dots.}
        \end{table*}

        \begin{table*}[!h]
            \centering
            \small
            \caption{Line properties of the lines in the $(5_K-4_K)$ and $(7_K-6_K)$ methanol bands.}
            \label{tab:poc_lines_hf}
            \begin{tabular}{lllllll}
                \hline
                \hline
                Source & $T_{MB,(5_{0}-4_{0})}$ & $T_{MB,(5_{-1}-4_{-1})}$ & $FWHM_{(5_K-4_K)}$ & $T_{MB,(7_{0}-6_{0})}$ & $T_{MB,(7_{-1}-6_{-1})}$ & $FWHM_{(7_K-6_K)}$ \\
                 & [K] & [K] & [km s$^{-1}$] & [K] & [K] & [km s$^{-1}$] \\
                \hline
                AGAL008.706-00.414 & \dots & $0.47 \pm 0.04$ & $4.84 \pm 0.49$ & \dots & $0.28 \pm 0.06$ & $3.07 \pm 0.4$ \\
                AGAL010.444-00.017 & $0.14 \pm 0.04$ & $0.36 \pm 0.04$ & $6.03 \pm 0.39$ & \dots & $0.19 \pm 0.06$ & $6.12 \pm 0.71$ \\
                AGAL010.472+00.027 & $1.89 \pm 0.15$ & $2.64 \pm 0.15$ & $9.1 \pm 0.58$ & $2.21 \pm 0.15$ & $2.33 \pm 0.15$ & $10.04 \pm 0.28$ \\
                AGAL010.624-00.384 & $2.26 \pm 0.15$ & $3.79 \pm 0.15$ & $8.34 \pm 0.5$ & $2.64 \pm 0.14$ & $4.14 \pm 0.14$ & $8.19 \pm 0.14$ \\
                AGAL012.804-00.199 & $1.04 \pm 0.09$ & $2.17 \pm 0.09$ & $5.92 \pm 0.39$ & $0.9 \pm 0.09$ & $1.52 \pm 0.09$ & $5.39 \pm 0.13$ \\
                AGAL013.178+00.059 & $0.59 \pm 0.09$ & $1.74 \pm 0.09$ & $5.1 \pm 0.43$ & $0.44 \pm 0.09$ & $1.3 \pm 0.09$ & $5.07 \pm 0.12$ \\
                AGAL013.658-00.599 & $0.47 \pm 0.1$ & $1.08 \pm 0.1$ & $6.36 \pm 0.59$ & $0.41 \pm 0.09$ & $0.71 \pm 0.09$ & $6.49 \pm 0.22$ \\
                AGAL014.114-00.574 & $0.19 \pm 0.04$ & $1.04 \pm 0.04$ & $3.02 \pm 0.23$ & $0.21 \pm 0.07$ & $0.69 \pm 0.07$ & $2.6 \pm 0.15$ \\
                AGAL014.194-00.194 & $0.67 \pm 0.13$ & $1.89 \pm 0.13$ & $6.16 \pm 0.52$ & $0.72 \pm 0.13$ & $1.7 \pm 0.13$ & $5.53 \pm 0.1$ \\
                AGAL014.492-00.139 & $0.18 \pm 0.06$ & $0.59 \pm 0.06$ & $5.48 \pm 0.5$ & \dots & $0.29 \pm 0.08$ & $6.42 \pm 0.49$ \\
                AGAL014.632-00.577 & $0.57 \pm 0.05$ & $1.89 \pm 0.05$ & $3.01 \pm 0.22$ & $0.3 \pm 0.08$ & $1.26 \pm 0.08$ & $3.07 \pm 0.08$ \\
                AGAL015.029-00.669 & $0.71 \pm 0.05$ & $1.8 \pm 0.05$ & $4.84 \pm 0.35$ & $0.51 \pm 0.08$ & $1.4 \pm 0.08$ & $5.38 \pm 0.12$ \\
                AGAL018.606-00.074 & $0.2 \pm 0.08$ & $0.66 \pm 0.08$ & $4.28 \pm 0.42$ & $0.18 \pm 0.08$ & $0.47 \pm 0.08$ & $5.09 \pm 0.28$ \\
                AGAL018.734-00.226 & $0.41 \pm 0.06$ & $1.18 \pm 0.06$ & $5.65 \pm 0.45$ & $0.51 \pm 0.09$ & $1.05 \pm 0.09$ & $5.87 \pm 0.14$ \\
                AGAL018.888-00.474 & $0.47 \pm 0.07$ & $1.32 \pm 0.07$ & $5.88 \pm 0.46$ & $0.3 \pm 0.11$ & $0.99 \pm 0.11$ & $5.0 \pm 0.19$ \\
                AGAL019.882-00.534 & $0.94 \pm 0.09$ & $2.29 \pm 0.09$ & $3.84 \pm 0.28$ & $1.68 \pm 0.13$ & $3.19 \pm 0.13$ & $4.15 \pm 0.03$ \\
                AGAL022.376+00.447 & $0.27 \pm 0.07$ & $0.96 \pm 0.07$ & $5.12 \pm 0.48$ & $0.23 \pm 0.05$ & $0.68 \pm 0.05$ & $5.05 \pm 0.16$ \\
                AGAL023.206-00.377 & $0.88 \pm 0.13$ & $1.63 \pm 0.13$ & $7.8 \pm 0.61$ & $1.26 \pm 0.11$ & $1.79 \pm 0.11$ & $6.99 \pm 0.09$ \\
                AGAL028.564-00.236 & \dots & $0.53 \pm 0.05$ & $6.83 \pm 0.66$ & \dots & $0.16 \pm 0.05$ & $7.78 \pm 0.91$ \\
                AGAL028.861+00.066 & N/A & N/A & N/A & $0.46 \pm 0.08$ & $0.88 \pm 0.08$ & $3.97 \pm 0.12$ \\
                AGAL030.848-00.081 & $0.11 \pm 0.05$ & $0.41 \pm 0.05$ & $6.86 \pm 0.57$ & \dots & $0.31 \pm 0.06$ & $7.53 \pm 0.68$ \\
                AGAL031.412+00.307 & $1.27 \pm 0.1$ & $1.67 \pm 0.1$ & $8.56 \pm 0.57$ & $2.14 \pm 0.13$ & $2.68 \pm 0.13$ & $7.51 \pm 0.2$ \\
                AGAL034.258+00.154 & $2.83 \pm 0.14$ & $4.54 \pm 0.14$ & $7.02 \pm 0.44$ & $4.22 \pm 0.17$ & $5.11 \pm 0.17$ & $6.97 \pm 0.12$ \\
                AGAL034.401+00.226 & $0.93 \pm 0.11$ & $2.53 \pm 0.11$ & $5.77 \pm 0.46$ & $0.82 \pm 0.09$ & $2.04 \pm 0.09$ & $5.58 \pm 0.06$ \\
                AGAL034.411+00.234 & $1.38 \pm 0.16$ & $2.99 \pm 0.16$ & $5.15 \pm 0.41$ & $1.96 \pm 0.22$ & $3.25 \pm 0.22$ & $4.98 \pm 0.06$ \\
                AGAL034.821+00.351 & $0.1 \pm 0.04$ & $0.45 \pm 0.04$ & $3.1 \pm 0.28$ & \dots & $0.17 \pm 0.06$ & $6.81 \pm 0.83$ \\
                AGAL035.197-00.742 & $0.99 \pm 0.08$ & $2.57 \pm 0.08$ & $4.82 \pm 0.35$ & $1.03 \pm 0.18$ & $2.26 \pm 0.18$ & $4.68 \pm 0.05$ \\
                AGAL037.554+00.201 & $0.37 \pm 0.06$ & $0.88 \pm 0.06$ & $5.27 \pm 0.48$ & $0.52 \pm 0.07$ & $0.59 \pm 0.07$ & $5.28 \pm 0.21$ \\
                AGAL049.489-00.389 & $5.47 \pm 0.36$ & $8.09 \pm 0.36$ & $8.95 \pm 0.54$ & $8.63 \pm 0.49$ & $10.38 \pm 0.49$ & $8.9 \pm 0.17$ \\
                AGAL053.141+00.069 & $0.49 \pm 0.08$ & $1.36 \pm 0.08$ & $4.82 \pm 0.35$ & $0.65 \pm 0.11$ & $1.48 \pm 0.11$ & $4.28 \pm 0.08$ \\
                AGAL059.782+00.066 & N/A & N/A & N/A & $0.44 \pm 0.09$ & $0.98 \pm 0.09$ & $2.93 \pm 0.11$ \\
                \hline
            \end{tabular}
               \tablefoot{Only one FWHM value is listed per band because the fit is performed forcing all lines to have the same width. The main-beam temperature of the lines is indicated as $T_{MB,(J_K-J'_{K'})}$. A non-detection is indicated with three dots, while missing data are indicated with ``N/A''.}
        \end{table*}
    \end{appendix}

\end{document}